\documentclass[a4paper,11pt]{article}
\pdfoutput=1 

\usepackage{jheppub} 
\usepackage[bottom]{footmisc}
\usepackage{amssymb}
\usepackage{amsmath}
\usepackage{amsthm}
\usepackage[usenames,dvipsnames]{xcolor}
\usepackage{epsfig}
\usepackage{dcolumn}
\usepackage{tikz}
\usetikzlibrary{shapes.geometric, arrows}
\usepackage{upgreek}
\usepackage{setspace}
\usepackage{subfig}
\usepackage{array,multirow,bigdelim,arydshln}
\usepackage{appendix}
\usepackage{xparse}
\usepackage{stmaryrd}
\usepackage[T1]{fontenc} 
\usepackage{mathtools}
\usepackage{physics} 
\usepackage{adjustbox}
\usepackage{multirow}
\usepackage{graphicx} 
\usepackage{float} 
\graphicspath{{./images/}}
\usepackage[nottoc]{tocbibind}
\usepackage{hyperref}
\usepackage[utf8]{inputenc}
\usepackage{CJK}
\hypersetup{
	colorlinks,
	urlcolor=Maroon,
	linkcolor=Maroon,
	citecolor=Maroon
	}

\NewDocumentCommand{\binomial}{omm}
 {%
  \genfrac(){0pt}{}{#2}{#3}%
  \IfValueT{#1}{_{\!#1}}%
 }
\NewDocumentCommand{\eulerian}{omm}
 {%
  \genfrac<>{0pt}{}{#2}{#3}%
  \IfValueT{#1}{_{\!#1}}%
 }

\usepackage{latexsym}
\usepackage{tikz}
\newcommand*\diff{\mathop{}\!\mathrm{d}}

\theoremstyle{plain}

\theoremstyle{definition}

\def\bea#1\eea{\begin{eqnarray}#1\end{eqnarray}}
\def\be#1\ee{\begin{equation}#1\end{equation}}
\def\ba#1\ea{\begin{align}#1\end{align}}

\usepackage{amsmath}
\usepackage{multicol}
\usepackage{bbm}
\usepackage{enumerate}

\usepackage{amsthm}
\usepackage{mathrsfs}
\usepackage{upgreek}
\usepackage{amssymb}
\usepackage{bm}
\usepackage{setspace}
\usepackage{array,multirow,arydshln}
\usepackage{bigdelim}
\usepackage{scalerel}
\usepackage{diagbox}

\usepackage{tabularx}

\usepackage{tikz}

\usetikzlibrary{shapes.geometric,arrows,arrows.meta,decorations.pathmorphing,decorations.markings,patterns}

\def\<{\langle}
\def\>{\rangle}

\newcommand{\Surf}[0]{\mathcal{S}}
\newcommand{\Sone}[0]{\mathcal{S}_1}
\newcommand{\Stwo}[0]{\mathcal{S}_2}

\usepackage[percent]{overpic}
\usepackage{multirow} 
\usepackage{slashed}

\title{All-order splits and multi-soft limits for particle and string amplitudes}

\author[a]{Nima Arkani-Hamed,}

\affiliation[a]{School of Natural Sciences, Institute for Advanced Study, Princeton, NJ, 08540, USA}
\author[b]{Carolina Figueiredo}
\affiliation[b]{Jadwin Hall, Princeton University, Princeton, NJ, 08540, USA}

\emailAdd{arkani@ias.edu}
\emailAdd{cfigueiredo@princeton.edu}

\abstract{The most important aspects of scattering amplitudes have long been thought to be associated with their poles. But recently a very different sort of ``split'' factorizations for a wide range of particle and string tree amplitudes have been discovered away from poles. In this paper, we give a simple, conceptual origin for these splits arising from natural properties of the binary geometry of the curve integral formulation for scattering amplitudes for Tr$(\Phi^3)$ theory. The most natural way of ``joining'' smaller surfaces to build larger ones directly produces a choice of kinematics for which higher amplitudes factor into lower ones. This gives a generalization of splits to all orders in the topological expansion. These splits allow us to access and compute loop-integrated multi-soft limits for particle and string amplitudes, at all loop orders. This includes split factorizations and multisoft limits for pion and gluon amplitudes, that are related to Tr$(\Phi^3)$ theory by a simple kinematical shift.}

\begin{document}
\maketitle
\addtocontents{toc}{\protect\setcounter{tocdepth}{2}}

\numberwithin{equation}{section}

		\tikzset{
		particles/.style={dashed, postaction={decorate},
			decoration={markings,mark=at position .5 with {\arrow[scale=1.5]{>}}
		}}
	}
	\tikzset{
		particle/.style={draw=black, postaction={decorate},
			decoration={markings,mark=at position .5 with {\arrow[scale=1.1]{>}}
		}}
	}
	\def  \layersep {.6cm}

\section{Splitting amplitudes from joining surfaces}
\label{sec:Intro}

The past year has seen an avalanche of progress in our understanding of scattering amplitudes for non-supersymmetric theories, moving ever closer to describing the real world~\cite{curveint,zeros, Gluons,NLSM,JaraSoftTheorem,CirclesNLSM,SongSplit,JaraZeros,Zeros_Li}. These developments have also exposed new qualitative properties of scattering amplitudes,  including surprising factorizations and associated hidden zeros of tree amplitudes for particles and strings on simple loci in kinematic space~\cite{zeros}. The existence of the zeroes in particular provided a powerful clue and motivation for the discovery of the surprising link between the amplitudes for Tr($\Phi^3$) theory, pions, and ``scalar-scaffolded'' gluons, seen to all arise from a single ``stringy'' amplitude with simple shifts of kinematic data. The kinematic shift was uniquely picked out by preserving the hidden zeros; while this connection was motivated by the zeros at tree-level, the shift generalizes to all loop orders~\cite{NLSM,Gluons}. 

The most general ``novel factorization'' phenomenon seen at tree-level is the factorization of the amplitudes in the ``split'' kinematics first defined in~\cite{FreddySplit} for the biadjoint scalar amplitudes, and more recently studied in greater generality for various stringy tree amplitudes in~\cite{SongSplit}. The other factorizations near zeros~\cite{zeros} follow from further specializations of these splits~\cite{SongSplit}. Given the crucial role of the tree-level factorizations and especially the associated zeros have played in recent developments, it is clearly important to understand the origin of these splits as transparently and conceptually as possible, with the hope of extending them to all orders in the topological expansion, for both integrands and loop integrated amplitudes. 

This is what we will do in this paper. We will find that the splits have an extremely simple, natural, and conceptual origin in ``surfaceology'' and the ``u-variables'' of the curve-integral formalism for scattering amplitudes~\cite{curveint}, allowing us to make general statements to all orders in the topological expansion. We will also see that these generalized splits lead to simple expressions for multi-soft limits of loop-integrated Tr$(\Phi^3)$ amplitudes at all orders in the topological expansion. Given the connection between shifted Tr$(\Phi^3)$, the NLSM, and scalar-scaffolded gluons, these provide all-loop order multi-soft results for all these theories. 

The story is so simple that we will explain it here with a single picture and a few accompanying sentences. The rest of the paper will provide a detailed exposition of the precise statements and give explicit examples, but nothing essential will be added beyond what is contained in the picture we now describe. 
\\ \\
\noindent{\bf The $u$ variables and curve-integrals} -- to begin with, we recall that for a general surface ${\cal S}$ defining a given order in the topological expansion, the ``stringy'' integral for amplitudes in the Tr$(\Phi^3)$ theory is expressed as~\cite{curveint,StringForms,StringyCanonicalForms} 
\begin{equation}
{\cal A} = \int_0^{+\infty}\prod_i \frac{dy_i}{y_i} \prod_X u_X^{\alpha^\prime X}(y_i),
\label{eq:curveInt}
\end{equation}
where $\alpha^\prime$ is the string constant that we will set to one in the rest of the paper. The integration variables, $y_i's$, are positive variables associated with the internal edges of a representative fat-graph for ${\cal S}$, and $u_X$ are the ``u-variables'' associated with every curve $X$ on the surface. Each curve on the surface is in one-to-one correspondence to a propagator appearing in the diagrammatic expansion. So $X$ denotes the squared momentum of the propagator associated with the curve that we draw on the surface. Ultimately, the $u_X's$ satisfy the ``u-equations'' 
\begin{equation}
u_X + \prod_Y u_Y^{{\rm int}(Y,X)} = 1,
\end{equation}
where ${\rm int}(Y,X)$ is the number of times the curves $Y$ and $X$ intersect on ${\cal S}$. There is a simple formalism for finding the space of solutions for these $u$ equations, which can be expressed as $u_X(y_i)$ in terms of the $y_i$ variables (see \cite{curveint}, or \cite{Gluons,tropL} for quick summaries of this procedure). 
\\ \\
\noindent{\bf Seeing surfaces inside surfaces} -- the $u$ variables have many remarkable properties that will be described at length in \cite{curvy}, but in this note, we will need one especially simple and natural one. Suppose we have a surface $\mathcal{S}$, and a subsurface $s$ contained inside it. We can concretely define a subsurface $s$ simply by giving its fat-graph as a subgraph of the parent fat-graph for ${\cal S}$. It is natural to wonder whether we can build out of the $u$ variables, $u^{(\mathcal{S})}_X$, for curves $X$ on the big surface $\mathcal{S}$, a set of objects $u^{(s)}_x$ that satisfy the $u$ equations labeled by curves $x$ on the smaller surface $s$. This question has a beautiful answer: we can build $u$ variables for the small surface out of simple monomials in the $u$ variables for the big one. Given a curve $x$ in $s$, we simply consider {\it all} ways of extending $x$ into curves on the big surface, $X_x$, and take the product over all these $u_{X_x}$ variables. In more detail, the curve $x$ can occur as a subcurve of $X_x$ in many ways, and we must include a factor of $u_{X_x}$ for every different way in which $x$ can be realized within $X$. Finally, we can write  
\begin{equation}
u_x = \prod_X u_X^{\#[x \subset X]}
\label{eq:extenduIntro}
\end{equation}
where $\#[x \subset X]$ is the number of different ways in which $x$ can occur as a subcurve of $X$; of course if $X$ is not an extension of $x$ we have $\#[x \subset X] = 0$. 

We can think of the $u$ equations themselves as a special case of this general fact. We simply identify a sub-surface corresponding to the four-point tree amplitude.  The $u$ equations for this subsurface are trivially $U + V = 1$. To reproduce the usual $u$ equations, we choose the four-point subsurface to have the property that the curve corresponding to $U$ starts and ends on actual boundaries of the big surface ${\cal S}$, while the curve corresponding to $V$ is extended into ${\cal S}$. The product formula for $V$ then gives the second term in the $u$ equation, with the exponents ${\rm int}(Y,X)$ arising from the $\#[x \subset X]$ exponents in \eqref{eq:extenduIntro}. If we choose a more general four-point subsurface we get ``generalized'' u equations, where both $U, V$ are interesting products over the $u$ variables of ${\cal S}$. The interested reader can refer to appendix \ref{sec:UAppendix} for a few worked examples of subsurface $u$ variables and generalized $u$ equations. 
\begin{figure}[t]
    \centering
    \includegraphics[width=\textwidth]{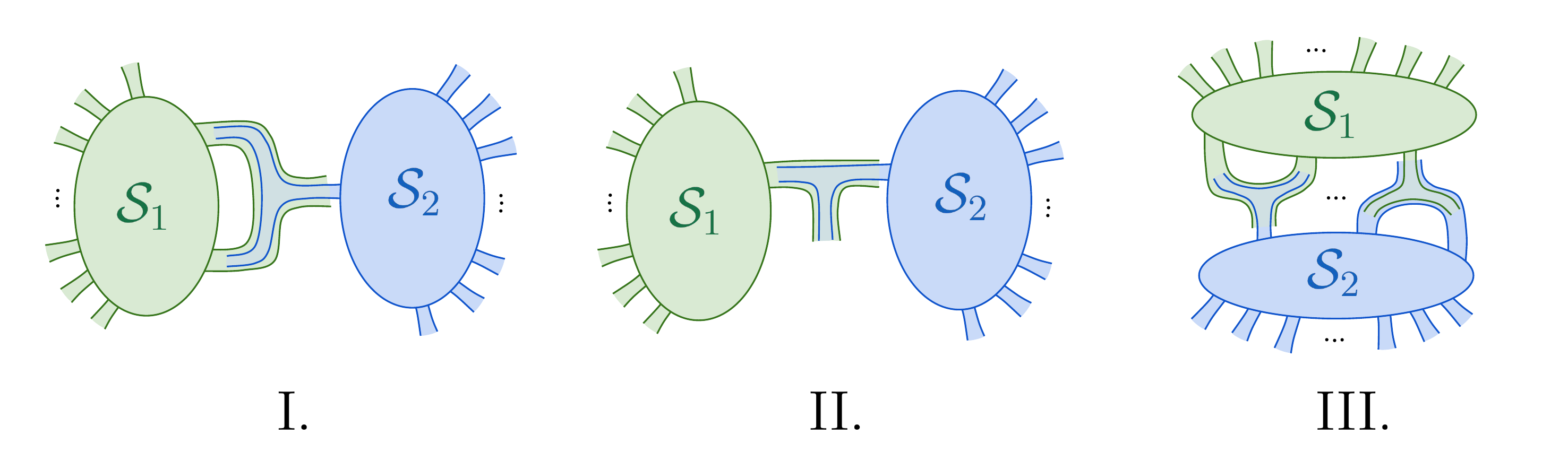}
    \caption{Possible ways of overlapping two surfaces $\mathcal{S}_1$ and $\mathcal{S}_2$ to obtain a bigger one, $\mathcal{S}= \mathcal{S}_1 \otimes \mathcal{S}_2$, without adding internal edges. Note that case II. is a special case of I., in which the bottom edge in the overlap I. is a boundary. }
    \label{fig:overlaps}
\end{figure}
\\ \\
\noindent{\bf Making bigger surfaces from joining smaller ones} -- Having understood how the $u$-variables for subsurfaces can be built from those of the full surface, it is trivial to {\it build} kinematics for a big surface that leads to factorizing amplitudes. Consider a surface ${\cal S}$, and let us identify two subsurfaces ${\cal S}_1$, ${\cal S}_2$, with the property that every internal edge of the fatgraph representative for ${\cal S}$ is an internal edge in exactly one of ${\cal S}_1$ or ${\cal S}_2$. 

In practice, instead of starting with ${\cal S}$ and identifying ${\cal S}_1, {\cal S}_2$, we can start with any ${\cal S}_{1,2}$ we like and naturally join them to build ${\cal S}$, that we indicate with the notation 
\begin{equation}
{\cal S} = {\cal S}_1 \otimes {\cal S}_2.
\end{equation}

We simply have to ensure that in joining ${\cal S}_1, {\cal S}_2$, we don't produce any new internal edges.
In terms of fatgraphs, this can be done by making the two surfaces overlap on a common vertex, as shown in \ref{fig:overlaps}. In panel I, the vertex has one boundary edge for ${\cal S}_1$ and two boundary edges for ${\cal S}_2$, so that no internal edges are created in overlapping the two fatgraphs. Said in the language of surfaces, ${\cal S}_1,{\cal S}_2$ are to made to overlap on a common triangle. In panel II we show a special case where the overlapping vertex/triangle contains two boundaries of each surface. Finally there is nothing special about overlapping on a single vertex/triangle. In panel III we show an example in which instead of a single overlap, $\mathcal{S}_1$ and $\mathcal{S}_2$ overlap on multiple triangles. Note that by performing this type of overlap, the resulting surface, $\mathcal{S}= \mathcal{S}_1 \otimes \mathcal{S}_2$, will be of higher genus. 

These are then the natural ways of joining two surfaces to build a bigger surface, without producing any new edges and having all internal edges of the big surface belong to exactly one of the two subsurfaces. 
\\ \\
\noindent {\bf Building split kinematics}--
The simple picture of joining surfaces simply and beautifully defines split kinematics on which the amplitude for the big surface is guaranteed to factorize into the product of the amplitude for the two smaller ones.  

\begin{figure}[t]
    \centering
    \includegraphics[width=0.9\textwidth]{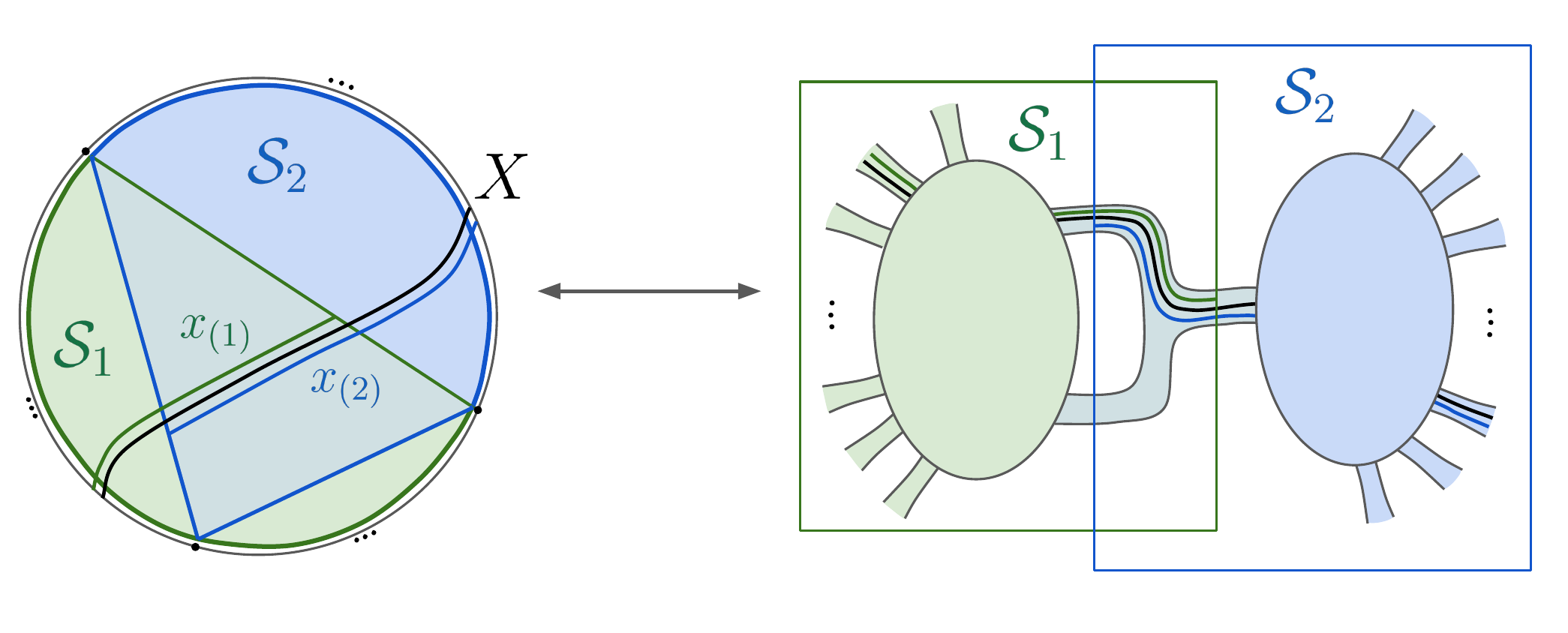}
    \caption{Two subsurfaces, $\mathcal{S}_1$ (green) and $\mathcal{S}_2$ (blue), overlapping on a triangle (left) and their fatgraph representation (right). A given curve on the big surface, $X$, can be decomposed into curves on the smaller surfaces $x_{(1)}$, $x_{(2)}$.}
    \label{fig:curvedecomp}
\end{figure}

Let's consider the product of the curve integrals for ${\cal S}_1$ and ${\cal S}_2$. We then express the corresponding $u$ variables $u_{x_{(1)}}, u_{x_{(2)}}$ as the product of the extension of the curves from ${\cal S}_1,{\cal S}_2$ into the full ${\cal S}$ respectively. This then automatically {\it defines} kinematics $X$ for the big surface, on which the curve integral over the big surface ${\cal S}$ factorizes into that for ${\cal S}_1$ and ${\cal S}_2$! 
Indeed, we have that 
\begin{equation}
\begin{aligned}
\prod_X u_X^X &= \left(\prod_{x_{(1)}\in \mathcal{S}_1} u_{x_{(1)}}^{x_{(1)}} \right) \times \left(\prod_{x_{(2)}\in \mathcal{S}_2} u_{x_{(2)}}^{x_{(2)}} \right)  \\
&= \left(\prod_{x_{(1)}\in \mathcal{S}_1} \prod_X u_X^{\#[x_{(1)} \subset X] x_{(1)}}\right) \times \left(\prod_{x_{(2)}\in \mathcal{S}_2} \prod_X u_X^{\#[x_{(2)} \subset X] x_{(2)}}\right), 
\end{aligned}
\end{equation}
which lets us read off the kinematics for ${\cal S}$ as 
\begin{equation}
X_{{\cal S} = {\cal S}_1 \otimes {\cal S}_2}[x_{(1)},x_{(2)}] = \sum_{x_{(1)}} \#[x_{(1)} \subset X] x_{(1)} + \sum_{x_{(2)}} \#[x_{(2)} \subset X] x_{(2)}.
\label{eq:kinMap}
\end{equation}

Said in words, the curve $X$ on the big surface ${\cal S}$ decomposes into certain curves (in general there may be more than one of them) $x_{(1,2)}$ on ${\cal S}_{1,2}$, which can occur with multiplicity $\#(x_{(1,2)} \subset X)$ respectively. Then the kinematics for $X$ is simply the sum over all the $x_{(1,2)}$ including their multiplicities. Figure \ref{fig:curvedecomp} illustrates the way curves on the joined surface decompose into curves on the smaller ones, defining the split kinematics. 

This then gives the general expression for split amplitudes, 
\begin{equation}
{\cal A}_{\cal S}\left[X_{{\cal S} = {\cal S}_1 \otimes {\cal S}_2} \right] = {\cal A}_{{\cal S}_1}\left[x_{(1)}\right] \times {\cal A}_{{\cal S}_2}\left[x_{(2)}\right].
\label{eq:split}
\end{equation}
Note that the total number of kinematic variables for ${\cal S}_1$ and ${\cal S}_2$ is smaller than those for ${\cal S}$, and so split kinematics imposing certain linear relations between the kinematic variables of ${\cal S}$. Note also that every curve on ${\cal S}$ is assigned a non-zero kinematic exponent, simply because every curve $X$ must obviously have a component in at least one of ${\cal S}_1$ or ${\cal S}_2$. 
\\ \\ 
\noindent{\bf Splits for the Surface Integrand}-- The split kinematics we have defined most obviously defines a locus in kinematic space for which the {\it surface integrands} factorize. The conventional notion of ``the'' loop integrand is famously defined only in the planar limit, and our split kinematics implies novel factorizations for the loop integrand in this case. 

These factorizations are inherited by any shift of the kinematic variables that leaves the linear relations imposed by split kinematics invariant.
A wide range of these shifts exist, importantly including the ``even-even, odd-odd'' $\delta$-shift of \cite{zeros, NLSM, Gluons} that relates Tr$(\Phi^3)$, pion and scalar-scaffolded gluon amplitudes, so these splits extend to all-loop order factorizations for these other theories as well. 

An aspect of the curve integral formalism \cite{curveint} is that for {\it any} surface, planar or not, there is a notion of an ``infinite loop integrand'', summing over all mapping-class-group images of triangulations of the surface, with factorization on poles correctly reflecting cutting the surface along the corresponding curves. Our splits also define factorizations on loci in kinemaic space for these interesting objects even beyond the planar limit. 
\\ \\ 
{\bf Splits for physical kinematics}--The surface integrand represents a generalization of the conventional loop integrand in momentum space. For instance, working in terms of standard momenta, some curves on the surface, such as those corresponding to tadpole propagators,  are assigned zero momentum. Instead in the surface integrand each curve on the surface is assigned its own kinematic $X$ variable. It is this feature of surface variables that crucially allows them to give canonically well-defined loop integrands with good properties, such as consistent factorization on cuts, as well as the Adler zero and gauge invariance for NLSM and YM integrands respectively \cite{curveint,Gluons,NLSM,CirclesNLSM}. But ultimately in order to compare surface integrand with physical amplitudes, we must map the surface kinematic variables to physical ones. This map imposes further restriction on the surface kinematics that we will explore in detail at one-loop. 
\\ \\
\noindent{\bf Splits and all-order loop-integrated multi-soft limits}--The locus in kinematic space defined by splits at loop level puts restrictions on the loop momenta, and hence doesn't extend to loop integrated statements. However if the kinematic variables in one of the subsurfaces are sent to zero, the constraints on loop momenta disappear and the split factorization holds post loop integration. This corresponds to a ``multi-soft'' limit where the particle momenta of the subsurface are taken to vanish. Of course in  massless Tr$(\Phi^3)$ theory this limit is divergent. Nonetheless we can make the limit well-defined by adding masses to the propagators of ${\cal S}_{1,2}$ which in turn defines a set of masses for ${\cal S}$.  This gives us remarkable statements about factorizations of loop-integranted amplitudes in generalized soft limits at all order orders in the topological expansion. Since mass shifts of field-theoretic Tr$(\Phi^3)$  give us amplitudes for the non-linear sigma model, we can derive new expressions for all-loop order multi-soft limits of the NLSM. These results hold directly for ``surface-soft'' limits of the surface integrand, and we will show how they non-trivially imply the same statements for the integrand with physical kinematics. We will also make some brief comments on soft-limits and splits for scalar-scaffolded YM from kinematic shifts of the stringy Tr$(\Phi^3)$ amplitudes. 

\section{Tree-level Splits}

Let's now see what the split translates to at tree-level. In this case, the kinematic locus for which we get a split is particularly simple since a curve on the big surface is always decomposable into at most a single curve in each subsurface, $i.e.$ $\#[x_{(1)}\in X],\#[x_{(2)}\in X]=0 \, {\rm or} \, 1$. In addition, as we will see momentarily, at tree-level, we can further interpret the split kinematics as setting to zero a collection of non-planar variables, $c_{i,j}=-2p_i \cdot p_j$ (with $i$ and $j$ non-adjacent indices). Doing this for a general split pattern we conclude that the kinematic mapping we obtain from asking for the surface split is precisely the one observed in previous works \cite{FreddySplit,SongSplit}.  

\subsection{Preamble on Fatgraphs and Curves}
To begin we review the most basic features of ``surfaceology'' relevant already at tree-level --- the definition of a surface via a fatgraph, and the identification of kinematic variables with curves on the surface or paths traveled in the fatgraph \footnote{More extensive and detailed explanations can be found in \cite{curveint,Gluons,tropL}}. 

To begin with, the picture of the ``momentum polygon`` for tree amplitudes is a very familiar one. Given a color ordering, we imagine drawing the $n$ momenta $p_1^\mu, \cdots, p_n^\mu$ tip to toe, with momentum conservation giving us a closed polygon. If the vertices of this $n$-gon are $x_i^\mu$, we can write $p_i^\mu = (x_{i+1}^\mu - x_i^\mu)$. The squared distance $X_{i,j} = (x_i - x_j)^2 = (p_i + p_{i+1} + \cdots p_{j-1})^2$ between the vertices $(i,j)$ of the polygon has the by now very familiar interpretation as giving the planar propagators seen for the amplitudes in this color ordering. We can also record this data in a two-dimensional picture of a polygon or what is the same, a disk with $n$ marked points on its boundary, identifying chords $(i,j)$ with propagators. Feynman diagrams are associated with a maximal set of $(n-3)$ chords that are mutually non-intersecting and triangulate this polygon/disk. Starting from the triangulation we can recover the Feynman diagram by drawing the dual diagram to the triangulation. 

But in the general story of surfaceology, and for our purposes in this paper, we will need a closely related but different way of thinking about labeling the kinematic data. Instead of ``chords'' that start and end on the vertices $(i,j)$, we will think of what are referred to in the mathematical literature as ``laminations'' but which we will more simply refer to as ``curves''. These start and end not on the vertices $i,j$ on the boundary on the disk, but instead on boundary segments $(i,i+1), (j,j+1)$. There is evidently a one-to-one correspondence between chords and curves: to get a curve from a chord, we simply push the two endpoints of the chord clockwise, to go from boundary vertices to boundary edges. This correspondence exists for any orientable surface, since the orientation consistently defines what pushing the endpoints ``clockwise'' means everywhere. From this connection, we can read off momenta associated with curves and also think of Feynman diagrams as being associated with maximal collections of non-intersecting curves. 

There is also a natural motivation for thinking about curves making no reference to their connection with chords. We can think of any double-line notation diagram, or more simply ``fatgraph'', as  {\it defining} a surface, by specifying a triangulation for it. The vertices of the fatgraph correspond to triangles of the triangulation, and color lines are the edges of these triangles. The propagators indicate the way the triangles are glued along their edges to build the triangulation. In this picture, it is natural to consider paths traversing through the fatgraph moving along the ``roads'' which are the edges. The paths that start and end on boundaries are just the ``laminations'' or ``curves''. 

At tree-level, it is natural to label the regions associated with each color line as $1,2,\cdots,n$, and every edge is bounded by two labels $(i,j)$. There is correspondingly a curve on the full surface associated with any edge $e$:   we simply walk out of the edge in both directions, and continuously turn left until we reach a boundary. At tree-level, one of these boundaries will be $(i,i+1)$ and the other $(j,j+1)$. More generally for any surface, the labeling of edges in a simple way analagous to ``$(i,j)$'' will not be possible, but every edge is canonically associated with a curve in exactly the same way. 

We can also define momenta associated with any curve in a simple way. We begin by assigning momenta to the edges of the fatgraph in the usual way. Now any curve $X$ begins in some boundary, and goes into the graph, turning left or right at the vertices it encounters. To obtain the momentum $p_X^\mu$, we start with the momentum of its starting edge and then walk into the graph. If we enter a vertex where we turn right, we add the momentum of the graph entering the vertex from the left, and if we turn left we do nothing. Abusing notation, we will refer to the kinematic variable associated with the curve $X$ with the same notation $X = p_X^2$. 
As a check note that since the curve associated with any edge begins from a boundary and turns right continuously till it hits the edge, this curve is assigned the sum of all the momenta encountered up to the edge, which is precisely the usual momentum assigned to the edge itself. 

\subsection{Examples: 6 and 5 point splits}
\begin{figure}[t]
    \centering
    \includegraphics[width=\textwidth]{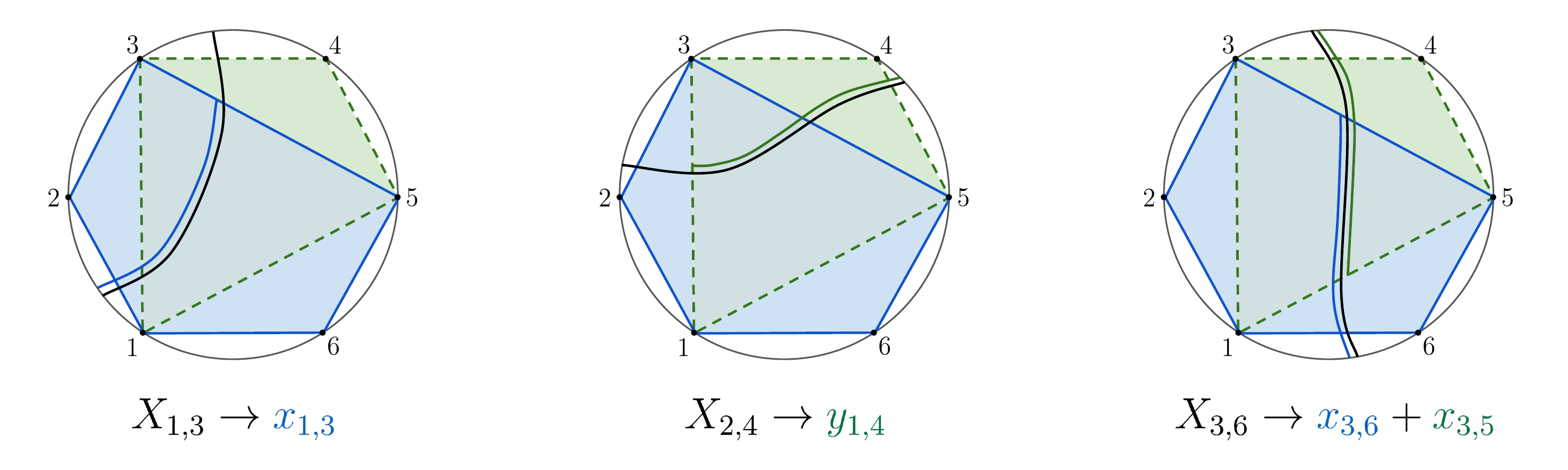}
    \caption{Kinematical mappings for the split of 6-point tree amplitude into a 4-point tree, $\mathcal{S}_1$ (green), times a 5-point tree, $\mathcal{S}_2$ (blue).}
    \label{fig:6ptTree}
\end{figure}

Let's now look at the simple case of a 6-point amplitude where the surface, $\mathcal{S}$ is given by the disk with 6 marked points on the boundary, and consider the split into two subsurfaces: $\mathcal{S}_1: (1,3,4,5)$ and  $\mathcal{S}_2: (1,2,3,5,6)$. In this case, the two subsurfaces intersect in triangle $(1,3,5)$ which does not contain any boundary (see figure \ref{fig:6ptTree}). 

Following the kinematic mapping defined previously, \eqref{eq:kinMap}, we get the following kinematic locus for this 6-point split:
\begin{equation}
   \begin{aligned}
    &X_{1,3} \to x_{1,3}, \\
    &X_{1,4} \to x_{1,3} + y_{1,4},\\
    &X_{1,5} \to  x_{1,5}, 
\end{aligned} \quad \quad \quad 
\begin{aligned}
    &X_{2,4} \to y_{1,4}, \\
    &X_{2,5} \to  x_{2,5},\\
    &X_{2,6} \to x_{2,6}, 
\end{aligned} \quad \quad \quad 
\begin{aligned}
    &X_{3,5} \to y_{3,5}, \\
    &X_{3,6} \to y_{3,5} + x_{3,6},\\
    &X_{4,6} \to x_{3,6},
\end{aligned} 
\label{eq:6ptTreeKin}
\end{equation}
where we use $y_{i,j}$ to denote the kinematics of the $\mathcal{S}_1$ and $x_{i,j}$ those of $\mathcal{S}_2$. A graphical explanation for some of these mappings can be seen in figure \ref{fig:6ptTree}. In this locus the stringy integral becomes:
\begin{equation}
    \mathcal{A}_6 \to \mathcal{A}_4 (y_{1,4},y_{3,5}) \times \mathcal{A}^{\alpha^\prime}_5 (x_{1,3},x_{1,5},x_{2,5},x_{2,6},x_{3,6}).
\end{equation}

In the field theory limit, the $6$-point amplitude is given by the sum of 14 diagrams:
\begin{equation}
\begin{aligned}
    {\cal A}_{6} = &\frac{1}{X_{1,3}X_{1,4}X_{1,5}} +  \frac{1}{X_{1,3}X_{3,6}X_{4,6}} + ({\rm cyclic}) + \frac{1}{X_{1,3}X_{3,5}X_{1,5}} + \frac{1}{X_{2,4}X_{4,6}X_{2,6}} .
\end{aligned}
\end{equation}
that under the kinematic limit \eqref{eq:6ptTreeKin} becomes
\begin{equation}
     \mathcal{A}_6 \to \left(\frac{1}{y_{1,4}} + \frac{1}{y_{3,5}} \right) \times \left(\frac{1}{x_{1,3}x_{1,5}} + \frac{1}{x_{2,5}x_{2,6}} + \frac{1}{x_{1,3}x_{3,6}}+ \frac{1}{x_{1,5}x_{2,5}} + \frac{1}{x_{2,6}x_{3,6}}\right) ,
\end{equation}
which is precisely the 4-point  times the 5-point of the two respective lower problems. 

Looking at \eqref{eq:6ptTreeKin}, we see that in this kinematical locus $X_{1,3},X_{1,4}$ and $X_{2,4}$ are related to each other, the same being true for $X_{3,5}, X_{3,6}$ and $X_{4,6}$, and these relations imply:
\begin{equation}
\begin{aligned}
    X_{1,3} + X_{2,4} - X_{1,4} =0 \Leftrightarrow c_{1,3}=0, \\
    X_{3,5} + X_{4,6} - X_{3,6} =0 \Leftrightarrow c_{3,5}=0, \\
\end{aligned}
\end{equation}
which is precisely one of the \textit{split kinematics} observed in~\cite{FreddySplit,SongSplit}.

Already at $6$-points, the amplitude has 14 terms so it is not trivial to track how the split kinematic locus implies factorization at the level of the algebraic expression for the amplitude. To illustrate more explicitly this phenomenon let's look at the $5$-point amplitude:
\begin{equation}
    \mathcal{A}_5 = \frac{1}{X_{1,3}X_{1,4}} + \frac{1}{X_{2,4}X_{2,5}} + \frac{1}{X_{1,3}X_{3,5}}+ \frac{1}{X_{1,4}X_{2,4}}+ \frac{1}{X_{2,5}X_{3,5}},
    \label{eq:5pt}
\end{equation}
and consider the split in which the two subsurfaces are two 4-point: $\mathcal{S}_1=(1,2,3,5)$ and $\mathcal{S}_2=(1,3,4,5)$, that overlap in triangle $(1,3,5)$. In this case the kinematic mapping is simply:
\begin{equation}
    X_{1,3} \to y_{1,3}, \quad   X_{1,4} \to y_{1,3}+x_{1,4}, \quad   X_{2,4} \to x_{1,4}, \quad   X_{2,5} \to y_{2,5}, \quad   X_{3,5} \to x_{3,5},
\end{equation}
where $y_{i,j}$ and $x_{i,j}$  stand for the kinematics of $\mathcal{S}_1$ and $\mathcal{S}_2$, respectively. Plugging this back into \eqref{eq:5pt}, we get:
\begin{equation}
\begin{aligned}
    \mathcal{A}_5 &\to \frac{1}{y_{1,3}(y_{1,3}+x_{1,4})} + \frac{1}{x_{1,4}y_{2,5}} + \frac{1}{y_{1,3}x_{3,5}}+ \frac{1}{(y_{1,3}+x_{1,4})x_{1,4}}+ \frac{1}{y_{2,5}x_{3,5}}, \\
    &= \frac{1}{y_{1,3}x_{1,4}}+ \frac{1}{x_{1,4}y_{2,5}} + \frac{1}{y_{1,3}x_{3,5}}+\frac{1}{y_{2,5}x_{3,5}}= \left(\frac{1}{y_{1,3}} + \frac{1}{y_{2,5}} \right) \times \left(\frac{1}{x_{1,4}} + \frac{1}{x_{3,5}} \right),
\end{aligned}
\end{equation}
which factorizes exactly as predicted.

\subsection{Split kinematics for general trees}

We now consider the general problem of an $n$-point tree amplitude, so the starting surface is a disk with $n$-marked points in the boundary. Let's look at two generic subsurfaces that overlap on triangle $\tau= (i,j,k)$, so
\begin{equation*}
    \mathcal{S}_1 = (i,i+1,\cdots,j,j+1,\cdots,k) \quad \text{and} \quad \mathcal{S}_2 = (i,j,k,k+1,\cdots, i-1).
\end{equation*} 

This automatically divides the set of indices into three disjoint subsets:
\begin{equation}
    B=(i,\cdots,j-1), \quad C=(j,\cdots,k-1), \quad A=(k,\cdots,i-1),
\end{equation} 
where $A$ and $B$ are indices entering in $\mathcal{S}_1$ while $C$ are those from $\mathcal{S}_2$. Easily generalizing what we saw in the 6-point example above, the split kinematics is given by 
\begin{equation}
X_{a,b} \to x_{k,b} + y_{a,i}, \, \quad  X_{a,c} \to x_{c,k} + y_{j,a},
\end{equation}
while the remaining $X's$ corresponding to curves that lives only in ${\cal S}_1$ or ${\cal S}_2$ are trivially given by the corresponding $x, y$ variables. 

\subsection{Split kinematics as a constraint on $X$'s}
\label{sec:splitsConstraintsTree}
We have built the split kinematics $X$ for ${\cal S}$ starting from the kinematics $x,y$ for ${\cal S}_1, {\cal S}_2$. But obviously, the $X$'s that are defined this way can not be generic, and live on a lower-dimensional plane in $X$ space. Said backwards, the $X$'s must obey certain linear relations to lie on the split kinematics. In the 5 and 6-point examples at the beginning of this section, we observed that these linear relations precisely corresponded to setting to zero some non-planar variables, and we will now study how this pattern generalizes for general $n$ and for arbitrary choices of surfaces ${\cal S}_{1,2}$.

\begin{figure}[t]
    \centering
    \includegraphics[width=\textwidth]{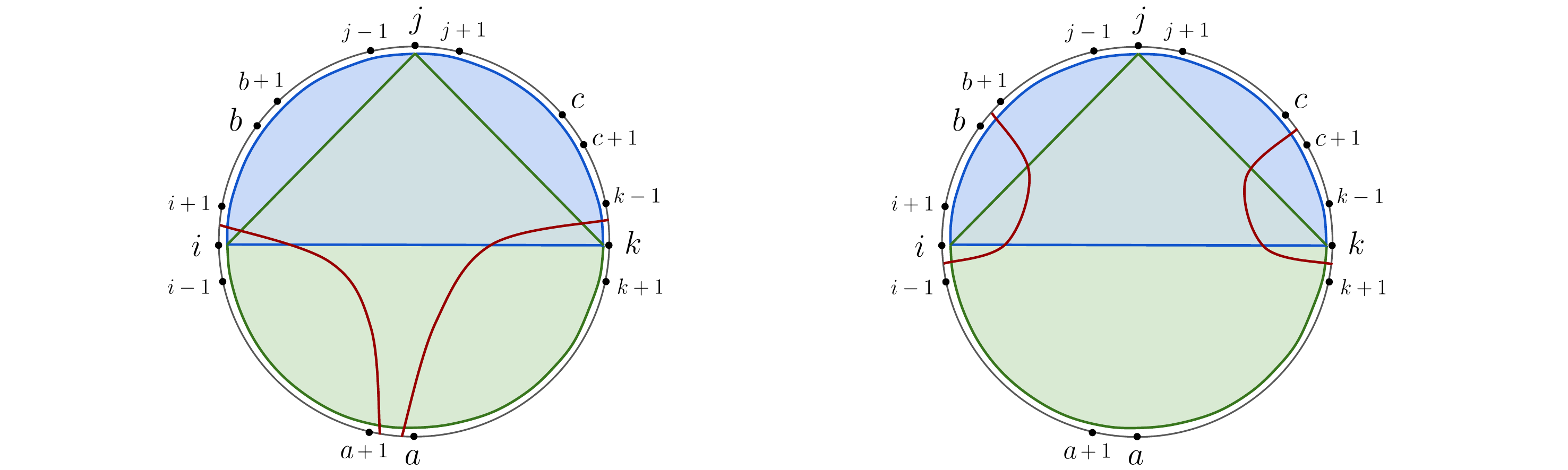}
    \caption{In the left panel, the curves $(a,i)$ and $(j,a)$ on the green surface are continued into $(a,i)$ and $(k-1,a)$ on the full surface, crossing the blue surface only on its boundaries $(k.i)$ and $(k-1,k)$. In the right panel, the curves $(k,b)$ and $(c,k)$ on the blue surface can be continued into $(i-1, b)$ and $(c,k)$ in the full surface, only crossing the green surface on the boundaries $(i-1,i)$ and $(j,k)$.} 
    \label{fig:treeconst}
\end{figure}

Consider a curve $x$ in ${\cal S}_1$. If this is already a full curve $X$ in ${\cal S}$, there is nothing to do, we can simply identify $x = X$. But if $x$ must be extended to be a full curve in ${\cal S}$, then there is always a way of doing this to some special curve $X_*$ which intersects ${\cal S}_2$ only in a boundary of ${\cal S}_2$; this means that the kinematic map will still tell us at $x=X_*$. The same is true for curves $y$ in ${\cal S}_2$. This determines the kinematic variables $x,y$ in terms of some special set of $X$'s for the big surface, $\mathcal{S}$. But then we have all the other curves, $X^\prime$, on $\mathcal{S}$, which are also expressed in terms of $x,y$. For every such $X^\prime$, substituting $x,y$ by the respective $X_*$'s, precisely gives the linear relations we seek. In this way, we both discover the linear relations, as well as determine the $x,y$ in terms of $X$'s on the locus where the linear relations are satisfied. 

Let's see how this works in our tree-level example, referring to figure \ref{fig:treeconst}. The non-trivial cases are of course only the curves on one of the smaller surfaces that end on boundaries shared with the other. 
Consider the curve $x_{k,b}$ in ${\cal S}_1$. It can be continued to $X_{i-1,b}$ intersecting ${\cal S}_2$ only in the boundary $(i-1,i)$; $x_{c,k}$ can be continued in the same way to $X_{c,k}$. Similarly $y_{a,i}$ in ${\cal S}_2$ can be continued into the big surface to $X_{a,i}$, crossing through ${\cal S}_1$ only on the boundary curve $(k,i)$; finally $y_{j,a}$ can be continued to $X_{k-1,a}$. 

Thus we have identified the $x,y$ in terms of the kinematics $X$ as 
\begin{equation}
x_{k,b} = X_{i-1,b}, \quad  x_{c,k} = X_{c,k}; \quad \quad  y_{a,i} = X_{a,i}, \quad  y_{j,a} = X_{k-1,a}.
\end{equation}

Having determined $x,y$, looking at the split kinematics for the remaining curves gives us the linear constraints we are looking for: 
\begin{equation}
\begin{aligned}
& X_{a,b} = x_{k,b} + y_{a,i} \Leftrightarrow   X_{a,b} - X_{a,i} - X_{i-1,b} = 0 ,\\ 
& X_{a,c} = x_{c,k} + y_{j,a} \Leftrightarrow X_{a,c} - X_{k-1,a} - X_{c,k} = 0.
\end{aligned}
\end{equation}

\begin{figure}[t]
    \centering
    \includegraphics[width=\textwidth]{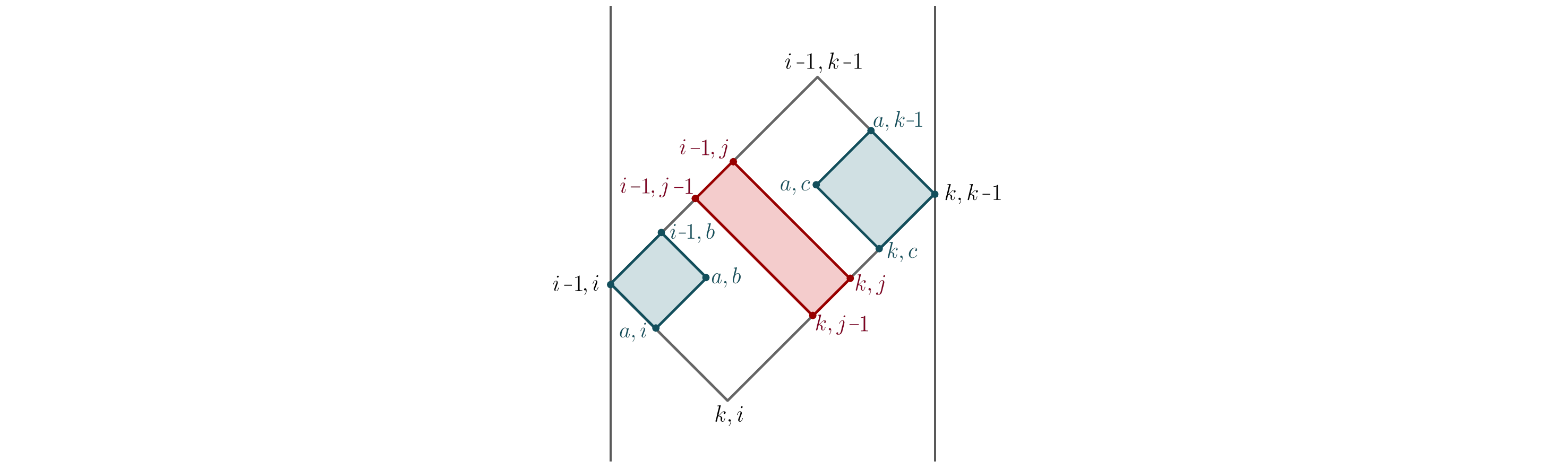}
    \caption{The linear relations on $X$ variables on split kinematics interpreted as the ``causal diamond'' relations in kinematic space. Since all the diamonds in each rectangle must vanish, we must have that all the mesh constants $c_{i,j}$ in each rectangle vanish}. 
    \label{fig:meshsplit}
\end{figure}

We can interpret these equations nicely in the ``kinematic mesh'' picture of \cite{GiulioClusters,zeros}, as in figure \ref{fig:meshsplit}. 
The big causal diamond touching the left boundary at $(i-1,i)$ and the right boundary at $(k,k-1)$ is divided into two smaller rectangles along the lines $(s,j-1)$ and $(s,j)$, for $s$  ranging from $k$ to $i-1$. The relations for $X_{a,b}$ is associated with a diamond inside the lower white rectangle with left and right corners on $(i-1,i)$ and $(a,b)$. The relations for $X_{a,c}$ are instead associated with the diamonds in the upper white rectangle, with left and right corners at $(a,c)$ and $(k,k-1)$, note that in this case, we use $X_{c,k} = X_{k,c}, \, \, X_{k-1,a} = X_{a,k-1}$ to see the relation as $X_{a,k-1} + X_{k,c} - X_{a,c} - X_{k,k-1} = 0$. Of course since these cover entire interior of the lower and upper rectangles, the relations imply that all the basic ``meshes'' $c_{i,j} = -2p_i \cdot p_j = X_{i,j} + X_{i+1,j+1} - X_{i,j+1} - X_{i+1,j} = 0$. We can also easily see this directly without the picture. The relations forcing the split kinematics imply 

\begin{equation}
   c_{a-1,c-1} = 0, \, \,  c_{b-1,c-1} = 0 \, \, \, {\rm for \, all } \, a \in A, b \in B, c \in C.
   \label{eq:cCondition}
\end{equation}
We simply write $c_{a-1,c-1}$ in terms of the $X$'s of the big surface and use the split kinematic mapping to find: 
\begin{equation}
\begin{aligned}
    c_{a-1,c-1} &= X_{a-1,c-1} +  X_{a,c} - X_{a-1,c} -  X_{a,c-1} \\ &= (x_{a-1,k} + y_{i,c-1}) + (x_{a,k} + y_{i,c}) - (x_{a-1,k} + y_{i,c}) - (x_{a,k} + y_{i,c-1})\\&=0.
\end{aligned}
\end{equation}

Of course, if $a-1 = i$ or $c-1=k$, we have $x_{a-1,k}=0$ or $y_{i,c-1}=0$, respectively, but the argument still holds. An analogous argument shows that $c_{b-1,c-1}=0$.

This is precisely the kinematic locus first introduced in \cite{FreddySplit} for which tree-level field-theory amplitudes split, and that was further shown to generalize to full string amplitudes in \cite{SongSplit}. In addition, as explained in \cite{SongSplit}, one can show that the factorization near zeros observed in~\cite{zeros} can be obtained by simply performing consecutive splits. 

\subsection{Factorization near zeros as sequential splits}

\begin{figure}[t]
    \centering
    \includegraphics[width=\textwidth]{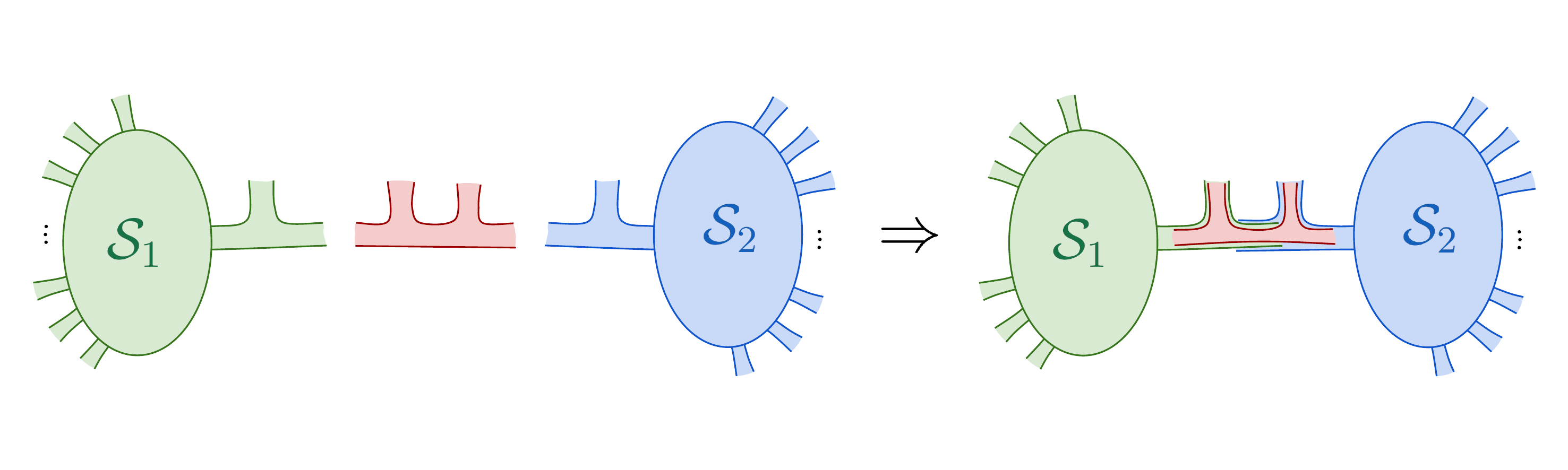}
    \caption{Factorization near zeros into $(4-$point$)\otimes \mathcal{S}_1\otimes\mathcal{S}_2$ as a sequential split. We have indicated one representative for the four-point fat-graph, any choice is allowed. }
    \label{fig:threesplit}
\end{figure}

Let us now see how to interpret the hidden zeros and near-zero factorizations of \cite{zeros} in the language of splits. We can join two surfaces ${\cal S}_1, {\cal S}_2$ and the surface for a four-point tree naturally by overlapping them in a way that does not introduce any new internal edges, as in figure \ref{fig:threesplit}. We can obviously think of this as simply doing our conventional split on a triangle twice, but our description of the ultimate configuration is more symmetrical. On this kinematics, the amplitude factorizes into the product of the four-point amplitude together with those of ${\cal S}_{1,2}$, precisely as we see happening in the factorization near zeros of \cite{zeros}. If we further take the two kinematic variables $x_{a,b},x_{c,d}$ of the four-point problem to be equal and opposite $x_{a,b} + x_{c,d} = 0$ the four-point amplitude vanishes. These are exactly the hidden zeros of \cite{zeros}. 

\subsection{Minimal kinematics as maximal splits}

We finally describe the ``minimal kinematics'' first described for trees in \cite{MinKin_FreddyNick,Nick_MinimalKin} (and recently studied in \cite{MinimalKin_EarlySturmfels}), in our language. We will discuss trees in this section but this construction applies more generally, and an avatar of minimal kinematics will make an interesting appearance associated with physical kinematics at loop level. 

Consider any tree fatgraph. We can associate every edge canonically with a four-point subsurface. Each of these is associated with the kinematical variables $(X_e, \tilde{X}_e)$, where $X_e$ is, as usual, the curve obtained from going outwards from $e$ and always turning left, while $\tilde{X}_e$ is instead what we get by always turning right. These are simply the $s$ and $t$ channel kinematics for each four-point factor.

The most extreme version of ``splits'' we can define is to join all these four-point subsurfaces together to define the full surface. On these kinematics, the amplitude is guaranteed to fully factorize into the product of the four-point amplitudes associated with the kinematical variables of each four-point problem. 

The kinematics mapping of any curve under this maximal split is very easy to describe. Every curve on the surface is associated with a word or ``mountainscape'' \cite{curveint}, starting at a boundary and recording the left and right turns it takes at the vertices along its path (recorded as up and down in the mountainscape).  Any peak or valley in this word corresponds to moving through either an $s$ or $t$ channel edge of a four-point problem. Thus, we have for any curve that 
\begin{equation}
X = \sum_{{\rm valleys}} X_{\text{valley}} + \sum_{{\rm peaks}} \tilde{X}_{\text{peak}}.
\end{equation}

On this kinematics, we get from the full tree-string amplitude the following simple product
\begin{equation}
{\cal A}|_{\rm min. kin.} = \prod_e \frac{\Gamma(X_e) \Gamma(\tilde{X}_e)}{\Gamma(X_e + \tilde{X}_e)}.
\end{equation}

An especially simple special case is the one where the fat graph for the tree is the standard ``half-ladder''. In this case, the kinematics associated with each 4 pt are those of the curves $(X_{1,3}, X_{2,n}),(X_{1,4},X_{3,n}), \cdots, (X_{1,n-1},X_{n-2,n})$. All the other kinematical variables $X_{i,j}$ are determined in terms of these via 
\begin{equation}
X_{i,j} = X_{i,n} + X_{1,j}.
\end{equation}

We will see these relations arise again in the next section, when we discuss splits for loop integrands satisfying ``physical kinematics''. 

\section{The Surface Loop Integrand}

Already at one-loop, the surface integrands (for a disk with a puncture) should be thought of as a generalization of the familiar integrands in momentum-space. The kinematic variables, $X$, on surfaces are naturally associated with every curve on the surface. But the assignment of momenta to the propagators associated with these curves is more restrictive. For instance there are ``external bubble'' variables $X_{i+1,i}$ that on the surface correspond to a curve starting from boundary $i$ and going to an adjacent boundary $i+1$, going around the puncture.  For conventional momentum-space integrands these are assigned the same on-shell momenta as the external particles and hence give a ``1/0'' factor for massless particles. A similar statement holds for the $X_{i,i}$ variables that start and end on $i$ going around the puncture; these correspond to tadpole propagators and also give $1/0$ factors for momentum integrands. 

The presence of these massless tadpoles and bubbles has long bedeviled attempts to define integrands for non-supersymmetric theories. It is of course permitted to throw out the massless tadpoles and bubbles since, in any case, the associated loop integrations are scale-less integrals that integrate to zero. But this obstructs other good properties expected of loop integands. For instance the one-loop integrand should have the property that on a loop-cut, it gives a tree amplitude in the ``forward limit'', adding two extra particles with equal and opposite momenta. The difficulty is that precisely in the forward limit the tree-amplitude can have $1/0$ singularities. 

Thus the most naive loop integrand is ill-defined, and even if we manually throw out tadpoles and external bubbles, loop cuts can't match (divergent) tree amplitudes in the forward limit. 

The surface integrands resolve this problem by assigning different ``kinematic'' variables to every curve on the surface, including the tadpoles and external bubbles. In addition, even curves that have (non-zero) momenta for conventional momentum-space loop integrands are in general assigned differing kinematics in the surface integrand; for instance, at one-loop, curves going around the puncture in two different ways are assigned the same momenta in momentum-space but are given different names in the surface integrand. This makes it possible for cuts of higher surfaces to be defined in terms of {\it completely generic} kinematics for the lower ones. For instance when we set a loop propagator to zero at one-loop, and look at all the curves on the loop surface compatible with it, we recover {\it all} the variables of the tree surface with two extra particles. This makes it possible to match cuts in a way that is impossible in the conventional forward limit. 

Thus surfaceology gives us a canonical notion of ``loop integrand'' that behaves correctly on cuts. As has recently been  seen in a number of settings, this object deserves to be called ``the'' integrand not only for Tr ($\Phi^3$) theory, but for the NLSM and YM as well. As we will prove in this paper, the surface NLSM integrands defined by kinematically shifting the Tr $(\Phi^3)$ theory have the Adler zero at the level of the integrand, something that has proven impossible with ordinary loop integrands \cite{JaraAdlerZeroRecursion,JaraSoftTheorem}. Similarly as indicated in \cite{Gluons} and will be more fully explored in \cite{gluonUpcoming}, there is a notion of ``the'' YM integrand which is gauge-invariant and factorizes correctly on cuts.

We can think of the surface integrand as generalizing the notion of kinematics away from what we usually associate with momenta, allowing all singularity properties to be manifest. This is roughly analogous to the generalization from Lorentzian momenta to general complex momenta, which allows massless three-particle amplitudes to be non-zero and four-particle factorization to play the amazingly powerful role it does in constraining consistently interacting theories of massless particles \cite{SMatrixMassless}. 

The factorization of amplitudes on split kinematics also happens automatically for the surface integrand. As we will see later in detail, a limit of split kinematics is also naturally associated with multi-soft limits. Thus splits will imply universal statements for multi ``surface-soft'' limits of the surface integrand, to all loop orders, some of which (in the case of the NLSM) were mentioned in \cite{CirclesNLSM}.

But to finally compare with physical amplitudes, we have to restrict the kinematic variables of the surface integrand to agree with those in momentum space. This means that we do ultimately have to set e.g. $X_{i+1,i}, X_{i,i} \to 0$ at one-loop, and more generally put $X_{i,j} = X_{j,i}$. For generic mass-deformed Tr $(\Phi^3)$ amplitudes, we will see that these conditions put further restrictions on the split kinematics, which are interestingly related to the ``minimal kinematics'' for trees studied in \cite{MinKin_FreddyNick} and described above. 

If we instead talk about the implications of splits for multi-soft limits of massless particles, we will present evidence for a natural conjecture: surface integrands and ``physical'' integrands agree up to scale-less terms in the multisoft limit. In this way, the conceptually straightforward factorization of the surface integrand following from splits in the multi-soft limit implies the same factorization at the level of integrated amplitudes also for physical momenta.

\subsection{The surface integrand one-loop integrand}
\label{sec:SurfaceIntegrand}
\begin{figure}[t]
    \centering
    \includegraphics[width=\textwidth]{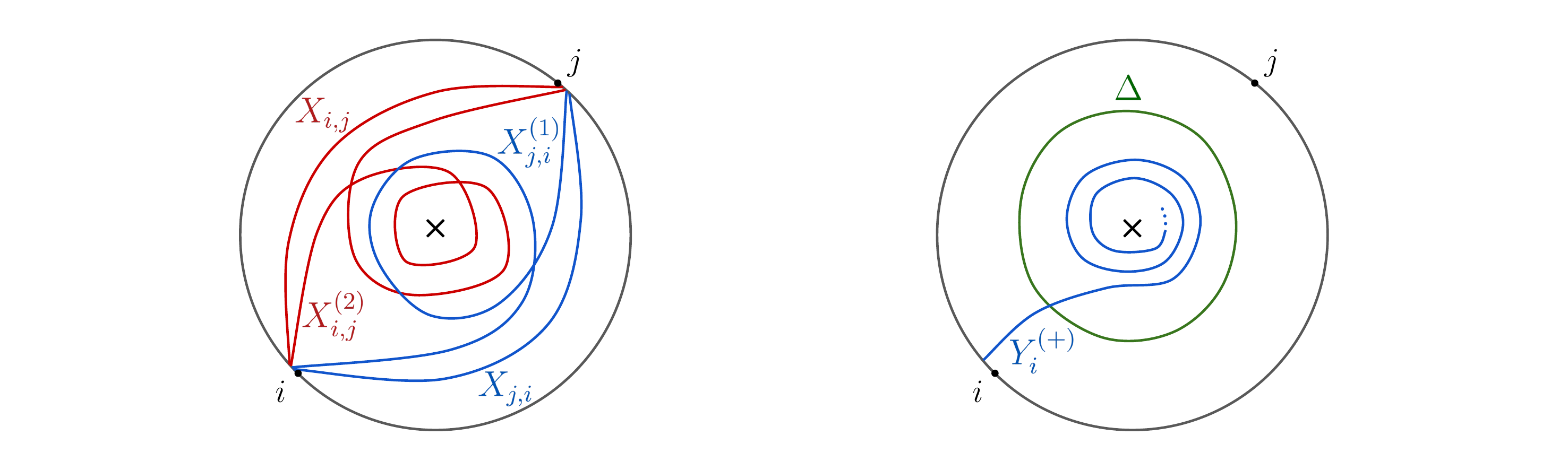}
    \caption{Possible curves on the punctured disk. Curves going from two marked points on the boundary $i$ and $j$, $X_{i,j}$ and  $X_{j,i}$, as well as their self-intersecting versions, $X_{i,j}^{(q)},X_{j,i}^{(q)}$.  On the left, we have the example of these two types of curves as well as their self-intersecting versions with self-intersection 2 and 1, respectively. On the right, we have an  example of a curve ending on the puncture which, drawn as a lamination spirals counterclockwise forever, $Y_i^{(+)}$; and the closed curve $\Delta.$} 
    \label{fig:1loopCurves}
\end{figure}

We now describe the salient features of the surface loop integrands of ~\cite{curveint,Gluons} at one-loop. As in the tree level case, the field theory integrand comes from extracting the low-energy limit of the surface integrals:
\begin{equation}
{\cal A}_\mathcal{S} = \int \prod_i \frac{dy_i}{y_i} \prod_{C \in \mathcal{S}} u_\mathcal{C}^{\alpha^\prime X_{\mathcal{C}}}(y_i),
\label{eq:curveInt_Loop}
\end{equation}
where we take the product over all the curves $C$ in the surface $\mathcal{S}$. Like in the tree-level case, each $y_i$ is associated to a propagator appearing in the base triangulation we choose, and all those corresponding to tree-like propagators are integrated from $0$ to $+\infty$, however those corresponding to loop propagators, $y_{i,p}$, are integrated over the region $0<\prod_p y_{i,p}<1$. 

At tree-level each curve on the disk is in one-to-one correspondence with the kinematics corresponding to planar propagators, $X_C = X_{i,j} = (p_i +p_{i+1}+ \cdots + p_{j-1})^2$, with $i/j$ labeling the marked points where the curve starts/ends. However, as mentioned above, at loop-level, not all the curves on the surface have a meaning in momentum space. In particular, already at one loop, where the surface is the punctured disk, the number of curves is infinite:

Analogously to the tree-level case, we have a curve going from the boundary-marked point $i$ to $j$, but now because of the presence of the puncture, this curve can either go through the left of the puncture or through the right. These two different possibilities correspond to different curves that we label as $X_{i,j}$ (left of the puncture) and $X_{j,i}$ (right of the puncture). Each of these is further associated with an infinite family of curves that self-intersect $q$ times while looping around the puncture, which we denote by $X_{i,j}^{(q)},X_{j,i}^{(q)}$ (see figure \ref{fig:1loopCurves}). In addition to curves that start and end on marked points, we have curves that start on a marked point, $i$, and end on the puncture which we label as $X_{i,p} \equiv Y_i$. Note that as laminations, $i.e.$ as curves on the fat graph, these correspond to curves that spiral around the puncture forever (see figure \ref{fig:1loopCurves}). In the integration domain considered, $0<\prod_p y_{i,p}<1$, we can only have spiraling counterclockwise, $Y_i^{(+)}$, as explained in \cite{curveint}. Finally, we have closed curves that go around the puncture that we label with the dimensionless parameter, $\Delta$.  

It turns out that at leading order in the low-energy expansion, only a finite number of curves contribute (with a finite number of self-intersections), and therefore to extract the field theory answer we can truncate the integrand by considering the product over only a finite set of curves. Regardless, to study the split properties of the loop integrand we must keep \textit{all} the curves, and of course the split limits we find for this object will also hold after taking the low energy limit. 

The integrand for field-theoretic Tr$(\Phi^3)$ amplitudes we get from the surface is simply the sum over all triangulations of the surface, corresponding to choosing a maximal set of non-intersecting curves. For each triangulation, we get a factor of 1 over the product of all the $X_C$ for the curves $C$ appearing in the triangulation, so that the final answer is simply:
\begin{equation}
    \mathcal{I}_n^{\mathcal{S}} = \sum_{\mathcal{T} \text{ triang. of } \mathcal{S}} \left(\prod_{C\in \mathcal{T}} \frac{1}{X_C}\right).
\end{equation}

Now just like in the tree-level case, each curve corresponds to a propagator appearing in cubic one-loop diagrams. This correspondence determines the map from $X_{C}$ to the physical momentum corresponding to the propagator with which $C$ is associated. When we apply this map to physical kinematics, we see that variables corresponding to different curves on the surface are set to be equal or zero. In particular we have that $X_{i,j}^{(q_1)} =X_{j,i}^{(q_2)} $ for any $q_1,q_2$. As a consequence, we have $X_{i+1,i} = X_{i,i+1}=0$ as well as $X_{i,i}=0$, this is the \textit{bubbles on external legs} as well as \textit{tadpoles}, see fig. \ref{fig:extBubTad}, are set to zero (for the massless case). This produces the problematic ''1/0'' s alluded to above, which are conventionally removed by hand in defining physical loop integrands. 
The more general object where we include all the curves is what we will refer to as the \textit{surface integrand}, $\mathcal{I}_n^\mathcal{S}$ \footnote{For simplicity, we will impose that the self-intersecting curves have the same kinematics as their non-self-intersecting versions i.e. $X_{i,j}^{(q_1)} =X_{i,j}\neq X_{j,i}=X_{j,i}^{(q_2)} $, for all $q_1,q_2$.}. As we have already stressed, this is the canonical object that enjoy good properties \cite{NLSM,CirclesNLSM,Gluons}, such as the Adler zero for the NLSM and gauge invariance for Yang-Mills.

Note that the surface integrand also gives us a canonical notion of loop integrated ``surface amplitude'' ${\cal A}^{\cal S}$. We simply use the assignment of loop momenta to define the loop propagators $X_{i,p} \equiv Y_i$, but we otherwise leave all $X_{i,j} \neq X_{j,i}$ general. We can then perform the loop integration to define  ${\cal A}^{\cal S}$.

In the rest of the paper, we will study the split properties of the surface integrand and how these connect with physical features that are obscured in the usual representation of integrands. 
In order to translate the results we find for the surface integrand/amplitude to the usual integrand/amplitude it's important to establish the precise relation between these two objects. As explained previously, the usual integrand is what we get by considering all the diagrams except those containing external bubbles or tadpoles, where we identify $X_{i,j}=X_{j,i}$, so we have:
\begin{figure}[t]
    \centering
    \includegraphics[width=\textwidth]{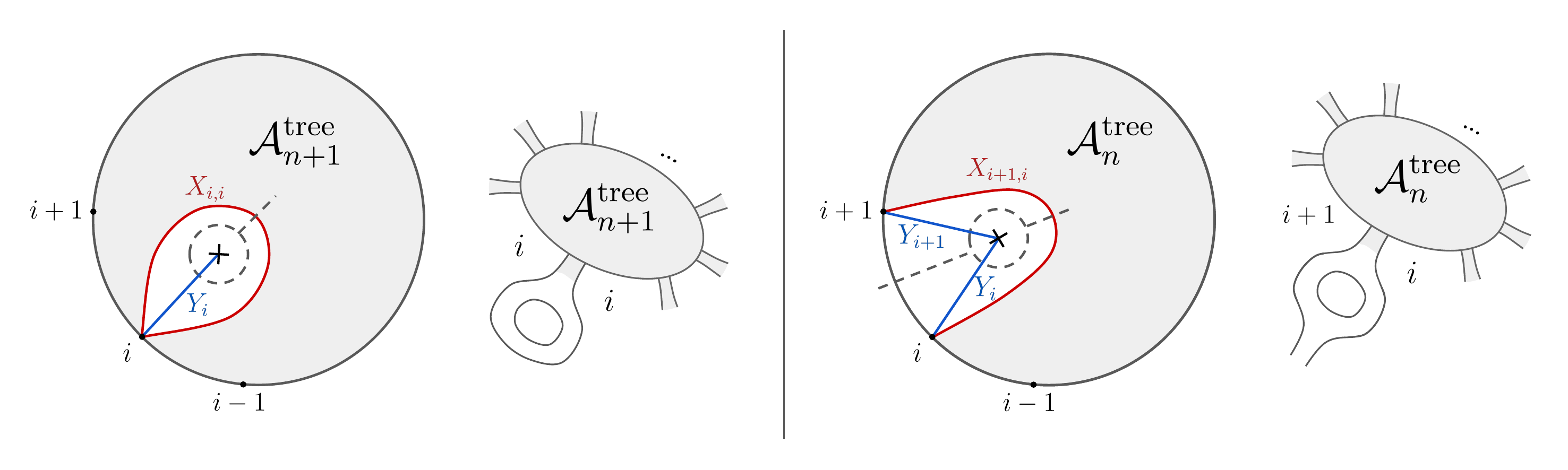}
    \caption{(Left) All triangulations containing the tadpole $X_{i,i}$. (Right) Triangulations containing the external bubble $X_{i+1,i}$ }
    \label{fig:extBubTad}
\end{figure}
\begin{equation}
    \mathcal{I}_n^{\text{standard}} = \left[\mathcal{I}_n^{\mathcal{S}} - \sum_{i=1}^n \left( \frac{\mathcal{A}^{\text{tree}}_{n+1}(i,i+1,\cdots,i-1,i)}{X_{i,i}Y_i} + \frac{\mathcal{A}^{\text{tree}}_{n}(i+1,\cdots,i-1,i)}{X_{i+1,i}Y_{i+1}Y_{i}}\right) \right]_{X_{j,i}=X_{i,j}},
    \label{eq:PhysIntegrand}
\end{equation}
where inside brackets we subtract all terms in $\mathcal{I}_n^{\mathcal{S}}$ which contain tadpoles, $X_{i,i}$, or external bubbles, $X_{i+1,i}$, which we can automatically determine using the fact that when we factorize on this poles we get just the tree-level amplitudes (see figure \ref{fig:extBubTad}). Eliminating all the dependence on these variables, we can then identify $X_{i,j}$ and $X_{j,i}$ without producing any ''1/0'' and this is precisely the standard physical integrand. 

Reading this expression as a formula for the surface integrand (on the support of $X_{i,j} = X_{j,i}$), the statement about factorization on splits for the surface integrand can be interpreted as a statement about factorization for the standard integrand where tadpoles and bubbles are added, and which are determined by the split kinematics in a specific way. 

We will have more to say about loop-integrated statements below, but end this section by making some quick comments about the relationship between the standard and surface amplitudes post loop-integration here. 

Note that for the case of massless particles, the terms corresponding to tadpoles or external bubbles always lead to scaleless integrals that vanish in dimensional regularization upon loop integration. Therefore, after loop integration, $\mathcal{A}_n^{\text{standard}}$ and $\mathcal{A}_n^{\mathcal{S}}$ agree as long as we set $X_{i,j}=X_{j,i}$:
\begin{equation}
    \mathcal{A}_n^{\text{standard}} = \left[\mathcal{A}_n^{\mathcal{S}}\right]_{X_{j,i}=X_{i,j}}. 
\end{equation}

If, instead, we are dealing with massive states, $X_{i,j} \to X_{i,j} + m_{i,j}^2$, then we still have  $X_{i,j}=X_{j,i}$ as well as $X_{i,i}\to m_{i,i}^2$, $X_{i+1,i}\to m_{i+1,i}^2$. So from these contributions, we obtain
\begin{equation}
\begin{aligned}
& T_i = T(m_{i,i}^2,m_{i}^2) \equiv \int \frac{d^{4 - 2 \epsilon} l}{(2 \pi)^{4 - 2 \epsilon}} \frac{1}{X_{i,i} Y_i} = \int \frac{d^d l}{(2 \pi)^{d}} \frac{1}{m_{i,i}^2 (l^2 + m_i^2)},  \\ 
& B_i = B(p_i^2,m_{i+1,i}^2,m_i^2,m_{i+1}^2) \equiv \int \frac{d^{4 - 2 \epsilon} l}{(2 \pi)^{4 - 2 \epsilon}} \frac{1}{X_{i+1,i} Y_i Y_{i+1}} \\
& \quad \quad \quad \quad \quad \quad  \quad \quad \quad \quad  = \int \frac{\diff^d l}{(2\pi)^d} \frac{1}{m_{i+1,i}^2 (l^2+m_i^2)((l+p_i)^2+m_{i+1}^2)},
\end{aligned}
\end{equation}
where the loop-integrals can be performed trivially in general dimensions $d$.

This gives us a relationship between integrated standard and surface amplitudes, as 
\begin{equation}
\begin{aligned}
    \mathcal{A}_n^{\text{standard}} = \left[\mathcal{A}_n^{\mathcal{S}} - \sum_{i=1}^n \right.&\left(\mathcal{A}^{\text{tree}}_{n+1}(i,i+1,\cdots,i-1,i)T_i  \right. \\
    &\left.\quad + \mathcal{A}^{\text{tree}}_{n}(i+1,\cdots,i-1,i) B_i \right) \left.\right]_{X_{j,i}=X_{i,j}}. 
\end{aligned}
\label{eq:PhysIntMass}
\end{equation}

This will allow us to relate statements about splits for the surface integrand, both pre- and post- loop integration, to those of the conventional physical integrands/amplitudes. 

\section{One-loop Splits}
Having understood splits from the surface perspective, the generalization to all orders in the topological expansion follows immediately, and in this section, we will describe the various patterns of splits that are possible at one-loop. However, in some cases, we will see that $X_{i,j}$ and $X_{j,i}$ (and their self-intersecting analogs) don't get mapped to the same kinematics in the lower surfaces. Therefore the surface integrand where we keep $X_{i,j} \neq X_{j,i}$ is the object that enjoys the most general splitting properties. In order to have splits defined for physical $X_{i,j}=X_{j,i}$, we need to live in restricted kinematics in the lower-point problems, that we will define at the end of this section. 

Before jumping to the general kinematical mapping at one-loop, we start by looking at our running example of the $6$-point factoring into ($4$-point)$\times$($5$-point), where now the $6$-point and the $5$-point are one-loop processes. In this example as well as in the rest of the section, we will present the explicit mapping of the curves with no self-intersection and explain how it generalizes to their self-intersecting versions. Even though in most cases self-intersecting curves don't contribute to the field theory limit\footnote{As far as this paper goes, the only context in which self-intersecting curves play a role is when describing scaffolded gluons, where we need to consider curves up to self-intersection one, as explained in \cite{Gluons}}, their mapping under the split is very simple and, as opposed to the mappings of $X_{i,j}$ and $X_{j,i}$, it does not introduce any subtleties in going to physical kinematics $X_{i,j}^{(q)} \equiv X_{i,j}$. 

\subsection{Example: (6-point one-loop)$\to$ (4-point tree) $\times$ (5-point one-loop)}

\begin{figure}[t]
    \centering
    \includegraphics[width=\textwidth]{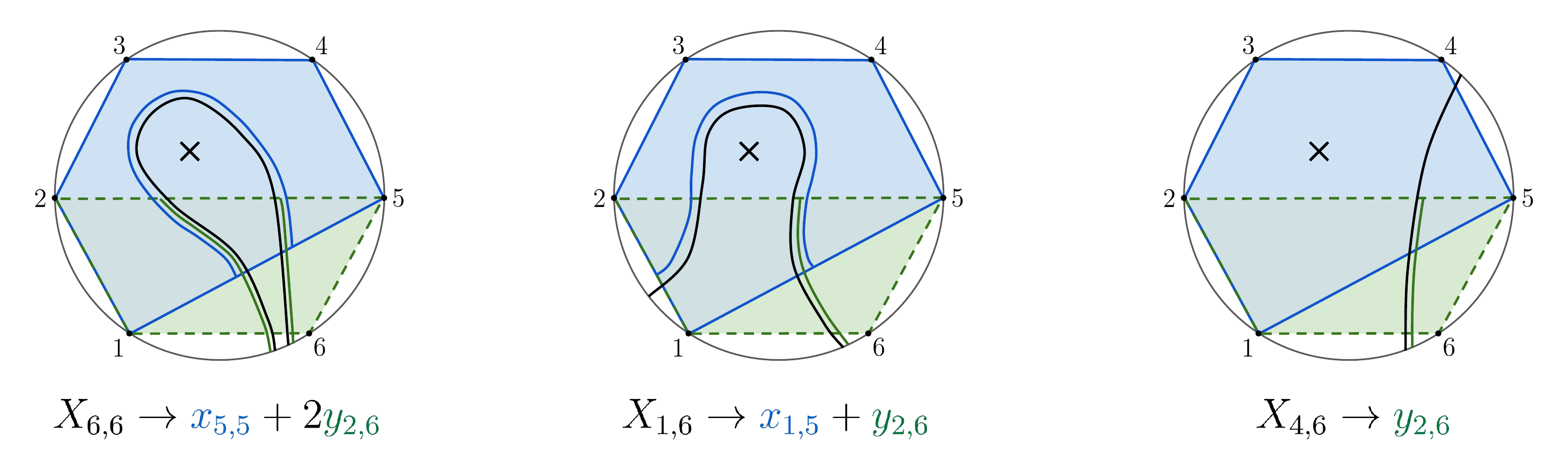}
    \caption{Kinematical mappings for the split of 6-point 1-loop integrand into a 4-point tree times a 5-point one-loop.}
    \label{fig:6pt1loop}
\end{figure}

We now explore a split of the $6$-point $1$-loop Tr($\Phi^3$) integrand. At one-loop, we are interested in studying the split pattern that leads to (tree)$\times$(one-loop). We will now consider the simplest case in which the tree part is a four-point problem, and so the one-loop is a five-point problem. As depicted in figure \ref{fig:6pt1loop}, $\mathcal{S}_1$, in green, is the $4$-point disk with vertices $(1,2,5,6)$, and $\mathcal{S}_2$, in blue, is the 5-point punctured disk with marked points $(1,2,3,4,5)$. We will denote the kinematics in $\mathcal{S}_1$ by $y_{i,j}$, and the kinematics in $\mathcal{S}_2$ by $x_{i,j}$ where we will use the index $p$ to label the puncture. 

Let's now build the kinematics that lands us on the split $\mathcal{S}_1 \otimes \mathcal{S}_2$ for this loop problem. Let's start by looking at the curves $X_{i,j}$ where $i,j$ live in $\mathcal{S}_2$. Most of these curves, only overlap $\mathcal{S}_2$, and thus they are mapped onto themselves, with the exception of $X_{5,1}$ which lives exclusively on $\mathcal{S}_1$:
\begin{equation}
\begin{aligned}
    X_{i,j}\to \begin{cases}
        x_{i,j} \, \text{for } i,j \in \{2,3,4,5,p\} \\
        x_{1,j}, \, \text{for } i=1,j\in\{1,2,3,4,5,p\} \\
        \end{cases}, \quad 
        X_{i,j}\to \begin{cases}
            x_{i,1}, \, \text{for } i\in\{2,3,4\},j=1 \\
        y_{1,5}, \, \text{for } i=5,j=1
        \end{cases}.
\end{aligned}
\end{equation} 
Finally, for the remaining curves, $i.e.$ those ending/starting in $6$, we have the following map:
\begin{equation}
\begin{aligned}
    &X_{6,p} \to x_{5,p} + y_{2,6}, \\
    &X_{1,6} \to x_{1,5} + y_{2,6},\\
    &X_{2,6} \to x_{2,5} + y_{2,6}, \\
    &X_{3,6} \to x_{3,5} + y_{2,6}, \\
    &X_{4,6} \to  y_{2,6}, \\
\end{aligned} \quad \quad \quad \quad \quad 
\begin{aligned}
    &X_{6,5} \to x_{5,5} + y_{2,6}, \\
    &X_{6,6} \to  x_{5,5} + 2y_{2,6},\\
    &X_{6,2} \to x_{5,2} + y_{2,6}, \\
    &X_{6,3} \to x_{5,3} + y_{2,6},\\
    &X_{6,4} \to x_{5,4} + y_{2,6}. \\
\end{aligned}
\label{eq:6pt1loop_Map}
\end{equation}

The mappings of curves $X_{6,6}$, $X_{1,6}$ and $X_{4,6}$, are illustrated explicitly in figure \ref{fig:6pt1loop}. Note that curve $X_{6,6}$ gets mapped to $x_{5,5}+2y_{2,6}$, where the factor of two comes from the fact that the tadpole curve crosses $\mathcal{S}_{1}$ twice, and thus decomposes into two $x_{2,6}$ curves (see figure \ref{fig:6pt1loop}, left).  

In this limit, the surface integrand becomes:
\begin{equation}
    \mathcal{I}_{6} \to \mathcal{A}_4(1,2,5,6) \times \mathcal{I}_{5}(1,2,3,4,5) = \left(\frac{1}{x_{1,5}}+\frac{1}{x_{2,6}} \right) \times \mathcal{I}_{5}(1,2,3,4,5). 
\end{equation}

As already mentioned,  for physical kinematics we must have $X_{i,j}=X_{j,i}$ and $X_{i,i}=0$ (setting external bubbles and tadpoles to zero), and we can see that by doing this in the lower surface, $i.e.$ $x_{i,j}=x_{j,i}$ and $x_{i,i}=0$ the split map is  going to impose some restrictions in the $y_{i,j}$ in particular from the mapping of $X_{5,1}$ and $X_{6,5}$ we have $y_{1,5},y_{2,6}\to 0$. So to make this split have meaning to the physical integrand we need to further impose these restrictions on the $y$'s. But with this further condition the split kinematic limit precisely matches the soft limit! Note that if we make particle $5$ soft, then $y_{2,6},y_{1,5} \to 0$, and the mapping in \eqref{eq:6pt1loop_Map} precisely matches the surface-soft limit (defined in \cite{CirclesNLSM}), in which $X_{5,j}\to X_{6,j}$.  Of course, for massless Tr($\Phi^3$) theory, this limit is divergent, but as we will see momentarily, this connection will be crucial to studying soft limits in the NLSM. 

\subsection{Split kinematics for $n$-point one-loop surface integrand}
\label{sec:SplitKinLoop}

Let's now consider the general $n$-point one-loop surface integrand, which is for the once-punctured disk with $n$ marked points on the boundary, $\mathcal{S}$. We want to understand the kinematical locus, $X \to X_{\mathcal{S}_1 \otimes \mathcal{S}_2}$, in which this object splits into lower-point objects, $\mathcal{S}_1$ and $\mathcal{S}_2$ ( which might be integrands or tree-level amplitudes). For the reasons explained previously, we will build these kinematics explicitly for the non-self-intersecting curves and explain how it generalizes for non-zero self-intersection.  

At loop-level, there are a few different factorization patterns. Considering two subsurfaces overlapping on a triangle, $(i,j,k)$, and so we have:
\begin{equation*}
    \mathcal{S}_1 = (i,i+1,\cdots,j,j+1,\cdots,k) \quad \text{and} \quad \mathcal{S}_2 = (i,j,k,k+1,\cdots, i-1),
\end{equation*} 
which naturally divides the indices into three different subsets:
\begin{equation}
 A=(i,\cdots,j-1), \quad B=(j,\cdots,k-1), \quad C=(k,\cdots,i-1).
\end{equation}

Now if the puncture is in $\mathcal{S}_1$, then it can either be in the region bounded $A$ or in the region bounded by $B$ (which are two morally equivalent cases), but can also have the puncture in $\mathcal{S}_2$, in which case it is inside the region bounded by $C$. Alternatively, we can have our two subsurfaces, $\mathcal{S}_{1,2}$, be both tree-level surfaces that overlap in two distinct triangles touching the puncture, $(i,j,p)$ and $(k,m,p)$. This last possibility corresponds to an overlap of type III. in figure \ref{fig:overlaps}.
Therefore, we have three different split possibilities:
\begin{enumerate}
    \item $\mathcal{S}_1$ and $\mathcal{S}_2$ overlap on a single triangle and the puncture is in $\mathcal{S}_2$, $i.e.$ the subsurface that contains two edges of the triangle;
    \item $\mathcal{S}_1$ and $\mathcal{S}_2$ overlap on a single triangle and the puncture is in $\mathcal{S}_1$, $i.e.$ the subsurface that contains only one edge of the triangle;
    \item $\mathcal{S}_1$ and $\mathcal{S}_2$ overlap on two triangles that share a common vertex -- the puncture. In this case, both subsurfaces, $\mathcal{S}_1$, $\mathcal{S}_2$, are tree surfaces. 
\end{enumerate}

We will now go through each case separately and proceed as we did at tree level to derive the constraints on the $X$'s that land us on the split kinematics \eqref{eq:kinMap}. 

\subsubsection{Split kinematics as constraints in $X$'s: Case 1.}

Let us start with case 1. So $\Sone$ and $\Stwo$ overlap on triangle $(i,j,k)$:
\begin{equation*}
    \mathcal{S}_1 = (i,i+1,\cdots,j,j+1,\cdots,k) \quad \text{and} \quad \mathcal{S}_2 = (i,j,k,k+1,\cdots, i-1),
\end{equation*} 
and the puncture is in $\mathcal{S}_2$. We will use $x_{i,j}$ to denote the kinematics of the tree problem defined by $\mathcal{S}_1$, and $y_{i,j}$ those of the loop problem from $\mathcal{S}_2$. Crucially, while $x_{i,j}=x_{j,i}$, $y_{i,j} \neq y_{j,i}$ since $\mathcal{S}_2$ is a loop surface. Once more let's define the three subsets of indices:
\begin{equation}
 A=(i,\cdots,j-1), \quad B=(j,\cdots,k-1), \quad C=(k,\cdots,i-1).
\end{equation}

Just like we did at tree-level, we start by identifying the curves $X$ on $\Surf$ that precisely agree with single curves $x$ and $y$ in $\Sone$, $\Stwo$, $i.e.$ curves that once we decompose into its $\Sone$, $\Stwo$ components have a single component. Starting for those in $\Sone$ we get:
\begin{equation}
\begin{aligned}
    &x_{a,b}=X_{a,b}, \\
    &x_{a_1,a_2}=X_{a_1,a_2},\\
    &x_{b_1,b_2} = X_{b_1,b_2},
\end{aligned} \quad \quad 
\begin{aligned}
    &x_{a,k}=X_{a,i-1}, \\
    &x_{b,k}=X_{b,k},\\
\end{aligned}
\label{eq:MapX}
\end{equation}
where $a_1<a_2$ and $a$ are indices in $A$, and similarly for $b_i$ and $b$ in $B$. For the curves in $\Stwo$ we get:
\begin{equation}
\begin{aligned}
    &y_{i,c}=X_{i,c}, \\
    &y_{j,c}=X_{k-1,c},\\
    &y_{c_1,c_2} = X_{c_1,c_2},
\end{aligned} \quad \quad 
\begin{aligned}
    &y_{c,i}=X_{c,i}, \\
    &y_{c,j}=X_{c,k-1},\\
    &y_{c_2,c_1} = X_{c_2,c_1},
\end{aligned}
\quad \quad 
\begin{aligned}
    &y_{c,p} = X_{c,p},\\
    &y_{i,p}=X_{i,p}, \\
    &y_{j,p}=X_{k-1,p},
\end{aligned}
\quad \quad 
\begin{aligned}
    &y_{c,c} = X_{c,c},\\
    &y_{i,i}=X_{i,i}, \\
    &y_{j,j}=X_{k-1,k-1},
\end{aligned}
\label{eq:MapY}
\end{equation}
where $c_1<c_2$ and $c$ are indices in $C$, and $p$ is the label of the puncture.

Now looking at the mapping of the remaining curves (which are no longer single component), using \eqref{eq:MapX} and \eqref{eq:MapY}, we can derive the constraints in $X$ space giving us split kinematics. Let's start by looking at the curves that go from $A/B$ to $C$:
\begin{equation}
\begin{aligned}
     &X_{a,c} = x_{a,k} + y_{i,c} \quad \to \quad  X_{a,c}-X_{a,i-1}-X_{i,c}=0, \\
    &X_{c,a} = x_{a,k} + y_{c,i}\quad \to \quad X_{c,a}-X_{a,i-1}-X_{c,i}=0,\\
    &X_{b,c} = x_{b,k} + y_{j,c}\quad \to \quad X_{b,c} -X_{b,k} -X_{k-1,c}=0, \\
    &X_{c,b} = x_{b,k} + y_{c,j}\quad \to \quad X_{c,b}-X_{b,k} -X_{c,k-1}=0, \\
\end{aligned} 
\end{equation}
as for the remaining curves we have:
\begin{equation}
\begin{aligned}
    &X_{a_2,a_1}= x_{a_1,k} + x_{a_2,k} + y_{i,i}  \\
    &X_{a,p}= x_{a,k} + y_{i,p} \\
    &X_{b_2,b_1}= x_{b_1,k} + x_{b_2,k} + y_{j,j}  \\
    &X_{b,p}= x_{b,k} + y_{j,p}   \\
    &X_{b,a} = x_{a,k} + x_{b,k} + y_{j,i}  
\end{aligned} \quad 
\begin{aligned}
&\to \quad X_{a_2,a_1}-X_{a_1,i-1}-X_{a_2,i-1} - X_{i,i}=0, \\
&\to \quad  X_{a,p}-X_{a,i-1}-X_{i,p}=0, \\
&\to \quad X_{b_2,b_1}- X_{b_1,k}-X_{b_2,k} - X_{k-1,k-1}=0,\\
&\to \quad X_{b,p}-X_{b,k} - X_{k-1,p}=0,\\
&\to \quad X_{b,a}- X_{a,i-1} - X_{b,k} - X_{k-1,i}=0.
\end{aligned}
\label{eq:oneloopsplit}
\end{equation}
\begin{figure}[t]
    \centering
    \includegraphics[width=\textwidth]{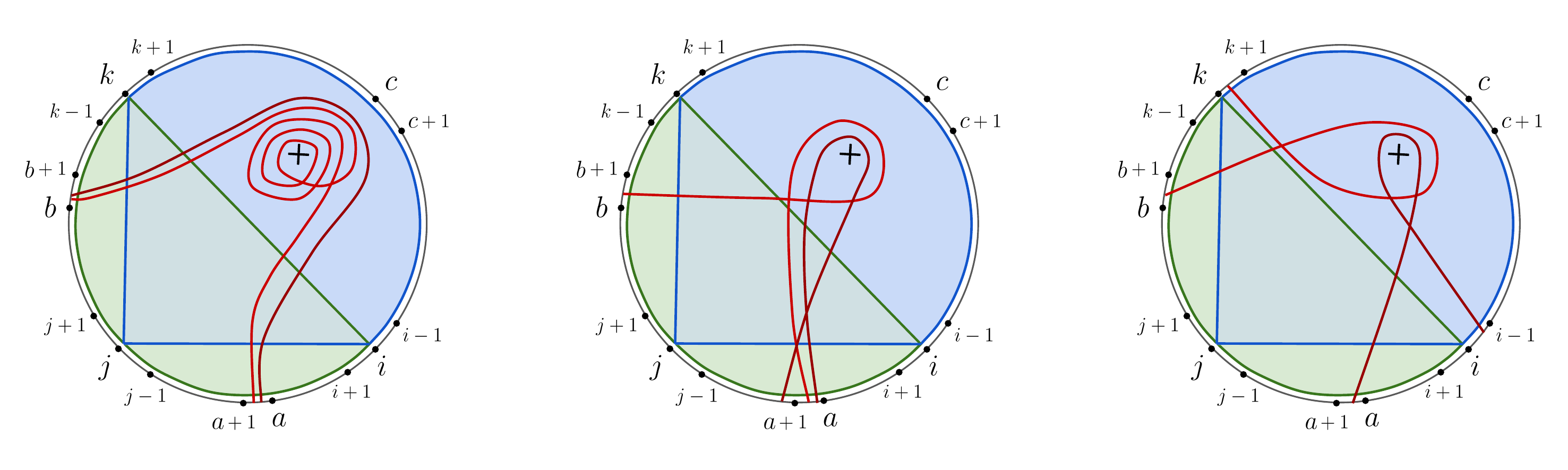}
    \caption{Split of surface one-loop integrand into $\mathcal{S}_1$(green)$\otimes$$\mathcal{S}_2$(blue). On the left, we have an example of curve $X_{b,a}$ and its self-intersecting version, for which the map is identical, simply needing the replacement of $y_{k,i} \to y_{k,i}^{(q)}$. On the center and right, we have the graphical representation of the mapping of curves $X_{a,a+1}^{(1)}, X_{a,b}^{(1)}$ (center) and $X_{i-1,a}^{(1)}, X_{b,k}^{(1)}$ (right), which, as highlighted in equation \eqref{eq:mappingSI}, is different then that of the respective non-self-intersecting curves.}
    \label{fig:selfint}
\end{figure}

Now, as illustrated in figure \ref{fig:selfint}, by allowing the curves to self-intersect, we either produce a curve of higher self-intersection in the loop subsurface and/or add boundary components. Therefore the mapping for the self-intersecting curves is precisely the same where we just replace the curves with its self-intersecting cousins: $X^{(q)} \to x+y^{(q)}$. The only exception is for the curves that map to curves exclusively in the tree problem $\Sone$, $X_{i,j}=x$. For these ones, with $q\neq 0$, we have (see figure  \ref{fig:selfint}):
\begin{equation}
\begin{aligned}
    &X_{a_1,a_2}^{(q)} = x_{a_1,k} + x_{a_2,k} + y_{i,i}^{(q-1)},  \\
    &X_{b_1,b_2}^{(q)} = x_{b_1,k} + x_{b_2,k} + y_{j,j}^{(q-1)},\\
    &X_{a,b}^{(q)} = x_{a,k} + x_{b,k} +y_{i,j}^{(q)}, 
\end{aligned} \quad \quad 
\begin{aligned}
      &X_{a,i-1} = x_{a,k}+y_{i-1,i}^{(q)},\\
    &X_{b,k}= x_{b,k}+y_{j,k}^{(q)} .  
\end{aligned}
\label{eq:mappingSI}
\end{equation}

When we map the curves on the surface to physical momentum, we have $X_{i,j}^{(q)}= X_{i,j}$, $X_{j,i}^{(q)}=X_{j,i}$, for any $q$, as well as $X_{i,j}=X_{j,i}$. These conditions will ultimately impose some further restrictions on the kinematics of $\Surf_{1,2}$. So far we are exclusively describing the locus under which the full surface integrand splits (where we allow all kinematics to be different), and we will discuss the connection to physically relevant limits in \ref{sec:PhyKin}.

\subsubsection{Split kinematics as constraints in $X$'s: Case 2.}
Let's now proceed to case 2. in which the puncture lives on subsurface $\mathcal{S}_1$, and in particular let's say it lives inside the subsurface associated with the indices in $A$. We omit the case in which the puncture is in the region bounded by $B$, simply because it's completely analogous to this case. 

Once more let's start by identify the curves $X$ that map to curves $x$ in $\Sone$:
\begin{equation}
\begin{aligned}
    &x_{a,k}=X_{a,i-1}, \\
    &x_{b,k}=X_{b,k}, \\
    &x_{a,b}=X_{a,b},\\
\end{aligned} \quad \quad 
\begin{aligned}
      &x_{k,a}=X_{i-1,a}, \\
    &x_{k,b}=X_{i-1,b}, \\
    &x_{b,a}=X_{b,a},
\end{aligned}\quad \quad 
\begin{aligned}
    &x_{a_1,a_2}=X_{a_1,a_2}, \\
    &x_{a_2,a_1}=X_{a_2,a_1}, 
\end{aligned}\quad \quad 
\begin{aligned}
    &x_{b_1,b_2}=X_{b_1,b_2}, \\
    &x_{b_2,b_1}=X_{b_2,b_1}, 
\end{aligned}
\label{eq:MapX2}
\end{equation}
as for curves ending on the puncture or tadpoles we get:
\begin{equation} 
\begin{aligned}
    &x_{a,p} = X_{a,p},\\
     &x_{b,p}=X_{b,p}, \\  
\end{aligned}
\quad \quad 
\begin{aligned}
    &x_{k,p}=X_{i-1,p},\\
    &x_{k,k}=X_{i-1,i-1}.
\end{aligned}
\label{eq:MapX22}
\end{equation}

For the curves mapping to curves purely in $\Stwo$, we have:

\begin{equation}
y_{i,c}=X_{i,c}, \quad y_{j,c}=X_{k-1,c},\quad y_{c_1,c_2} = X_{c_1,c_2}.
\label{eq:MapY2}
\end{equation}

Once more to derive the split kinematical locus we use \eqref{eq:MapX2}, \eqref{eq:MapX22} and \eqref{eq:MapY2} to write the non-trivial constraints in the remaining variables, $X$. 
\begin{equation}
\begin{aligned}
     &X_{a,c} = x_{a,k} + y_{i,c}, \\
    &X_{c,a} = x_{k,a} + y_{i,c}, \\
    &X_{b,c} = x_{b,k} + y_{j,c}, \\
    &X_{c,b} = x_{k,b} + y_{i,c}, \\
&X_{c,p} = x_{k,p} + y_{i,c},\\
&X_{c_2,c_1} = x_{k,k}+ y_{i,c_1} + y_{i,c_2}. 
\end{aligned} \quad \quad 
\begin{aligned}
     &\to \quad \quad  X_{a,c} - X_{a,i-1} - X_{i,c}=0, \\
    &\to\quad \quad  X_{c,a} - X_{i-1,a} - X_{i,c}=0, \\
    &\to \quad \quad X_{b,c} - X_{b,k} - X_{k-1,c}=0, \\
    &\to\quad \quad  X_{c,b} - X_{i-1,b} - X_{i,c}=0, \\
&\to\quad \quad X_{c,p} -X_{i-1,p} - X_{i,c}=0,\\
&\to \quad \quad  X_{c_2,c_1} -X_{i-1,i-1}- X_{i,c_1} -X_{i,c_2}=0. 
\end{aligned}
\end{equation}

In addition, like in case 1., the mapping for the self-intersecting curves is precisely what we get by replacing curves with its self-intersecting cousins in the previous mappings: $X^{(q)} \to x^{(q)}+y$. The only exception is now for the self-intersecting versions of curves $X_{b,c}$ and  $X_{c_1,c_2}$, for which we get:
\begin{equation}
X_{b,c}^{(q)} \to x_{b,k}^{(q)} + y_{i,c},  \quad \quad X_{c_1,c_2}^{(q)} \to x_{k,k}^{(q-1)}+ y_{c_1,i} + y_{c_2,i},
\end{equation}
with $q\neq0$. 

\subsubsection{Split kinematics as constraints in $X$'s: Case 3.}
\label{sec:TreeTreeSplit1loop}
Finally, let's look at case 3. for which the split is into two tree processes. In this case, the subsurfaces overlap in two triangles: $\tau_1 =(p,i,j)$ and $\tau_2 = (p,k,m)$, where once more the index $p$ is labeling the puncture; and they are defined as follows:
\begin{equation*}
    \mathcal{S}_1 = (i, j, p,k,m,m+1\cdots, i-1), \quad \mathcal{S}_2 = (i, i+1, \cdots, j, j+1, \cdots, k, k+1, \cdots, m, p),
\end{equation*}
let us additionally define four different subsets of indices:
\begin{equation}
\begin{aligned}
    &A=(i, i+1, \cdots, j-1), \\
    &B=(j,j+1, \cdots, k-1), 
\end{aligned} \quad \quad 
\begin{aligned}
    &C=(k, k+1, \cdots, m-1), \\
    &D=(m,m+1, \cdots, i-1). 
\end{aligned}
\end{equation}

We now want to understand what are the curves on the big surface $\Surf$ that map to curves of $\Sone$. As we will see momentarily, to find such curves we need to consider curves that end in the puncture but that can spiral both counter-clockwise ($+$), as we have been considering so far, as well as those spiraling anti-clockwise ($-$). Previously, we didn't run into this because one of the smaller surfaces was always a loop surface, which, by consistency, only contained curves spiraling around the puncture counter-clockwise (which converge in the integration region 
in $y$-space in which we define the loop integrals \cite{curveint}). However, when we overlap the two tree-level surfaces just like in this case, the resulting object is integrated over the full integration domain, and therefore we to account for both types of spirals. 

Now in this case we have four different subsets of indices, so the general mapping that gives us split kinematics is slightly more complicated than the one described in the previous two cases. For this reason we describe the general kinematic constraint for this split in appendix \ref{sec:TreeTreeSplit}, while here we go through the 6-point one-loop example in detail.

\subsubsection{Case 3. split example: 6-point 1-loop integrand}

Consider the two-triangle split for the 6-point 1-loop integrand in which we have two tree-level surfaces that overlap two triangles $(6,1,p)$ and $(4,5,p)$ so that:
\begin{equation*}
    \mathcal{S}_1 = (1,2,3,4,5,p,6), \quad \mathcal{S}_2 = (1,p,4,5,6),
\end{equation*}
where $p$ stands for the puncture label. So we have that $\mathcal{S}_1$ is a 7-point tree-level amplitude while  $\mathcal{S}_2$ is a 5-point one. We will use $x_{i,j}$ to denote the kinematics of the former and $y_{i,j}$ those of the later; since these are both describing tree kinematics we have, $x_{i,j}=x_{j,i}$ and $y_{i,j}=y_{j,i}$. 

By finding the decomposition of the curves of the 1-loop surface in terms of curves of the smaller surfaces, we land on the following split kinematic mapping:

\begin{equation}
\begin{aligned}
    &X_{1,i}\to \begin{cases}
        y_{1,p} + y_{1,5}, \, \text{for } i=1 \\
        y_{1,i}, \, \text{for } i\in[p,3,4] \\
        y_{1,5}+x_{i,p}, \, \text{for } i\in[5,6]
    \end{cases},\\
     &X_{2,i}\to \begin{cases}
        y_{2,i}, \, \text{for } i\in[p,4] \\
        y_{2,5} + y_{i,p}, \, \text{for } i\in[1,2] \\
        y_{2,5}+x_{i,p}, \, \text{for } i\in[5,6]
    \end{cases},\\
    &X_{3,i}\to \begin{cases}
        y_{3,p}, \, \text{for } i=p \\
        y_{3,5} + y_{i,p}, \, \text{for } i\in[1,2,3] \\
        y_{3,5}+x_{i,p}, \, \text{for } i\in[5,6]
    \end{cases},
\end{aligned} \quad \quad 
\begin{aligned}
    &X_{4,i}\to \begin{cases}
         y_{4,p}, \, \text{for } i=p \\
        x_{1,4} + y_{i,p} , \, \text{for } i\in[1,2,3,4] \\
        x_{4,6}, \, \text{for } i=6
    \end{cases},\\
     &X_{5,i}\to \begin{cases}
        x_{5,p}, \, \text{for } i=p \\
        x_{1,5} + y_{i,p}, \, \text{for } i\in[1,2,3,4] \\
        x_{1,5}+x_{5,p}, \, \text{for } i=5
    \end{cases},\\
    &X_{6,i}\to \begin{cases}
        x_{6,p}, \, \text{for } i=p \\
        y_{i,6} , \, \text{for } i\in[2,3,4] \\
        y_{5,6}+x_{i,p}, \, \text{for } i\in[5,6]
    \end{cases},
\end{aligned}
\end{equation}

where with $X_{j,p}$ we are denoting the curve that spirals counterclockwise around the puncture, as usual. However, as explained previously, in each tree-level problem, the integrals defining the amplitude run over the positive coordinate, $y$ going from $0$ to $+\infty$. To reproduce this domain of integration on the loop problem, need the add the loop contribution that converges in the region $1<\prod_{i}y_{i,p}<\infty$, and this the one in which we include the curves $X_{i,p}$ spiraling around the puncture clockwise, $X_{i,p}^{-}$. Evidently changing the spiraling of curves $X_{i,0}$ only affects the mapping of these curves, so for clockwise spiraling we obtain instead:
\begin{equation}
    X_{i,p}^- \to \begin{cases}
        y_{i,5},\, \text{for } i\in[1,2,3,6] \\
        x_{1,i},\, \text{for } i\in[4,5] 
    \end{cases} .
\end{equation}

Then we have:
\begin{equation}
    \left[\mathcal{I}_6(X_{i,j},X_{i,p}^+) +  \mathcal{I}_6(X_{i,j},X_{i,p}^-)\right]\bigg\vert_{\text{split kinematics}} \to \mathcal{A}_7(y_{i,j}) \times \mathcal{A}_5(x_{i,j}). 
\end{equation}

\subsection{Splits for physical kinematics and the physical integrand}
\label{sec:PhyKin}

All the split patterns described in this section are true for the surface integrand, $i.e.$ the object we get directly from the surface defined in section \ref{sec:SurfaceIntegrand}, in which we the kinematic variables associated to all curves distinct, $i.e.$ $X_{i,j} \neq X_{j,i}$, as well as their self-intersecting analogs. To connect this object to the physical integrand we need to do the identifications $X_{i,j}^{(q)}=X_{i,j} =X_{j,i} = X_{j,i}^{(q)}$. It turns out that asking for the self-intersecting curves to map to the same as their non-self-intersecting versions, simply amounts to also setting $x^{(q)}=x$ in the lower loop-problem we're splitting into. Therefore, the further constraints on the kinematics really come from asking for $X_{i,j} = X_{j,i}$. 

We will now understand what these constraints imply in general for the surface integrand for the split in which we get a (tree)$\otimes$(loop). The reason to focus on these is that they are the split patterns that allow us to access interesting physical limits, namely multi-soft limits, that lead to factorizations of the amplitude post-loop integration. In particular let's focus on the case in which $\mathcal{S}$ splits into $\mathcal{S}_1\otimes \mathcal{S}_2$, where $\mathcal{S}_1 = (2,3,\cdots,n_1-1,n_1)$ is an $(n_1-1)-$point one-loop surface and $\mathcal{S}_2 = (n_1-1,n_1,\cdots,n, 1,2)$ is an $(n-n_1+4)-$point tree, that overlap in triangle $(2,n_1-1,n_1)$ --- we are picking this particular pattern as this will be our running example later on to derive multi-soft statements in sections \ref{sec:SoftScalars}, \ref{sec:SoftNLSM} and \ref{sec:Softgluons}. Let's use $x_{i,j}$ to label the kinematics of $\mathcal{S}_1$ and $y_{i,j}$ those of $\mathcal{S}_2$, as usual. So going on physical kinematics on the smaller loop problem associated with $\mathcal{S}_1$ just means that we set $x_{i,j}^{(q)} = x_{i,j} = x_{j,i} = x_{j,i}^{(q)}$. However, we get more interesting relations by asking for the $X_{i,j}$'s of the big surface in which $i$ is an index of $\mathcal{S}_1$ and $j$ from $\mathcal{S}_2$, or when both $i$ and $j$ are in $\mathcal{S}_2$. For the particular split pattern we chose, these can be summarized as follows:
\begin{equation}
    \begin{cases}
        y_{n_1-1,j} = x_{n_1-1,n_1} + y_{2,j} \\
        y_{n_1-1,1} = x_{n_1-1,n_1}
    \end{cases} \, \begin{cases}
        y_{j_1,j_2} = y_{2,j_1} + y_{2,j_2} + x_{n_1,n_1} \\
        y_{1,j} = x_{n_1,n_1} + y_{2,j}
    \end{cases} \, y_{2,j} = x_{2,n_1} + y_{2,j} 
    \label{eq:RestrictionPhyKin}
\end{equation}
which forces $x_{2,n_1}=0$. 
In addition, the bubbles and tadpoles of the loop surface must also clearly be set to zero. 

The restriction to $X_{2,1} = x_{2,n_1} = 0$ would appear to force on a singular locus for the split kinematics, where both the full and the lower-point loop integrands have a $1/X_{2,1} = 1/0$ singularity. But with our discussion of the relation between the full surface integrand and the more conventional loop integrands, we can make an interesting well-defined statement by isolating and subtracting this external bubble factor. For any $n$ point one-loop integrand ${\cal I}_{1\, {\rm loop}}$, we define an ``external bubble subtracted'' integrand $\hat{{\cal I}}_{1\,{\rm loop}}$ via 
\begin{equation}
\hat{\cal I}_{1 \, {\rm loop}} = {\cal I}_{1 \, {\rm loop}} - \frac{1}{X_{2,1}} \left(\frac{1}{X_{p,1} X_{p,2}} + \frac{1}{X_{p,1} X_{1,1}} + \frac{1}{X_{p,2} X_{2,2}}\right) {\cal A}_{{\rm tree}}.
\end{equation}

By construction $\hat{\cal I}$ is completely independent of $X_{2,1}$ and is in particular well-defined as $X_{2,1} \to 0$. 
But now we have a nice point. On the split kinematics, $X_{2,1}$ as well as $X_{p,1},X_{p,2},X_{1,1},X_{2,2}$ are unshifted. Furthermore, the split factor on the lower-tree is exactly the same for ${\cal A}_{{\rm tree}}$. We thus learn that $\hat{{\cal I}}_{1\, {\rm loop}} $ factors on split kinematics in exactly the same pattern as ${\cal I}_{1 \, {\rm loop}}$, 
\begin{equation}
\hat{{\cal I}}_{n, {\rm split \, kin. }} = \hat{{\cal I}}_{\mathcal{S}_1,n_1}[x_{i,j}] \times {\cal A}_{\mathcal{S}_2,n_2}[y_{i,j}],
\end{equation}
with $n=n_1 + n_2 - 3$ as usual. 
This takes care of the singular behavior associated with setting $X_{2,1} = x_{2,n_1} \to 0$ to achieve the physical kinematics $X_{i,j} = X_{j,i}$ conditions. All that remains is to impose a constraint encountered above on the $x,y$ kinematics to ensure that $X_{i,j} = X_{j,i}$ for the big surface. 
We now see that these conditions precisely correspond to a certain \textit{minimal kinematics} for the tree-problem of $\mathcal{S}_2$.

\begin{figure}
    \centering
    \includegraphics[width=\textwidth]{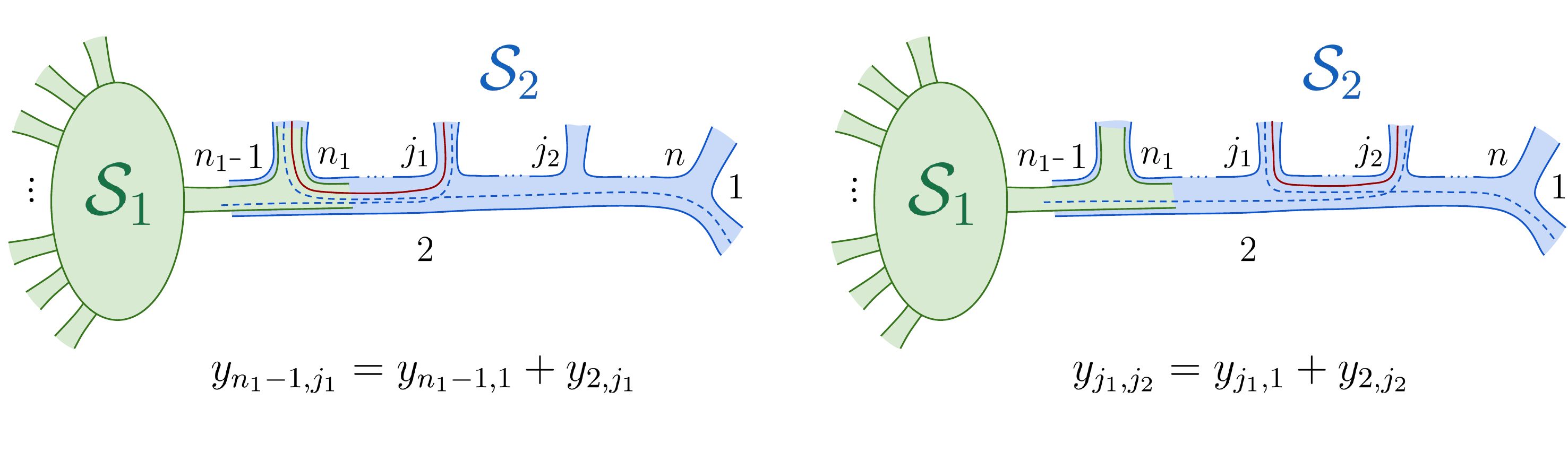}
    \caption{Fat-graph of $\mathcal{S}_2$ associated to the minimal kinematics that reproduce kinematical locus \eqref{eq:RestrictionPhyKin} giving physical kinematics for the one-loop split. Linear constraints on the kinematics of $\mathcal{S}_2$, $y_{n_1-1,j_1}$ on the left and $y_{j_1,j_2}$ on the right. }
    \label{fig:PhysKin}
\end{figure}
To understand which minimal kinematics this limit corresponds to, it is useful to consider the triangulation of the full surface $\mathcal{S}$ for which the triangulation of $\mathcal{S}_2$ contains chords $X_{2,n_1-1}, X_{2,n_1}, \cdots X_{2,n}$, which is a triangulation containing the triangle in which the two surfaces, $\mathcal{S}_{1,2}$ overlap (see figure \ref{fig:PhysKin}). So the minimal kinematics we land on precisely corresponded to the product of all the 4-points appearing in this triangulation/fat-graph. Let's now see that the kinematic limit in which we get this maximal split precisely agrees with \eqref{eq:RestrictionPhyKin}. Take curve $y_{n_1-1,j}$ (see figure \ref{fig:PhysKin}, left), now in the first 4-point problem it decomposes into a non-boundary curve corresponding to the $t$-channel of that 4-point, which is exactly the curve that $y_{n_1-1,1}$ decomposes into -- this curve is a boundary in all the remaining 4-point problems except for this one. Then in all the 4-points before the one involving propagator $X_{2,j_1}$ it decomposes into boundaries, whereas for the 4-point involving $X_{2,j_1}$ it maps to the $s$-channel of this 4-point. In this case the curve in $\mathcal{S}_2$ that maps to this curve is $y_{2,j_1}$ as it decomposes into boundaries in all the remaining 4-points it goes through other than the one involving $X_{2,j_1}$ in which it maps to the $s$ variable. A similar argument can be used to derive the physical kinematics for $y_{j_1,j_2}$, see figure \ref{fig:PhysKin} (right). Thus the split for physical kinematics can be written as:
\begin{equation}
\hat{{\cal I}}_{n, {\rm split \, kin. }} = \hat{{\cal I}}_{\mathcal{S}_1,n_1} [x_{i,j}] \times  \prod_{i=n_1}^{n}{\cal A}_{4}(y_{2,i},y_{i-1,1}), 
\end{equation}
where ${\cal A}_4(x,y) = (\frac{1}{x} + \frac{1}{y})$ is the 4-point tree amplitude.  

\section{All-loop splits}

Having seen how splits work in detail at tree-level and one-loop, in this section we make some general comments on splits at all orders in the topological expansion. There are many interesting aspects of splits to explore at all orders and we will not do a systematic study of all of them here. Instead, we will indicate some of the interesting novelties that arise, and define the simplest cases of ``loop $\otimes$ tree'' splits where the statements at all loop orders are essentially identical to what we have seen at one loop, allowing us to make all-order predictions for the planar integrand, as well as for integrated amplitudes in multi-soft limits even away from the planar limit. 

\subsection{Infinite integrands and the mapping class group}
There is one essentially new feature associated with ``surfaceology'' that does not show up at tree-level or planar one loop but exists for all other surfaces: for general surfaces, there are infinitely many curves that differ by ``winding'', or more formally by the action of the mapping class group (MCG) of the surface. In the field theory limit the curve integral most directly gives us an ``infinite integrand'', summing over {\it all} triangulations of the surface, which include infinitely many copies of the same diagrams, which differ only by the action of the mapping class group. 

Mathematically this infinite integrand is a perfectly canonical object.  There are infinitely many curves but depending on how the exponents $X$ are scaled at ``large winding'', the sums can be perfectly convergent. Physically, however, we want to keep just one copy of all the diagrams and so we mod out by the action of the mapping class group. In \cite{curveint,curveint2} this is done using the ``Mirzakhani trick'', adding a simple kernel to the curve integral essentially applying a cousin of the familiar Fadeev-Popov trick to the curve integral. 

It is interesting and non-trivial that a canonical ``infinite integrand'' exists for any surface, even beyond the planar limit. On the other hand, we have long appreciated that conventional integrands can be defined in the planar limit but not beyond. This has a nice interpretation in the language of the infinite integrand and the mapping class group. A planar $L$-loop amplitude is associated with a surface which is a disk with $L$ punctures. Nicely, for these surfaces, any two curves differing by the action of the MCG are assigned exactly the same momentum, since the punctures carry no momentum. Thus, the curve integral is invariant under the MCG, and we can mod out by the mapping class group before loop integration, giving us the well-defined loop integrand. Instead beyond the planar limit, two curves differing by the action of the MCG are {\it not} in general assigned the same momentum. This is seen in the simplest example of the annulus, where a curve winding around the inner circle of the annulus $w$ times is assigned momentum $(l + w q)$, where $l$ is the loop momentum assigned to the $w=0$ curve, and $q$ is the sum of all momentum on (say) the inner circle of the annulus. Thus beyond the planar limit, we can't mod out by the MCG to define an integrand. Nonetheless, the momenta differs only by a shift of the loop momentum and hence integrates to the same result. So we can first integrate over loop momenta, and then mod out by the MCG, giving the story of the ``surface symanzik polynomials'' in \cite{curveint,curveint2}.

\subsection{Splits for the infinite integrand} 

Returning to splits, our argument began with the manifestly factorized product of the curve integrals for the surfaces ${\cal S}_{1,2}$, then simply expressed the $u$ variables for each surface by the extension formula into the full surface ${\cal S}_1 \otimes {\cal S}_2$ to define kinematics for the full curve integral. Thus most directly, this defines kinematics for an infinite integrand of the big surface, that factorizes into a product of integrands of the two smaller ones. An illustration of this for the simple case of the annulus is given in appendix \ref{sec:annulusSplit}. 

Of course, if each surface has its own mapping class group action MCG$({\cal S}_{1,2})$, then we are free to mod out by each of them. But the kinematic data defined for the big surface will {\it not} in general be invariant under the full MCG of the join, instead we clearly have that MCG$({\cal S}_1 \otimes {\cal S}_2) \in {\rm MCG}({\cal S}_1) \times {\rm MCG}({\cal S}_2)$. So we get a split involving kinematics invariant only under this smaller group, and the full curve integral for the big surfaces can only be modded out by this product, not the full MCG of the big surface. This makes it obvious that joining two surfaces will not in general give us kinematics for the loop momenta that be interpreted in terms of an infinite integrand for the big surface. 

\subsection{Tree $\otimes$ Loop splits}
However, there is an important special case where split kinematics are directly physically interpretable: when one of the surfaces is a tree and the other contains all the non-trivial loop structure. In this case, the full mapping class group is obviously just the same as that of the surface containing the loops. This nice fact is one of the important ingredients in the ``loop-tree factorization'' of \cite{curveint2}, allowing all-multiplicity curve integrals at any loop-order, once those for ``tadpole'' diagrams with one particle per color trace factor are computed. 

\subsection{Splits for the planar integrand and all-order multi-soft limits}

In the planar limit, we can mod out by the MCG of the surface with the loops, so the ``tree $\otimes$ loop'' split kinematics gives us factorization at the level of the planar integrand. Beyond the planar limit, we can only mod out by the MCG post-loop integration. As we have already seen at one-loop, in general, split kinematics will restrict the values of certain loop propagators, and so can't extend to loop integrations. But as we will discuss in detail in the coming section, if the momenta in the tree-level surface are taken to be soft, the split restriction on kinematics can be interpreted as dictating masses for the propagators, rather than restricting loop momenta, and so the splits give rise to loop integrated statements in these multi-soft limits, at all orders in the topological expansion. 

\subsection{Physical kinematics at all orders}

Already at one-loop, we saw that requiring the locus of split kinematics to agree with the assignment of physical momenta puts a restriction on the pattern of splits. The constraints come from curves $(i,j)$ that start and end on the tree part of the fatgraph. These curves can either take the ``short'' path from $i$ to $j$--staying purely within the tree part of the fatgraph--or can take ``long'' way starting from $i$ traversing into the loop surface and returning back to $j$. The pictures have already been given in \ref{fig:PhysKin}, and indeed the restrictions on the $y$ variables of the tree surface are exactly the same as the one-loop ones. The only difference is that for general surfaces, we will have many $x_{n_1,n_1-1}, x_{n_1,n_1},x_{2,n_1}$ variables corresponding to all these external-bubble and tadpole-type curves that can wind around in the big surface, so all these variables must take the same value for physical kinematics. And once again, we force all the $x_{2,n_1} \to 0$, which demands all the external bubbles $X_{2,1} \to 0$ as well, but as at one-loop, we can canonically subtract all these external bubble factors, and use the universality of the split factor to make the same statement of factorization for the well-defined, bubble-subtracted objects.

\subsection{Characterization of split kinematics for general surfaces}
We close this short section by explaining how to characterize the split kinematics in terms of constraints on the $X$ variables of the big surface for general surfaces, in the same we saw at tree-level and one-loop. The key idea is what we already encountered in these earlier examples: it is easy to identify curves on the full surface whose kinematics are just the same as those on the smaller surfaces. The split kinematics of the remaining curves then give us the constraints we are imposing on the curves of the big surface. 

\begin{figure}[t]
    \centering
    \includegraphics[width=\textwidth]{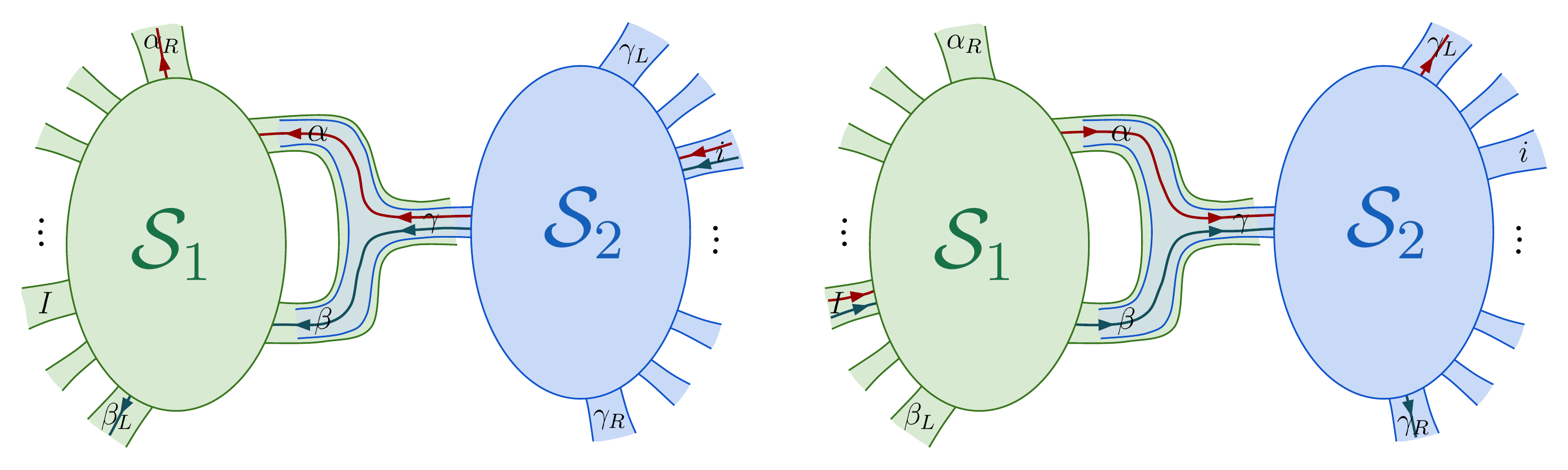}
    \caption{In the left panel, we see that curves in ${\cal S}_2$ can be continued into curves in all of ${\cal S}$ by continually turning left or right, so that they are a  boundary curve when restricted to ${\cal S}_1$. In the right panel we see the analagous fact for curves in ${\cal S}_1$.}
    \label{fig:extbound}
\end{figure}

This is illustrated in figure \ref{fig:extbound}.
Let's pick the non-trivial case of a curve that begins on external road $i$ in ${\cal S}_2$ and ends in the external road $\alpha$, getting to $\alpha$ by making a right turn from $\gamma$. Then we can extend this to a curve on all of ${\cal S}$ by simply continuously turning right after $\alpha$. 
Similarly for curves that make a left out of $\gamma$ into $\beta$, for which we turn continually left after $\beta$. In the same way for curves $x_{(1)}$ in ${\cal S}_1$ and end up going into the boundary road $\gamma$, turning left onto $\gamma$ from $\alpha$, we can keep turning left continually after $\gamma$ going into $\gamma_L$, and similarly for those going into $\gamma$ by making a right from $\beta$, we keep turning right exiting in $\gamma_R$. In this way we can identify the variables $x_{(1),(2)}$ of the two surfaces as 
\begin{equation}
\begin{aligned}
&x_{(1); (I, \alpha \to \gamma)} = X_{(I, \gamma_L)}, \,  x_{(1); (I, \beta \to \gamma)} = X_{(I, \gamma_R)}, \\ 
&x_{(2);(i,\alpha)} = X_{(i, \alpha_R)}, \, x_{(2);(i,\beta)} = X_{(i, \beta_L)}.
\end{aligned}
\end{equation}

Inserting these expressions into the equation specifying split kinematics given in the introduction, 
\begin{equation}
X_{{\cal S} = {\cal S}_1 \otimes {\cal S}_2}[x_{(1)},x_{(2)}] = \sum_{x_{(1)}} \#[x_{(1)} \subset X] x_{(1)} + \sum_{x_{(2)}} \#[x_{(2)} \subset X] x_{(2)},
\label{eq:again}
\end{equation}
then gives us the characterization of the split kinematics in terms of linear relations on the kinematic $X$ variables of ${\cal S}$. 

\section{Masses, splits and the $\delta$ shift}
We now want to understand how the split generalizes for the case of massive kinematics. To go from the massless case to the massive one all we do is $X_{i,j} \to X_{i,j}+m_{i,j}^2$, so that the poles are now located at $X_{i,j} = - m_{i,j}^2$. Of course, by allowing a different mass to each propagator we are considering the most general mass pattern a theory could have, which obviously contains the simpler cases in which most masses are the same. 

As we will see, given a particular split, there will be a family of solutions of $m_{i,j}^2$ for which the split holds, and we can determine this family systematically for each split. However, asking for which set of $m_{i,j}^2$ accommodate \textit{all} the possible splits, already at tree-level reduces it down to the simple: 
\begin{equation}
    m_{i,j}^2 \equiv \delta_{i,j} = \begin{cases}
        +\delta, \quad \text{ if } $(i,j)$ \text{ are both even}, \\
        -\delta, \quad \text{ if } $(i,j)$ \text{ are both odd}, \\
        0, \quad \, \,\, \,\text{ otherwise. } 
    \end{cases}
    \label{eq:DeltaShift}
\end{equation}
this is precisely the $\delta$-shift \cite{zeros} that allows us to go from scalars to pions and gluons. For the case of pions, this shift can be applied directly at the level of the field theory amplitudes giving us the non-linear sigma model (NLSM) at low energies \cite{NLSM,CirclesNLSM}, while for the case of gluons, applying this kinematic shift with $\delta = 1/\alpha^\prime$ in the full-surface integral gives us scalar-scaffolded gluons at low energies \cite{Gluons}.

\subsection{General masses}
\label{sec:genMasses}
Let's start by understanding what pattern of masses is compatible with a given split at tree-level as the procedure to determine the same thing at loop-level is completely analogous. 

Say we start with a given split pattern in which an $n$-point tree, $\mathcal{S}$, splits into $\mathcal{S}_1\otimes\mathcal{S}_2$, which are, respectively, $n_1$ and $n_2$-trees with $n_1+n_2-3=n$, and let's label the kinematics in $\mathcal{S}_1$ by $x_{i,j}$ and $y_{k,m}$ those of $\mathcal{S}_2$. Then, building the split kinematics, we get that a total of $n_1(n_1-3)/2+ n_2(n_2-3)/2$ out of the $n(n-3)/2$ $X$'s in $\mathcal{S}$ map into a single curve in the lower point surfaces, $i.e.$
\begin{equation}
    X_{i^\star,j^\star} = x_{i,j}, \quad \quad X_{k^\star,m^\star} = y_{k,m},
\end{equation}
where $X_{i^\star,j^\star}$ and $X_{k^\star,m^\star}$ are the curves in $\mathcal{S}$ that decompose into solely $x_{i,j}$ and $y_{k,m}$ respectively, for each $(i,j)\in \mathcal{S}_1$ and $(k,m)\in \mathcal{S}_2$. Now we can give any masses, $m_{i^\star,j^\star}^2$, $m_{k^\star,m^\star}^2$, to these $n_1(n_1-3)/2$ $X_{i^\star,j^\star}$'s and $n_2(n_2-3)/2$ $X_{k^\star,m^\star}$'s and this automatically determines the masses of the propagators in the subsurfaces $\mathcal{S}_1$, $\mathcal{S}_2$. 

Now as derived in \ref{sec:splitsConstraintsTree} (for the tree case) and in \ref{sec:SplitKinLoop} (for the loop-case), in the split kinematics all remaining $X$'s in $\mathcal{S}$ become linear functions in the $X_{i^\star,j^\star}$'s and $X_{k^\star,m^\star}$'s. Therefore we can use these linear relations to determine the masses of the remaining propagators of $\mathcal{S}$ in terms of $m_{i^\star,j^\star}^2$, $m_{k^\star,m^\star}^2$ -- and the final result gives us a mass pattern compatible with the split. So we have that, given split, there is a family of solutions of masses compatible with the split that is parametrized by the masses, $m_{i^\star,j^\star}^2$, $m_{k^\star,m^\star}^2$, of the curves in $\mathcal{S}$ that map into single component curves in the lower surfaces. 

Recall that, at tree level, the linear relations between the $X$'s that land us on split kinematics are equivalent to setting a collection of non-planar variables $c_{i,j}$ to zero, as explained in \ref{sec:splitsConstraintsTree}. So another way of phrasing the mass assignments compatible with the split is by asking that the mass shift  $X_{i,j} \to X_{i,j}+m_{i,j}^2$ preserves the $c_{i,j}$'s that are set to zero on the split kinematics. 

From this point of view, it is trivial that, at least at tree-level, the only mass shift that is compatible with \textit{all} split patterns is then the one that preserves all the $c_{i,j}$, and as argued in \cite{zeros} the only such shift is precisely the $\delta$-shift \eqref{eq:DeltaShift}. At loop-level, as we will see momentarily, the $\delta$-shift does accommodate all possible split patterns however this is not proven whether it is the only mass shift that does it. 

\subsection{The $\delta$ shift: the NLSM and scalar-scaffolded gluons}
\label{sec:deltaShift}
It has recently been understood that Tr$(\Phi^3)$ theory secretly contains both the NLSM and (in its stringy form) scalar-scaffolded Yang-Mills amplitudes via the simple kinematic shift $\delta_{i,j}$ in \eqref{eq:DeltaShift} which can phrase directly in terms of the kinematics as:
\begin{equation}
    X_{i,j} \to X^{(\delta)}_{i,j} = X_{i,j} + \delta_{i,j}, \quad \text{where } X^{(\delta)}_{e,e} = X_{e,e} + \delta, \, \, X^{(\delta)}_{o,o} = X_{o,o} - \delta.
    \label{eq:deltaShift}
\end{equation}

The connection to the non-linear sigma model can be made just with the field-theoretic Tr$(\Phi^3)$ amplitudes, and has a simple Lagrangian derivation that connects to the NLSM for the $U(N) \to U(N-k) \times U(k)$ symmetry breaking pattern \cite{NLSM}, establishing the link between kinematically shifted Tr$(\Phi^3)$ amplitudes and this theory to all orders in the topological expansion. The amplitudes for the conventional chiral Lagrangian associated with the $U(N) \times U(N) \to U(N)$ symmetry breaking pattern turn out to co-incide with those of the $U(N) \to U(N-k) \times U(k)$ in the planar limit.

Now, we have already explained why the $\delta$-shift trivially accommodates all the splits at tree-level, so in this section, we will understand how this phenomenon generalizes to loop-level.  We will focus on the $\delta$ shifts in the field theory limit, and so will only look at the split kinematics defined for non-self-intersecting curves. 
\\ \\
\textbf{Pions} 
To arrive at the NLSM, we consider kinematics $X^{(\delta)}_{r,s}$ for Tr$(\Phi^3)$ where $X_{r,s} = \delta_{r,s} + X^{\text{NLSM}}_{r,s}$ and we look at low energies  $X^{\text{NLSM}}_{r,s} \ll \delta$. Now we have seen that the split kinematics forces certain linear constraints on $X_{r,s}$, and hence in order to be able to reach the split kinematics in the NLSM, setting $X_{r,s} \to \delta_{r,s}$ must satisfy all these constraints. If this happens, for the general kinematics $X_{r,s} = \delta_{r,s} + X^{\text{NLSM}}_{r,s}$, the linear constraints on $X_{r,s}$ simply turn into the same linear constraints on the $X^{\text{NLSM}}_{r,s}$, defining split kinematics for the NLSM. 

We have seen many different linear relations that the $X_{r,s}$ must satisfy at tree and loop level, but quite remarkably, $\delta_{r,s}$ automatically satisfies all of them! Indeed, everything follows from the simple fact that the $\delta_{r,s}$ satisfy a ``mesh relation'', associated with a square of four indices: 

\begin{equation} 
\delta_{r,s} + \delta_{v,w} - \delta_{r,v} - \delta_{s,w} = 0
\end{equation}
for all $r,s,v,w$. This can trivially be verified by looking at all the possible parities for $r,s,v,w$. 

Let us see how this relation guarantees that the $\delta_{r,s}$ satisfy all the split kinematic relations. We begin at tree-level, where the linear constraint was that $X_{a,b} = X_{a,i} + X_{i-1,b}$, $X_{a,c} = X_{a,k-1} + X_{k,c}$. These are satisfied by the $\delta_{r,s}$ since $\delta_{i-1,i} = 0$ given that $i-1,i$ have opposite parity so 

\begin{equation}
0 = \delta_{a,b} + \underbrace{\delta_{i-1,i}}_{0} - \delta_{a,i} - \delta_{b,i-1}
\end{equation}
and similarly for the the $\delta_{a,c}$ relation. The same thing happens at loop level, in most cases the mesh relation directly gives the needed relation. There are some very slightly more interesting cases. For instance in case 1. at one-loop, we had the relation $X_{a,a^\prime} = X_{a,i} + X_{a^\prime, i} + X_{i-1,i-1}$; this has three plus signs and doesn't look like a ``mesh'' relation. Nonetheless, using the obvious fact that $\delta_{r,s} = - \delta_{r-1,s-1}$, we have that 
\begin{equation}
\delta_{a, a^\prime} - \underbrace{\delta_{i-1,i-1}}_{+\delta_{i,i}} - \delta_{a,i} - \delta_{a^\prime, i} = \delta_{a,a^\prime} + \delta_{i,i} - \delta_{a^\prime, i} - \delta_{a,i} = 0.
\end{equation}

Another amusing example arose in case 2. Here we had the relation $X_{b,a} = X_{a,i-1} + X_{b,k} + X_{k-1,i}$. Here not only do we have three plus signs, the relation involves six indices $(a,b,i-1,i,k-1,k)$, not four! But we have already seen that $\delta_{a,b} = \delta_{a,s-1} + \delta_{b,s} - \delta_{s-1,s} = \delta_{a,s-1} + \delta_{b,s}$ since $\delta_{s-1,s} = 0$. Hence 
\begin{equation}
\delta_{a,i-1} + \underbrace{\delta_{b,k} + \delta_{k-1,i}}_{\delta_{b,i}} = \delta_{a,i-1} + \delta_{b,i} = \delta_{a,b},
\end{equation}
exactly as needed. 

Thus the $\delta_{r,s}$ satisfies all the linear relations of the split kinematics, and hence we can define split kinematics for NLSM amplitudes \footnote{Here we are referring to the split patterns that give us factorization into (tree)$\otimes$(loop) as this is the one that gives us access to physically interesting limits.}. Note that we also learn that the NLSM amplitudes will factor onto shifted Tr$(\Phi^3)$ amplitudes, where we can read off the shifts $\delta^{(x)}, \delta^{(y)}$ for the surfaces from our general expressions for $x,y$ in terms of $X$, specialized to the case where $X_{r,s} =\delta_{r,s}$. In other words, the shift $\delta^{(x)}$ for any $x$ in ${\cal S}_1$ is given by the $\delta$ shift for the curve $X$ that either coincides with $x$ or extends $x$ into ${\cal S}$ only intersecting ${\cal S}_2$ in boundaries of ${\cal S}_2$, and similarly for the shifts $\delta^{(y)}$ -- exactly in the same way we determine the general mass shifts for the kinematics of the subsurfaces. 

Some special cases are worth commenting on, where one or both of  $\delta^{(x)},\delta^{(y)}$ shifts are the same as $\delta$, which will imply that one or other subsurface is also shifted to give NLSM integrands. For instance, at tree-level, we have that 
\begin{equation}
\delta^{(x)}_{k,b} = 
\delta_{i-1,b}, \quad  \delta^{(y)}_{j,a} = \delta_{k-1,a},
\end{equation}
with all other shifts the same as the $\delta$ shift. We conclude that 
\begin{equation}
\begin{aligned}
& \delta^{(x)}_{r,s} = \delta_{r,s} \, {\rm when \,} k, i \, {\rm have \, opposite \, parity} \\ 
& \delta^{(y)}_{v,w} = \delta_{v,w} \, {\rm when \,} k, j \, {\rm have \, opposite \, parity}
\end{aligned}.
\end{equation}

Now if $i,k$ have opposite parity and the surface ${\cal S}_1$ has an even number of vertices, then the indices of ${\cal S}_1$, $(k,i,i+1,\cdots, k-1)$ are alternating in parity, and therefore the $\delta^{(x)}$ shift gives the NLSM for ${\cal S}_1$. Similarly if $k,j$ have opposite parity and ${\cal S}_2$ has an even number of indices, then the indices for ${\cal S}_2$, $(i,j,k,k+1, \cdots, i-1)$ are also alternating if $(i,k)$ have the same parity. So summarizing, 
\begin{equation}
\begin{aligned}
& {\cal S}^{\rm (even)}_1 \, \text{ is the NLSM  when $(k,i)$ have opposite parity} \\ 
& {\cal S}^{\rm (even)}_2 \, \text{ is the NLSM  when $j$ has  opposite  parity to  $(k,i)$}
\end{aligned}.
\end{equation}

Note that we can never get that {\it both} ${\cal S}_{1,2}$ are the NLSM, for the trivial parity reason that $n_1 + n_2 = n+3$, and hence if $n$ is even, we can't have both $n_1,n_2$ be even. 
Nonetheless, by doing a further split on the odd-point non-NLSM factor into two even factors,  we discover the ``near-zero'' factorization of \cite{zeros} as  (4-point Tr $(\Phi^3)$) $\times$ (NLSM) $\times$ (NLSM). 
\\ \\
\textbf{Scalar-scaffolded gluons}
We have learned that sending $\alpha^\prime \delta \to 1$ on the surface integrals gives us scalar-scaffolded gluon amplitudes at low-energies \cite{Gluons}. At loop level, in order to correctly describe Yang-Mills amplitudes, it is important to include curves with one self-intersection per puncture in the curve integral.  However, as was observed previously, the mappings associated to the self-intersecting curves are completely analogous to the non-self-intersecting ones, and are compatible with the $\delta$ shift.

In order to discover implications for gluon amplitudes from splits, we must further take ``scaffolding residues'' on these scalar amplitudes. We will some some brief comments about this in section \ref{sec:ScalarScaffGluons}, leaving a fuller exploration to future work \cite{gluonUpcoming}.  

\section{Multi-soft limits for all-loop-integrated massive Tr$(\Phi^3)$ theory}
\label{sec:SoftScalars}

The factorization associated with split kinematics is true for finite values of the kinematic variables, $x,y$, of surfaces ${\cal S}_1,{\cal S}_2$. For the ``tree $\otimes$ loop'' splits, there is an especially simple limit where splits are obviously related to soft limits, where all the kinematic variables associated with the ``tree'' factor are set to zero. In this limit, splits turn into soft limits, which we will examine in this section. 

One important feature of taking soft limits is that they make it possible for split kinematics to be realized without putting any restrictions on loop momenta. We have already seen at 1-loop, that split kinematics relates the $X_{i,p}$ variables for different $i$. For instance, in our first 6-point one-loop example, we found, as shown in equation \eqref{eq:6pt1loop_Map}, that split kinematics enforced $X_{6,p} = X_{5,p} + y_{2,6}$. 
Now, if we associate the curve $(5,p)$ with momentum $l^\mu$, then $X_{5,p} = l^2 + m_5^2, \, X_{6,p} = (l+p_5)^2 + m_6^2$, where $m_{5,6}^2$ are masses for the corresponding loop propagators. Thus the split kinematics demands that 
\begin{equation}
(l + p_5)^2 + m_6^2 = l^2 + m_5^2 + y_{2,6} \to 2 l \cdot p_5 + p_5^2 + m_6^2 = m_5^2.
\end{equation}

Evidently in general this equation puts restrictions on the loop momentum, which means that the splits give information for some fixed loop momenta but not for all loop momenta, and hence don't give rise to loop-integrated statements. But this is not the case if we take the soft limit where $p_5 \to 0$. In this case the split kinematics only dictates a relationship between the masses and the kinematics of the tree-factor $m_6^2 = m_5^2 + y_{2,6}$, which holds for all $l$ and hence gives rise to a split factorization statement that holds post-loop-integration. 

There are in fact two related but different sorts of soft limits we will be discussing here. The first is soft limits for massless Tr$(\Phi^3)$ theory. Already at tree-level, what we might call the ``leading soft theorem'' for these amplitudes is a trivial consequence of standard factorization near collinear poles, but we will point out that split factorization makes interesting predictions for an infinite number of subleading terms in the soft-expansion. We will however leave a fuller exploration of this aspect of splits to future work, and next turn to a different kind of soft limit, for amplitudes with internal masses at loop level, where the split kinematics are interpreted as fixing relationships between the internal masses as we saw in our example above.

\subsection{Splits, leading and subleading soft theorems and beyond}

Before going back to splits, let us recall that for all theories and in particular for Tr($\Phi^3$), leading soft theorems follow trivially from the more basic statement of factorization on collinear poles. If particle $n$ goes soft, so that the point $n$ is on top of the point $1$, we have that $X_{1,n-1} \to 0$ and $X_{2,n} \to 0$, with all other $X_{n,j} \to X_{1,j}$. In order to better understand this limit, we can divide the amplitude into the part that has poles as $X_{2,n} \to 0$ and $X_{1,n-1} \to 0$ and the rest. Working at tree-level for simplicity, we can thus write 
\begin{equation}
{\cal A}(1, \cdots, n) = \frac{{\cal A}(2,3,\cdots, n)}{X_{2,n}} + \frac{{\cal A}(1,2 \cdots, n-1)}{X_{1,n-1}} + {\cal R},
\end{equation}
here the coefficients of $\frac{1}{X_{2,n}}, \frac{1}{X_{1,n-1}}$ are simply determined by factorization, while ${\cal R}$ is independent of $X_{2,n}$ and $X_{1,n-1}$. Already we see that the leading soft theorem -- associated with the poles as $X_{2,n},X_{1,n-1} \to 0$ -- is entirely dictated by the conventional factorization of the amplitudes on poles. But with this knowledge alone, we don't know anything about ${\cal R}$ and  can't predict anything about the subleading terms in soft expansion. 

This is where splits appear, as they tell us there is a kinematical locus on which the amplitudes factorize even at finite values of the kinematics, not only in the soft limit. Indeed, we will see that splits imply that if we expand $X_{n,j} = X_{1,j} + s_j$ and think of ${\cal R}$ as a function  of the soft $s_j$ via
\begin{equation}
{\cal R} = {\cal R}_{(0)} + {\cal R}_j s_j + \frac{1}{2} {\cal R}_{j,k}  s_j s_k + \cdots,
\end{equation}
that the leading term ${\cal R}_{(0)}$ is determined by derivative of the lower amplitude as 
\begin{equation}
{\cal R}_{(0)}= \sum_j \frac{\partial {\cal A}(1, \cdots, n-1)}{\partial X_{1,j}}.
\end{equation}

Meanwhile, the higher order terms in the soft expansion can not be predicted in terms of operators acting on the lower-point amplitude, they do satisfy nice sum rules of this character, such as
\begin{equation}
\sum_j {\cal R}_{j} = \sum_{j,k} \frac{\partial^2 {\cal A}(1, \cdots, n-1)}{\partial X_{1,j} \partial X_{1,k}}\, \sum_{j,k} {\cal R}_{j,k} = \sum_{j,k,l} \frac{\partial^3 {\cal A}(1, \cdots, n-1)}{\partial X_{1,j} \partial X_{1,k} X_{1,l}}, \cdots.
\end{equation}

The derivation of these statements from splits is straightforward. Consider the simplest split where the surfaces overlap on the triangle $(2,n-1,n)$. Then on split kinematics we have that $X_{n,j} = X_{1,j} + X_{2,n}$, which implies 
\begin{equation}
    {\cal A}_n \left( X_{n,j} = X_{1,j} + X_{2,n} \right) = \left(\frac{1}{X_{1,n-1}} + \frac{1}{X_{2,n}} \right) \times M(1, \cdots, n-1).
\end{equation}

Combined with the definition of ${\cal R}$ and simple relabelling
\begin{equation}
\left.{\cal A}(2, 3,\cdots, n)\right|_{X_{n,j} = X_{1,j} + X_{2,n}} = \left.{\cal A}(1,2, \cdots, n-1)\right|_{X_{1,j} \to X_{1,j} + X_{2,n}},
\end{equation}
this tells us that 
\begin{equation}
{\cal R} \left(X_{n,j} = X_{1,j} + X_{2,n} \right) = \frac{1}{X_{2,n}} \left({\cal A}(1, \cdots, n-1) - \left.{\cal A}(1, \cdots, n-1)\right|_{X_{1,j} \to X_{1,j} + X_{2,n}} \right).
\end{equation}

Hence while conventional factorization on poles tells us nothing about ${\cal R}$, the split factorization pattern allows us to predict ${\cal R}$ along a line in the $s_j = X_{n,j} - X_{1,j}$ space, where $s_j = X_{2,n}$ are all equal to $X_{2,n}$. Taylor expanding the above statement in powers of the $X_{2,n}$ gives us the expressions for ${\cal R}_{(0)}$ as well as the sum rules on ${\cal R}_j, {\cal R}_{j,k}, \cdots$ given above. 

It is easy to extend this analysis to the case of several particles going soft, and, in this way, using the $\delta$-deformations to pions and gluons, we can access to leading and subleading soft limits for these various theories of massless particles. We will have more to say about the non-linear sigma model and scalar-scaffolded gluons below, but we now move on to describing another sort of soft limits for Tr $\Phi^3$, where masses are added but momenta are still taken to be soft. 

\subsection{Massive multi-soft limits 
at tree-level} 
\begin{figure}[t]
    \centering
    \includegraphics[width=\textwidth]{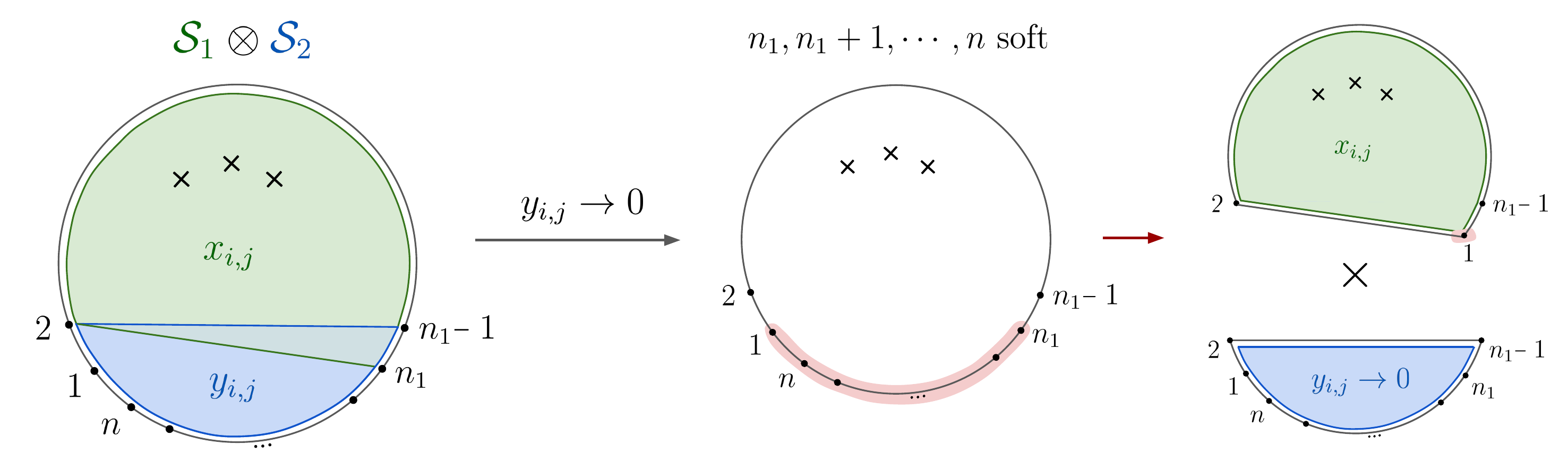}
    \caption{Split into $\mathcal{S}_1$(loop)$\otimes \mathcal{S}_2$(tree). In the limit where the kinematical variables of the tree $y_{i,j}$ go to zero, we precisely get the multi-soft limit in which particles $n_1,n_1+1,\cdots,n$ go soft. In this limit $\mathcal{S}_1$ is the lower-point amplitude while the  $\mathcal{S}_2$ amplitude is giving us the usual soft-factor.}
    \label{fig:softTree}
\end{figure}

Let's start by understanding how the splits give us access to multi-soft limits in the simplest case of tree-level. Consider an $n$-point tree, $\mathcal{S}$, splitting into $\mathcal{S}_1\otimes \mathcal{S}_2$, where $\mathcal{S}_1 = (2,3,\cdots,n_1-1,n_1)$ is an $(n_1-1)-$point tree and $\mathcal{S}_2 = (n_1-1,n_1,\cdots,n, 1,2)$ is an $(n-n_1+4)-$point tree, that overlap in triangle $(2,n_1-1,n_1)$ (see figure \ref{fig:softTree}). Let's use $x_{i,j}$ to label the kinematics of $\mathcal{S}_1$ and $y_{i,j}$ those of $\mathcal{S}_2$, as usual.

Now the split kinematics tell us that:
\begin{equation} 
\begin{cases}
    X_{k,1} = x_{k,n_1} \\
    X_{k,j} = x_{k,n_1} + y_{2,j}
\end{cases},\quad  X_{n_1-1,j} =  y_{n_1-1,j}, \quad X_{n_1,j} =  y_{n_1,j},  \quad X_{2,n_1} = y_{2,n_1},
\end{equation}
for  $k\in \mathcal{S}_1 \setminus \{n_1-1,n_1\}$, and $j\in \mathcal{S}_2 \setminus \{2,n_1-1\}$. So if we consider the limit in which all the kinematics of $\mathcal{S}_2 $ go to zero, $i.e.$ $y_{i,j}\to 0$, then the split kinematics becomes:
\begin{equation} 
X_{k,j} \to X_{k,1},\quad X_{n_1-1,j}\to 0, \quad X_{1,j} \to 0, \quad  X_{n_1,j} \to 0 , \quad X_{2,n_1} \to 0,
\end{equation}
which is precisely the limit we get if we make particles $n_1,n_1+1,\cdots, n$ soft -- the multisoft limit in which we make $n-n_1$ particles soft. At the level of the surface, this corresponds to making the indices $n_1,n_1+1,\cdots, n-1$ and $1$ all lie on top of each other so that we are left with an $(n_1-1)-$point problem. 

As we have discussed above in this limit, the massless Tr$(\Phi^3)$ amplitudes diverge as dictated by the leading soft expansion associated with massless poles. The same is not true if we add masses. As explained in \ref{sec:genMasses}, given a split pattern, we are free to give arbitrary masses, $m_{i^\star,j^\star}^2$, to the curves $X_{i^\star,j^\star}$ on the big surface $\mathcal{S}$ that map to a single curve on one of the smaller surfaces. In the split kinematics, all the remaining curves on $\mathcal{S}$ are given as linear combinations of these $X_{i^\star,j^\star}$'s, and therefore their masses$^2$ are also linear combinations of $m_{i^\star,j^\star}^2$. So to make the multi-soft limit finite all we need to do is give masses to the curves in $\mathcal{S}$ that map to curves solely living in $\mathcal{S}_2$ under the split kinematics. In the case we are studying this corresponds to having:
\begin{equation}
    X_{k,m} \to X_{k,m} + m^2_{k,m}, \Rightarrow  X_{k,m} + m^2_{k,m}= y_{k,m} +m^2_{k,m}, \quad \text{for } k,m \in \mathcal{S}_2
\end{equation}

So in the split limit, we obtain:
\begin{equation}
    \mathcal{A}_{\mathcal{S}} \to \mathcal{A}_{\mathcal{S}_1}(x_{i,j}+m^2_{i,j}) \times \mathcal{A}_{\mathcal{S}_2}(y_{k,m}+m^2_{k,m}),
\end{equation}
where we are allowing the curves that decompose into single curves in $\mathcal{S}_1$ to also have general mass$^2$, $m_{i,j}^2$ (but, evidently, we can also have them be massless). Now in this massive case, we can take the soft-limit in which all the kinematic invariants in $\mathcal{S}_2$ go to zero, $y_{k,m} \to 0$, yielding:
\begin{equation}
    \mathcal{A}_{\mathcal{S}} \to \mathcal{A}_{\mathcal{S}_1}(x_{i,j}+m^2_{i,j}) \times \underbrace{\mathcal{A}_{\mathcal{S}_2}(m^2_{k,m})}_{\text{multi-soft factor}}. 
\end{equation}

\subsection{All-loop massive multi-soft limits}

Now that we have understood how the split connects to the multi-soft limit at tree level, we can proceed to study the loop case which as we will see momentarily is completely analogous. The only subtlety is that the object we get from the surface --- the surface integrand --- does not agree with the standard physical integrand, as explained in section \ref{sec:SurfaceIntegrand}. Therefore all the loop-integrated statements derived here apply to $\mathcal{A}_n^{\mathcal{S}}$, in the end, we will explain how they generalize to $\mathcal{A}_n^{\text{standard}}$.

To access the multi-soft limit we need to study exactly the same split that we look at in the tree case, with only the difference that now $\mathcal{S}_1$ is a loop surface. For simplicity let's consider the case in which $\mathcal{S}_1$ is a one-loop surface, however, it should be clear that all the arguments follow analogously for the multi-loop case. The main thing we need to bear in mind is how the physical kinematics affect the split result. 

So if we start with the surface integrand we can go on the split kinematical locus for which the integrand splits into the tree-factor, $\mathcal{S}_2$, times the loop factor, $\mathcal{S}_1$. Now once more when we make all the kinematic invariants in the tree problem go to zero, the split mapping reduces to the multi-soft limit where particles $n_1, n_1+1, \cdots, n$ go soft (just as we saw in the tree-level example). Evidently, for this limit to be well-defined we need to give masses to all the chords, $X_{i,j}$ in $\mathcal{S}$ that map to curves exclusively in $\mathcal{S}_2$, under split kinematics, $X_{i,j} = y_{i,j}$. Doing this ensures that the kinematics in $\mathcal{S}_2$ are also $y_{i,j} + m^2_{i,j}$, and therefore sending $y_{i,j}\to 0$ does not produce any new divergences. Naively it seems like we are free to give different masses to each curve but as we will see in a moment this is not the case if we further impose physical kinematics. 

At last, we want to connect this result to a statement about the physical integrand as defined in \eqref{eq:PhysIntegrand}, more specifically for the case of massive Tr$(\Phi^3)$ \eqref{eq:PhysIntMass}. To do this we need to impose the condition $X_{i,j}=X_{j,i}$, $i.e.$ go to physical kinematics as described in section \ref{sec:PhyKin} (for one-loop), as well as ensure that by doing this we don't produce any new divergences. 

As explained in section  \ref{sec:PhyKin}, the physical kinematics under the split imply $x_{2,n_1}=X_{2,1}=0$, however since we are precisely subtracting the external bubbles to obtain the standard integrand, we can go on this locus without producing any new divergences (see sec. \ref{sec:PhyKin}). 

In addition from the remaining constraints coming from imposing $X_{i,j}=X_{j,i}$, \eqref{eq:RestrictionPhyKin}, we conclude that the masses of the tree problem are also restricted in the following way:
\begin{equation}
        m^2_{n_1-1,j} -m^2_{2,j}= m^2_{n_1-1,1} , \, \quad  
        m^2_{j_1,j_2} -  m^2_{2,j_1} - m^2_{2,j_2} = m^2_{1,j} - m^2_{2,j},
    \label{eq:MassRestriction}
\end{equation}
with $j,j_1,j_2 \in\mathcal{S}_2 \setminus \{2,n_1-1\}$ and $m^2_{i,j}$ denotes the mass associated to propagator $y_{i,j}$ in $\mathcal{S}_2$.

Now by further going on the multi-soft limit which implies $y_{i,j} \to 0$, the physical kinematics constraints yields $x_{n_1-1,n_1} \to 0$ and  $x_{n_1,n_1} \to 0$. Now the curves in the big surface that map to these are $X_{1,n_1-1}=x_{n_1-1,n_1}$ and $X_{1,1}=x_{n_1,n_1}$, so neither of them corresponds to an external bubble so the argument used for $X_{2,1}\to 0$ does not hold. However, in both as long as these propagators have a mass then we get no divergence in the limit they are set to zero. From \eqref{eq:RestrictionPhyKin}, the mass$^2$ of $x_{n_1-1,n_1}$ must be that of $y_{n_1-1,1}$, $m^2_{n_1-1,1}$, and therefore there is no divergence when $x_{n_1-1,n_1}\to 0$; and similarly the mass$^2$ of $x_{n_1,n_1}$ is equal to the difference of the mass$^2$ of $y_{1,j}$ and $y_{2,j}$, $m^2_{1,j}-m^2_{2,j} $, so as long as these are different we also don't have any divergence upon setting $x_{n_1,n_1} \to 0$.

Thus we can finally give a statement on splits for massive Tr$(\Phi^3)$ amplitudes. On split kinematics, the physical integrand is related to the surface integrand via 
\begin{equation}
    \mathcal{I}^{\text{standard}}_{n} =  \left[  \mathcal{I}^{\mathcal{S}_1} - \mathcal{A}^{\mathcal{S}_1,\text{tree}} B_1 \right] \times {\cal A}_{{\cal S}_2}(m^2_{k,m}) - \text{(remaining bubbles + tadpoles)}\vert_{\text{split}} ,
\end{equation}
where $B_1$ is the loop-integrated bubble function defined in eqn. \ref{eq:PhysIntMass} and the remaining bubbles and tadpoles are the rest of the tadpole/bubble terms from eqn.\ref{eq:PhysIntMass}. Furthermore since the tree amplitude ${\cal A}_{{\cal S}_2}$ is evaluated on minimal kinematics, it collapses to the extremely simple factor
\begin{equation}
{\cal A}_{{\cal S}_2} = \prod_{i=n_1}^n \left(\frac{1}{m^2_{2,i}} + \frac{1}{m^2_{i-1,1}} \right).
\end{equation}

\section{Multi-soft limits for all-loop-integrated NLSM amplitudes}
\label{sec:SoftNLSM} 

Pion multi-soft limits have been long-studied. In addition to the trivial behavior in the single-soft case, starting with the double-soft case, where two adjacent pions go soft, the result is already non-zero and has been understood from a number of points of view \cite{simplest,SingleDouble,DoubleSoftShift,Cachazo_DoubleSoft,JaraNLSM}. Generic multi-soft statements have also been computed using recursive techniques at tree-level \cite{multisoft}. 

In this section, we explore how splits in the NLSM surface integrand give us access to multi-soft limits at all orders. It turns out that all the statements derived for the NLSM surface integrand are consequences of the more primary object: the $\delta$-shifted surface integrand. For a fixed parity of the punctures, we can show that the $\delta$-shifted surface integrand directly vanishes when we take an \textit{odd} number of particles soft. On the other hand, if the number of adjacent particles taken soft is \textit{even}, say $2m$, then this surface integrand precisely factorizes into:
\begin{equation}     \mathcal{I}^{\delta,\mathcal{S}}_{2n} \to \mathcal{A}_{2m+3}(y_{i,j} + \delta_{i,j})\bigg\vert_{y_{i,j}\to 0} \times \mathcal{I}^{\delta,\mathcal{S}}_{2n-2m} ( x_{i,j} +\delta_{i,j}).
\end{equation}

In the second factor the $\delta$-shift is the usual one for an even number of particles. This gives us the statement of multisoft factorization for the $\delta$ shifted surface integrand. Exactly the same factorization holds once we sum over parity assignments for punctures to get NLSM amplitudes at all orders in the topological expansion for the $U(N) \to U(N-k) \times U(k)$ model, which are the same as those of the familiar chiral Lagrangian for $U(N) \times U(N) \to U(N)$ in the planar limit. Recalling that we can loop-integrate surface integrands with no $1/0$ ambiguities, we have a clean statement for multi-soft factorization of surface NLSM amplitudes as $2m$ goldstones are taken soft
\begin{equation}
{\cal A}_{2n,{\rm NLSM}}^{{\cal S}} \to {\rm Soft}^{{\rm NLSM}}_{2m} \times {\cal A}_{2n - 2m},
\end{equation}
where 
\begin{equation}
{\rm Soft}^{{\rm NLSM}}_{2m} = {\cal A}^{{\rm Tr}(\Phi^3), {\rm tree}} \left(1,2,(2n-2m-1),(2n-2m),\cdots 2n \right) \left( y_{i,j} + \delta_{i,j} \right) \bigg\vert_{y_{i,j}\to 0}.
\end{equation}

Note that since there are an odd number of particles in the tree amplitude defining ${\rm Soft}^{{\rm NLSM}}_{2m}$, the $\delta$ shift in the $y_{i,j} \to 0$ limit can be interpreted as a mixed amplitude, with the $\pi$'s being the particles taken soft, and $\phi$'s the remaining three particles entering this factor \cite{NLSM,Cachazo_ExtensionSoft}. So we can also recognize
\begin{equation}
{\rm Soft}^{{\rm NLSM}}_{2m} = 
\mathcal{A}_{2m+3}(1^\phi,2^\phi,(2n-2m-1)^\phi, (2n-2m)^\pi, \cdots, (2n)^\pi) .
\end{equation}

These expressions for the multi-soft limit of the NLSM surface integrands and amplitudes had been presented without derivation in \cite{CirclesNLSM}, together with the observation that the Adler zero was also manifest directly at the level of the surface integrand; our discussion here provides a simple proof of these all-order statements following from splits and the shifted Tr$(\Phi^3) \leftrightarrow$ NLSM connection. 

As with our previous discussions of Tr $(\Phi^3)$ theory, the simplest statements about splits hold for the natural surface integrand, but it is also interesting to understand what can be said about the conventional physical integrand. We will see that for an odd number of (adjacent) particles taken soft, as a consequence of splits in physical kinematics, the physical integrand vanishes at all orders in the $1/\delta$ expansion. Instead for an even number of soft particles, there is a more interesting statement. We will consider two different integrands: 1. the low-energy $\delta$-shifted surface integrand -- the object we get from expanding $X\ll \delta$ the $\delta$-shifted surface integrand for a given parity assignment; 2. The low-energy physical  $\delta$-shifted integrand -- the object we get from expanding $X\ll \delta$ the $\delta$-shifted standard integrand as defined in \eqref{eq:PhysIntegrand} (where we subtract the contributions from tadpoles and external bubbles), for the same parity assignment of the punctures. We conjecture that 1. and 2. match in the multi-soft limit up to terms giving scaleless contributions that vanish upon loop-integration. 

This result is well-known for the single-soft case, and we present an explicit proof for the double-soft case at one-loop. If this natural conjecture holds at all loop orders, the soft-factors derived for the surface integrand directly from splits automatically generalize to the physical integrand. In particular, summing over the possible parities of the punctures, they also give us directly the multi-soft factors of the physical integrand for pion amplitudes, that hold post-loop integration.  

\subsection{Multisoft limits for the $\delta$-shifted integrand}
Now that we have studied how we can derive multi-soft limits for massive Tr$(\Phi^3)$, it is easy to understand how these generalize to the case in which the masses are precisely those of the $\delta$-shift. As explained in section \ref{sec:deltaShift}, while a given split might not be present for a generic mass pattern, the tree $\otimes$ loop splits are always there for the $\delta$-shift, $i.e.$ the $\delta$-shift is a particular mass assignment compatible with all the tree $\otimes$ loop split kinematics. 

Once more, we will be looking at the type of split that gives us access to multi-soft limits in which  $\mathcal{S}$ splits into $\mathcal{S}_1\otimes \mathcal{S}_2$, where $\mathcal{S}_1 = (2,3,\cdots,n_1-1,n_1)$ and contains all the punctures/non-trivial topology and $\mathcal{S}_2 = (n_1-1,n_1,\cdots,2n, 1,2)$ is an $(n-n_1+4)-$point tree, that overlap in triangle $(2,n_1-1,n_1)$ (see figure \ref{fig:softTree}). As previously we will denote the kinematics of $\mathcal{S}_2$ by $y_{i,j}$ and those of $\mathcal{S}_1$ by $x_{i,j}$. 

Since the $\delta$-shift is compatible with the split, we just need to make sure that we don't get any divergences when we go on the multi-soft limit. Before proceeding to multi-soft limits, let's start by looking at the single-soft case. 

\subsubsection{Single-soft and the split approach to the Adler zero}

In this case $\mathcal{S}_2$ is simply a $4$-point tree. So we have $n_1 =2n$ and
\begin{equation*}
    \mathcal{S}_1 = (2, 3,\cdots, 2n-1,2n), \quad \mathcal{S}_2 = (1,2,2n-1,2n). 
\end{equation*}

Let's say that $\mathcal{S}_1$ describes an $N$-loop process, and so has $N$-puncture. To obtain the NLSM integrand, we have to sum over all the possible parities of the punctures (even or odd) and this is what gives us the NLSM integrand at low energies. 
However let's look at a certain parity assignment of the punctures, then going on the split kinematics yields:  
\begin{equation}
    \mathcal{I}^{\delta, N-\text{loop}}_{2n} = \left(\frac{1}{y_{1,2n-1} -\delta } + \frac{1}{y_{2,2n} + \delta } \right) \times \mathcal{I}^{\delta, N-\text{loop}}_{2n-1} ( x_{i,j}),
\end{equation}
where $\mathcal{I}^{\delta, N-\text{loop}}_{2n-1}$ stands for the integrand associated to $\mathcal{S}_1$, with a fixed parity for the punctures. Note that, while $x_{2,j}$ and $x_{2n-1,j}$ map to curves $X_{2,j}$ and $X_{2n-1,j}$ on $\mathcal{S}$, for any $j\in \mathcal{S}_1$, $x_{2n,j} = X_{1,j}$. Therefore the $\delta$ shift in $\mathcal{S}_1$ is the usual $\delta-$shift except that now label $2n$ should be considered \textit{odd}. Therefore, at low-energies, the object we get from $\mathcal{S}_1$ will not be pure NLSM. 

At tree-level, this object has the same simple interpretation we saw above, as a mixed amplitude with three adjacent $\phi$'s and remaining $\pi$'s. So we see that the coefficient of the Adler zero, when approached in the direction of this split kinematics, is given by a single mixed amplitude. This can be compared with the observation of \cite{Cachazo_ExtensionSoft}, which showed that the coefficient of the Adler zero reached by scaling $p_n \to t p_n$ with $t \to 0$ is instead a sum over such mixed amplitudes. This highlights the contrast between splits and soft limits. Splits define a lower-dimensional linear subspace of kinematic space and are defined arbitrarily far from soft limits, so in this sense, they tell us much more than the behavior in the soft limit. On the other hand, they only let us approach the soft limit along this lower-dimensional subspace, and so the full subleading structure in the expansion away from the soft limit is not the most general one. Instead, as we saw in our brief discussion of the sub$^k$ soft factors for massless Tr$(\Phi^3)$, the splits give us sum rules that capture the soft expansion along the split directions. In the present context of approaching the Adler zero, this means that the more complicated sum over mixed amplitudes collapses into a single term when the zero is approached along the split direction. 

Since the leading soft limit is well-defined, we can access it using splits without losing any information. Therefore going on the soft-limit where particle $n$ goes soft corresponds to setting $y_{i,j} \to 0$, and so the 4-point factor manifestly vanishes and thus the surface integrand $\mathcal{I}^{\delta, N-\text{loop}}_{2n}$ vanishes. This makes the Adler zero manifest directly at the level of the surface integrand, before loop-integration. We would like to note that interestingly the vanishing on the soft limit is true \textit{for each parity assignment of the punctures individually}, while the NLSM amplitude is obtained after summing over parity assignments. 

Now in order to translate this statement to the physical integrand, we need to go to physical kinematics which in this simple case only amounts to:
\begin{equation}
    y_{2n-1,1} = x_{2n-1,n}, \quad  y_{2,2n} = x_{2,2n} +y_{2,2n},\quad 0 = 2 y_{2,2n} + x_{2n,2n}
\end{equation}
where the second equation, even before going on the soft limit, forces $x_{2,2n} =0$ which, as explained in \ref{sec:PhyKin} does not cause any divergences in the standard integrand precisely because we have subtracted the contribution coming from the external bubbles. Now when we go on the soft-limit we set $y_{i,j}\to 0$, which does not cause any divergences in the tree factor because both $y$'s have a mass, and similarly since the index $n$ in $\mathcal{S}_1$ should be regarded as odd, $x_{2n-1,2n}$ and $x_{2n,2n}$ are both shifted with $-\delta$, and so the result is finite when we set them to zero. 

So have that in split kinematics the standard integrand becomes:

\begin{equation}
\begin{aligned}
    \mathcal{I}^{\text{standard}, \delta}_n =  \left(\frac{1}{y_{1,2n-1} -\delta } + \frac{1}{y_{2,2n} + \delta } \right) &\times \left[  \mathcal{I}^{\mathcal{S}_1} -\frac{\mathcal{A}^{\mathcal{S}_1,\text{tree}}}{x_{2,2n} x^{(\delta)}_{2,p} x_{2n,p}^{(\delta)}} \right] \\
    &- \text{(remaining bubbles + tadpoles)}\vert_{\text{split}},
\end{aligned}
\end{equation}
where everything is evaluated in physical kinematics $X_{i,j} = X_{j,i}$. In the soft-limit, the standard integrand reduces to simply the tadpole contribution which for finite $\delta$ can be computed in dim. reg. (see sec. \ref{sec:SurfaceIntegrand}). However, at low-energies, $i.e.$ expanding in $X\ll \delta$, the tadpole integrals become scaleless and integrate to zero. Note that this is true already for each parity assignment for the punctures, so when we sum over possible parities we obtain the familiar Adler statement. In particular, we see that while for the surface integrand the Adler zero is directly manifest at the level of the integrand, for the standard physical integrands it is only true \textit{after} loop integration, due to the presence of \textit{scaleless} integrals.

\subsubsection{Double-soft limits} Before proceeding to the general multi-soft statement, let's study the case in which particles $2n$ and $2n-1$ go soft. To access this limit we look at the split pattern:
\begin{equation*}
    \mathcal{S}_1 = (2, 3,\cdots, 2n-2,2n-1), \quad \mathcal{S}_2 = (1,2,2n-2,2n-1,2n), 
\end{equation*}
where $\mathcal{S}_2$ is now a $5$-point tree and $\mathcal{S}_1$ involves $2n-2$ particles and contains all the non-trivial topology. Once more, fixing a parity assignement to the punctures, going on split kinematics yields:
\begin{equation}
     \mathcal{I}^{\delta, N-\text{loop}}_{2n} = \mathcal{A}_5(y_{i,j} + \delta_{i,j}) \times \mathcal{I}^{\delta, N-\text{loop}}_{2n-1} ( x_{i,j} +\delta_{i,j}),
\end{equation}
where now the $\delta$-shift in $\mathcal{S}_1$ is simply the usual $\delta$-shift. Again at low-energies the factor from $\mathcal{S}_2$ gives a  \textit{mixed} amplitude of pions and scalars~\cite{NLSM}:
\begin{equation}
    \mathcal{A}_5(1,2,2n-2,2n-1,2n)[y_{i,j} + \delta_{i,j}] \xrightarrow{y_{i,j}\ll \delta}  \mathcal{A}_5(\phi,\phi,\phi,\pi,\pi)[y_{i,j}].
\end{equation}

Note that in the limit where the split kinematics match the soft-limit of particles $2n-1$ and $2n$, we have $y_{i,j} \to 0$ and thus $y_{i,j} \ll \delta$. Therefore in the multi-soft limit, even at finite delta the $\mathcal{S}_2$ factor is given by this mixed amplitude. 

Now going to physical kinematics together with the soft limit implies:

\begin{equation}
    x_{2,2n-1} =0, \quad x_{2n-1,2n-2} =0, \quad x_{2n-1,2n-1} =0.
\end{equation}

As opposed to what happened in the single soft case, while $x_{2,2n-1} = X_{2,1}$ is an external bubble (and thus cancels in the standard integrand), and $x_{2n-1,2n-1} = X_{1,1}$ which is shifted by $-\delta$ so setting it to zero doesn't cause any divergences, in the double-soft case $x_{2n-1,2n-2} = X_{1,2n-2}$ which is \textit{not} an external bubble and is also not shifted by $\delta$, so setting it to zero would produce a genuine divergence in the standard integrand. 

We will see that this is a general feature of even-point soft limits: the limit as the $y$'s go to zero is not trivially well-defined. Relatedly the the observed soft factors have $y/y$ behavior as $y \to 0$. Thus it is not trivial that the specific approach to the soft limit defined by split kinematics correctly captures even the leading soft factor away from the split approach. But in a moment we will show that at least at one-loop and for the double-soft limit, we can nonetheless define a soft factor for the physical integrand using the splits-derived soft factor for the surface integrand.  

\subsubsection{General multi-soft: odd versus even points} The feature just observed on how the single-soft and the double-soft have different behaviors -- while for the single-soft case, everything is finite, for the double-soft at finite $\delta$ we have a divergence when going to physical kinematics -- turns out to be completely general to any \textit{odd} vs. \textit{even} number of particles going soft. Looking back at the general split pattern that gives us access to multi-soft limits:
\begin{equation}
    \mathcal{S}_1 = (2,3,\cdots, n_1 -1, n_1), \quad  \mathcal{S}_2 = (1,2, n_1 -1, n_1, \cdots, 2n),
\end{equation}
which $\mathcal{S}_1$ containing all the non-trivial topology, by going on  physical kinematics together with the multi-soft limit in which $y_{i,j} \to 0$, we have
\begin{equation}
    x_{2,n_1} =0 \Leftrightarrow X_{2,1} =0 , \quad x_{n_1, n_1 -1} =0 \Leftrightarrow X_{1,n_1-1} =0, \quad x_{n_1,n_1} =0 \Leftrightarrow X_{1,1} =0 .
\end{equation}

Now while the first and last limits can be reached without producing any new divergences at the level of the standard integrand, the second one, $X_{1,n_1-1}=0$, will lead to a divergence precisely when $n_1$ is \textit{odd}, as in this case this propagator is not $\delta$-shifted and also does not correspond to any bubbles/tadpoles, so it will lead to an honest singularity. This precisely corresponds to the case in which an \textit{even} number of particles is going soft.

So while the even-soft limit is a bit more subtle, this argument allows us to conclude that for a given parity assignment of the puncture(s), the $\delta$-shifted splitted integrand will vanish in the odd-soft limit. Going on split kinematics as well as imposing $X_{i,j} = X_{j,i}$ to obtain the standard integrand yields:
\begin{equation}
    \mathcal{I}^{\text{standard}, \delta}_{2n} =  \mathcal{A}^{\delta}_{2n - n_1 +3}[y_{i,j}]\times \left[  \mathcal{I}^{\mathcal{S}_1} - \frac{\mathcal{A}^{\mathcal{S}_1,\text{tree}}}{x_{2,n_1} x^{(\delta)}_{2,p} x^{(\delta)}_{n_1,p}} \right] - \text{(remaining bubbles + tadpoles)}\vert_{\text{split}},
\end{equation}
but now precisely because $n_1$ is \textit{odd}, $2n - n_1 +3$ is even and so the tree-factor in the limit $y_{i,j} \ll \delta$ gives us the NLSM amplitude which, simply by units, vanishes when $y_{i,j} \to 0$! 

Clearly, when we take the low energy limit, $X\ll \delta$, the terms coming from the tadpoles become scaleless, and so we have that $ \mathcal{I}^{\text{standard}, X\ll \delta}_{2n}$ vanishes for the case in which the number of adjacent particles taken soft is \textit{odd}! Note that this result is for a given parity assignment to the puncture, and consequently the same holds when we sum over puncture parities meaning that the NLSM integrand also vanishes for an \textit{odd} number of adjacent particles going soft.

\subsection{Even-soft limits for the NLSM integrand}

We will now focus on the even-soft limits for the $1$-loop integrand. We have so far been looking at this limit via the splits. Even though it is a reasonable way of studying this limit, asking for physical kinematics on the split imposes further constraints on the tree-subsurface (explained in \ref{sec:PhyKin}) and so gives a particular way of approaching the limit which is a not the most general one. In particular we know that for the surface integrand, in the surface multi-soft limit the soft-factor is well-defined for completely generic soft kinematics.

The goal in this section is to show that exactly the same soft-factor is there for the standard physical integrand -- this is a non-trivial result as naively if we first identify $X_{i,j}=X_{j,i}$ in the multi-soft limit we could obtain new divergences that are not there when $X_{i,j}\neq X_{j,i}$ as these variables map to different curves in the surface soft limit. So to show that the soft-factor is the same, we need compare the following two objects: 1. the surface integrand in the surface multi-soft limit, where after taking this limit we set $X_{i,j} = X_{j,i}$; 2. the standard physical integrand in the multi-soft limit; and show that the difference between these two terms in the multisoft limit reduces to scaleless integrals that vanish upon loop-integration. If this is the case then we have that the soft-factor that we derive for the surface integrand generalizes to the standard physical one. 

Let's look at what happens in the simplest case of the double-soft limit we studied before, and use this example to recall the differences between the surface soft and the standard soft-limit for the physical integrand. As we will see, this proof heavily relies on the detailed features of the terms that appear on the numerators of the NLSM integrand. Luckily we have a very simple interpretation for the numerator factors coming from the Catalan rule presented in \cite{CirclesNLSM}, which is crucial to let us easily identify the terms that lead to scaleless integrals that we are happy to throw away in dim. reg.

In the double-soft case where we take particles $2n$ and $2n-1$ soft, the surface soft-limit is given by the following kinematic map:
\begin{equation}
\begin{aligned}
    &X_{2n-2,1} \to t X_{2n-2,1},\\  &X_{2n-1,1} \to t X_{2n-1,1},
\end{aligned} \quad 
\begin{aligned}
&X_{2n-1,2} \to t X_{2n-1,2}, \\ &X_{2n,2} \to t X_{2n,2},  
\end{aligned}\quad 
X_{2n-2,2n} \to t X_{2n-2,2n },
\end{equation}
$i.e.$ all the curves living inside the pentagon $(1,2,2n-2,2n-1,2n)$ (not containing the puncture), are rescaled by $t$ and we send $t\to 0$, as for the remaining curves on the surface we have:
\begin{equation}
\begin{aligned}
    &X_{2n,j} \to X_{1,j}, \\
    &X_{j,2n} \to X_{j,2n}, 
\end{aligned}\quad  \quad \quad 
\begin{aligned}
&X_{2n-1,j} \to X_{1,j},\\
&X_{j,2n-1} \to X_{j,2n-1},.
\end{aligned}
\end{equation}

While the double-soft limit for the physical integrand where $X_{i,j} = X_{j,i}$ is the same but now $X_{2n,j}$ and $X_{j,2n}$ map to the same variable, and similarly for $X_{2n-1,j}$ and $X_{j,2n-1}$, therefore some curves outside the pentagon but for which the end-points are in the pentagon are rescaled by $t$ and sent to zero. In particular, while in the surface soft limit curves such as $X_{1,2n-2}$ or $X_{2n,2n-2}$ get mapped to $X_{1,2n-2}$ which is, in the soft-limit, an external bubble, in the physical soft-limit $X_{1,2n-2} = X_{2n-2,1}$ and
$X_{2n,2n-2} = X_{2n-2,2n}$ are both sent to zero. This is precisely the reason why naively the soft limit in the two cases could differ, in principle even by something divergent. 

To understand that the difference is always scaleless we need to study precisely these terms that have a different mapping in the soft-limit. Let's start by considering the case in which the parity of the puncture is odd. Among the two curves that we pointed out the case in which we could obtain an extra divergence for the physical integrand is by setting $X_{2,2n-1}=0$ as this is a pole of the NLSM integrand, while $X_{2,2n}$ only appears in the numerators. So let's understand the coefficient of $X_{2,2n-1}$ in the physical integrand. Recall, according to the Catalan rule presented in \cite{CirclesNLSM}, the $X_{1,2n-2}$ multiplies a sum over all the possible even-angulations of the tree problem: $(1,2,\cdots, 2n-2)$, where for each even-angulation the numerator factor is the product over all the even-gons of the sum of the $X_{e,e}/X_{o,o}$ inside that each even-gon; and similarly for the loop-problem $(1,2n-2,2n-1,2n).$ In particular, since we care about the part of this coefficient that does \textit{not} give a scaleless integral on the loop part, we care about the terms corresponding to even-angulations of the loop problem containing at least two \textit{different} loop-propagators, which in this case are $X_{2n,p}$ and $X_{2n-2,p}$ -- since $p$ is odd these propagators appear as poles in the NLSM integrand. So $(X_{2n,p},X_{2n-2,p})$ already forms a full even-angulation that divides the loop-part into the two quadrilaterals: 
\begin{equation}
    \mathcal{Q}_1 =(1,2,2n-2,p), \quad \mathcal{Q}_2 =(2n,p,2n-2,2n-1).
\end{equation}

Then the contribution to the loop integrand coming from this even-angulation is:
\begin{equation}
   \frac{1}{X_{1,2n-2}} \times \frac{(-X_{2n,2n-2} -X_{1,p})(-X_{2n-1,p} - X_{2n-2,2n})}{X_{2n,p} X_{2n-2,p} } \times \text{(even-ang. tree)}
\end{equation}
where evidently we need to multiply by the sum of the possible even-angulations of the tree problem. Now if we expand the numerator factors, we see that any term having either $X_{1,p}$ or $X_{2n-1,p}$ turns into a scaleless integral in the soft-limit since, in this limit we have $X_{2n-1,p} = X_{2n,p} = X_{1,p}$ and so one of the poles cancels and what is left becomes scaleless. Therefore the only term we need to consider is the one proportional to $X_{2n,2n-2} X_{2n-2,2n}$. In the surface integrand, under the surface soft mapping, this term becomes:
\begin{equation}
    \frac{1}{X_{1,2n-2}} \times \frac{X_{2n,2n-2}  X_{2n-2,2n}}{X_{2n,p} X_{2n-2,p} } \to  \frac{1}{X_{1,2n-2}} \times \frac{X_{1,2n-2}  (tX_{2n-2,1})}{X_{1,p} X_{2n-2,p} } =  \frac{ (tX_{2n-2,1})}{X_{1,p} X_{2n-2,p} }
\end{equation}
while in the physical integrand this contribution reads
\begin{equation}
    \frac{1}{X_{1,2n-2}} \times \frac{X_{2n-2,2n}^2}{X_{2n,p} X_{2n-2,p} } \to  \frac{1}{t X_{1,2n-2}} \times \frac{t^2 X_{2n-2,2n}^2}{X_{1,p} X_{2n-2,p} } =  \frac{1}{X_{1,2n-2}} \times \frac{t X_{2n-2,2n}^2}{X_{1,p} X_{2n-2,p} }
\end{equation}
so, even though they are different, they manifestly both vanish in the $t\to 0$ limit. 

Now the other curve that is also set to zero in the soft limit is $X_{2,2n-1}$ for the physical integrand. (We are not considering here curves like $X_{2n,2n-1}$, as these correspond to external bubbles which are subtracted in the physical integrand.) Once again, we can use the Catalan rule to write down the coefficient of $X_{2,2n-1}$ in the NLSM integrand. The argument proceeds exactly as above simply cycling the indices. 

If instead, we consider the parity of the puncture to be \textit{even}, the argument is even more trivial. This is simply because in this case, to complete the even-angulations inside the loop problem, in both cases with curves $X_{1,2n-2}$ and $X_{2,2n-1}$, we need to pick the loop propagators $X_{1,p}$ and $X_{2n-1,p}$ in order to obtain contribution which is not scaleless. However, in the soft-limit we have $X_{1,p} = X_{2n-1,p}$, and so these terms also become scaleless when we make particles $2n$ and $2n-1$ soft. 

Therefore we conclude that for each parity separately, the leading order in the $\delta$-expansion for the physical integrand in the double-soft limit becomes
\begin{equation}
    \mathcal{I}^{X\ll \delta}_{2n} \to \mathcal{M}_5(1^{\phi},2^{\phi},(2n-2)^{\phi}, (2n-1)^\pi,(2n)^\pi) \times  \mathcal{I}_{2n-2}^{X\ll \delta},
\end{equation}
summing over both parities then tells us that this prefactor is also there for the NLSM physical integrand in the double-soft limit. In particular, since the tree-factor does \textit{not} depend on the loop momentum this soft-factor is what we get after loop integration as well so that the $1$-loop NLSM amplitude in the double soft limit becomes:
\begin{equation}
    \mathcal{M}^{\text{NLSM}}_{2n} \to \mathcal{M}_5(1^{\phi},2^{\phi},(2n-2)^{\phi}, (2n-1)^\pi,(2n)^\pi) \times  \mathcal{M}^{\text{NLSM}}_{2n-2},
\end{equation}

It is natural to conjecture that this connection between the surface and physical integrands in the even-multisoft limit persists to all loop orders; if it does, the loop integrated statements for the physical integrand also follow in the same way.

\subsection{Soft universality: multi-soft factors are amplitudes}

We close with a nice interpretation of the (multi) soft factors we have found for the NLSM, which will also extend to our brief discussion of gluons below. We have seen that the soft factor for $2m$ soft pions is given by a mixed amplitude for three adjacent scalars $\phi$, enjoying a Tr($\phi^3$) interaction, together with $2m$ $\pi's$. 

Thus, our derivation of the multi-soft limit from splits is really directly telling us, not the conventional statement that the amplitude factorizes into ``lower-point amplitude $\times$ soft factor'', but more directly that the NLSM amplitude factors into the product of different amplitudes:
\begin{equation}
\left.{\cal A}_{2n} \right|_{2m \, 
 {\rm soft} \, \pi} \to {\cal A}_{2n - 2m} \times {\cal A}^{{\rm tree}}(\phi, \phi, \phi, \underbrace{\pi, \cdots, \pi}_{2m}).
\end{equation}

This is actually an example of the universality of the multisoft factor; we expect that for the amplitude with any particle species coupled to the pions we factor out exactly the same multisoft factor in the limit where $m$ momenta become soft. The case of 3 $\phi$'s coupled to $2m$ pions is special simply because in the soft limit where all $2m$ pions are taken soft, the remaining ``hard'' amplitudes for the three $\phi$'s is simply the cubic coupling for the three $\phi$'s! So this is how we can interpret the amplitude factor in the splits as giving us the multi-soft factor, via 
\begin{equation}
\left.{\cal A}^{{\rm tree}}(\phi, \phi, \phi, \underbrace{\pi, \cdots, \pi}_{2m})\right|_{2m \, {\rm soft} \pi} \to {\cal A}^{{\rm tree}}(\phi, \phi, \phi) \times {\rm Soft}_{2m} = {\rm Soft}_{2m}.
\end{equation}

\section{Scalar-Scaffolded Gluons and Splits}
\label{sec:ScalarScaffGluons}

Having understood some of the implications of split kinematics for pion amplitudes, we now turn to discussing gluon amplitudes. As proposed in \cite{zeros,Gluons}, scalar-scaffolded gluon amplitudes are obtained by unit integer $\delta$ shifts of the ``stringy'' Tr $(\Phi^3)$ amplitudes. This gives us $2n$-point amplitudes, and we can further take the ``scaffolding residue'' where pairs of scalars fuse to on-shell gluons, to obtain gluon amplitudes. In this section we will very briefly explore some of the simplest consequences of split kinematics for these gluon amplitudes, where the kinematics is arranged to be trivially compatible with taking this scaffolding residue, so that the split patterns for the $2n$ scalar amplitudes are directly inherited by those of the gluon amplitudes. We leave a wider investigation of these and related novel aspects of gluon amplitudes to future work \cite{gluonUpcoming}. 

\subsection{Multigluon soft limits}
\label{sec:Softgluons}

The simplest application of splits is to derive gluon soft limits, in the same way we derived NLSM soft limits. However, now to make one gluon we first need the two scalars that fuse to produce it soft. Hence we can study the amplitude for $m$ soft gluons starting from one where the $2m$ scaffolding scalars are taken to be soft. 

Let's begin with the simplest case of $m=1$, where we take $2$ scalars to be soft. As with our discussion for pions, this will involve the split giving us the multi-soft limit in which the tree factor is a 5-point amplitude $(1,2,2n-2,2n-1,2n)$, with the $\delta=1$ shift. 

Thus on the split kinematics we have for the full $2n$ scalar amplitude
\begin{equation} 
{\cal A}_{2n}\bigg \vert_{\text{split. kinematics}} \rightarrow {\cal A}_{2n - 2} \times {\cal S}_2({y_{i,j}=X_{i,j}})
\end{equation}
where ${\cal S}_2$ is denoting the tree-factor with kinematics $y_{i,j}$ that in this case precisely agree with the kinematics of $\mathcal{S}$ living inside the pentagon. Explicitly this term looks like:
\begin{equation}
\begin{aligned}
{\cal S}_2 &= \int \omega  \, u_{1,2n-2}^{X_{1,2n-2}} u_{1,2n-1}^{X_{1,2n-1}-1} u_{2,2n-1}^{X_{2,2n-1}} u_{2,2n}^{X_{2,2n} + 1} u_{2n-2,2n}^{X_{2n-2,2n}+1} \\ 
&= \int \frac{dy_s}{y_s^2} \frac{dy}{(1+y)} y_s^{X_{1,2n-1}} y^{X_{1,2n-2}} (1 + y_s)^{X_{2,2n} - X_{2,2n-1} - X_{2n-2,2n}} \\ & \quad \quad  (1 + y)^{X_{1,2n-1} - X_{1,2n-2} - X_{2,2n-1}} (1 + y_s + y_s y)^{X_{2,n-1} - X_{2,n} - X_{1,2n-1}} 
\end{aligned}.
\end{equation}

To extract the gluon amplitude we simply take the residue on the pole where $X_{1,2n-1} \to 0$, which is computed by taking the residue of the above at $y_s=0$, and performing the resulting integral over $y$ yielding a sum of Beta functions: 
\begin{equation}
\begin{aligned}
& {\cal S}_{2,{\rm gluon}} = 
-\frac{(\Gamma[X_{1, 2 n-2}] \Gamma[
      X_{2, 2n -1}] (X_{1, 2n - 2} (X_{2, 2 n} - X_{2, 2 n-1}) +
        X_{2, 2 n-1} X_{2 n-2, 2 n})}{
   \Gamma[1 + X_{1, 2n - 2} + X_{2, 2 n - 1}]}.
\end{aligned}
\end{equation}

Finally, to take the soft limit, we simply take the field-theory limit of the above, sending $\alpha^\prime X \to 0$ to obtain the soft factor 
\begin{equation}
{\rm Soft}^{(1)} = 1 - \frac{X_{2, 2 n}}{X_{2, 2 n - 1}} - \frac{X_{2 n - 2, 2 n}}{X_{1, 2n - 1}}.
\end{equation}

This is the gluon soft factor written in the scaffolding $X$ variables. 

As for the case of the NLSM, we can interpret the split factorization derivation of the soft factor more deeply as a derivation of the universality of the soft gluon limit.  The tree-split factor is a full tree-amplitude in its own right, for three colored scalars $\phi$, enjoying the same Tr $\phi^3$ interaction, coupled to gluons and the two scalars scaffolding a gluon. After taking the scaffolding residue to get an amplitude with a gluon, the soft limit factors out the gluon soft factor multiplying the constant $(\phi ,\phi, \phi)$ amplitude. We can interpret the multi-soft factor for $m$ gluons in the same way, as coming from the amplitude for $3 \phi$'s plus $2m$ scalar-scaffolding scalars, on the support of the scaffolding residue. This is given by the field theory limit of the $\alpha^\prime \delta = 1$ shifted $(1, 2, 2n - 2m, \cdots, 2n)$
amplitude.

\subsection{Scaffolding splits and new zeroes for gluon amplitudes}

\begin{figure}[t]
    \centering
    \includegraphics[width=\textwidth]{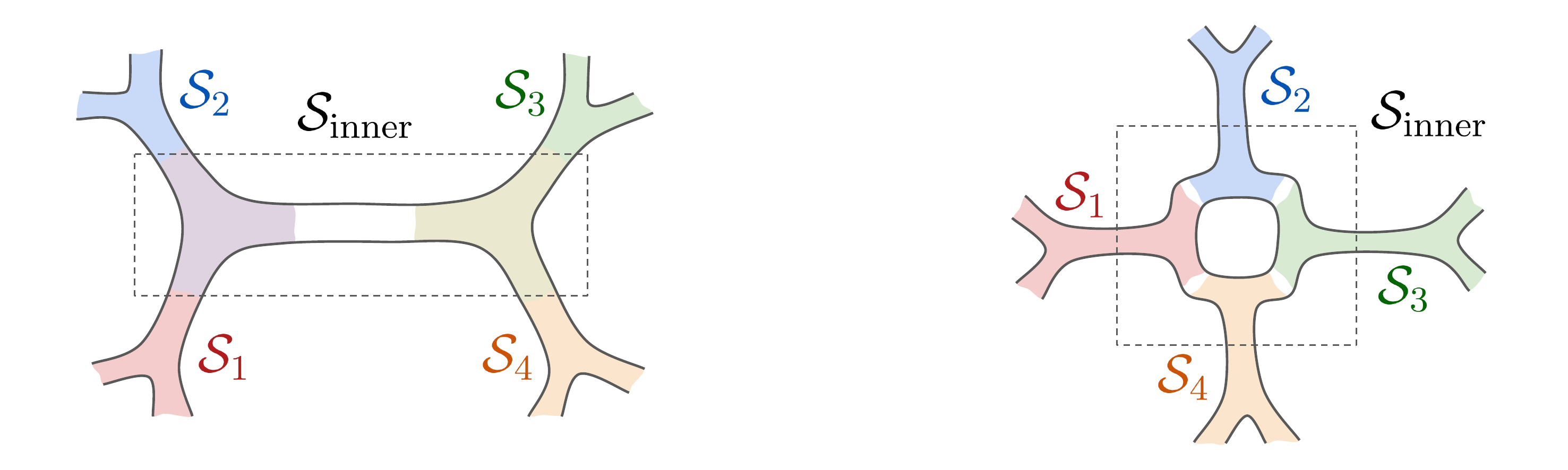}
    \caption{Two examples of ``scaffolding-split'' kinematics, where we join 4 pt trees including the external scaffolded scalars, with a fatgraph associated with the inner $n$-gluon problem, for four-gluon scattering at tree-level in the left and one-loop in the right.}
    \label{fig:ScaffSplits}
\end{figure}

There is a natural choice of multi-splits for gluon amplitudes, that will lead also lead us to novel sorts of zeros. Since we must take the scaffolding residue to extract gluon amplitudes, it is natural to choose splits overlapping on four-point tree surfaces, factoring out terms that manifest the scaffolding pole and making it trivial to take the scaffolding residue. We can call this ``scaffolding splits''. A simple choice at tree-level is to join the four-point trees to a half-ladder fat graph for the internal $n$-point tree (see figure \ref{fig:ScaffSplits}, left, for the case $n=4$). 

In the split kinematics we pull out a factor $\prod_j \frac{X_{2j,2j-2}}{X_{2j-1,2j+1}}$ from each 4 point factor, and after taking the scaffolding residue on each $X_{2j-1,2j+1}$ we are left with the factor $\prod_j X_{2j,2j-2}$. This multiplies the integral for a lower-$n$ point problem. This amplitude does not vanish. However, the lower $n$ points problem can produce at most $(n-3)$ poles, and hence the units of the final result will be at least $\sim X^n/X^{n-3} \sim X^3$. But the tree-amplitude for pure Yang-Mills theory, without the addition of higher-dimension operators like $F^3$ etc., have units of $X^2$. Thus the leading Yang-Mills amplitudes actually vanish on this split kinematics! This is a different sort of zero from the ones discussed in \cite{zeros}, which are present for the full string amplitude including all higher-order corrections.
Indeed, it is enough to split off only $(n-1)$ four point factors; this pulls out a factor of $\sim X^{n-1}$, but it is easy to see that taking the remaining scaffolding residue necessarily pulls down another factor of $\sim X$, so that the same units argument tells us the gluon amplitude vanishes. 

The analogous split at one-loop is more interesting. We can for instance perform the split on 4-point factors with the internal fat-graph being that of an $n$-gon  (see figure \ref{fig:ScaffSplits}, right). Again, on split kinematics we can trivially take the scaffolding residue, giving us the same factor $\prod_j X_{2j-2,2j}$. Now the rest of the amplitude can give us $n$ poles so that the full amplitude can scale as $X^n/X^n \sim 1$, which is the correct units for the YM integrand. But it is easy to see that the only way to get only $1/X$ factors with no numerators, is to use the $u_\Delta^\Delta$ factor in the gluon integrand and expand to leading order in $\Delta$. Recall from \cite{Gluons} that $\Delta$ is the dimensionless parameter that introduces the dependence on the spacetime dimension, $d$, via: $\Delta= 1-d$. This tells us that on this split kinematics, the full 1-loop YM integrand becomes simply 
\begin{equation}
{\cal I}^{{\rm one \, loop}}_{{\rm scaffolding \, split}} = \Delta \frac{\prod_j X_{2j-2,2j}}{\prod_j Y_{2j +1}}.
\end{equation}

Note that this loop-level split does not preserve $X_{i,j} = X_{j,i}$ unless we further set the $X_{2j-2,2j} \to 0$.  So this interesting factorization statement holds for the full surface gluon integrand. 

\section{Outlook}

Since time immemorial, the most important qualitative features of scattering amplitudes have been thought to be seen in their singularity structure, with consistent factorization on poles at tree-level and cuts for loop integrands being prime examples. Recent investigations have revealed a new striking feature of amplitudes for a wide class of theories--factorization not in the neighborhood of poles, but at entirely different non-singular loci in kinematic space \cite{zeros,JaraZeros}. These properties are not a mere academic curiosity--the observation of zeros associated with these  ``split'' factorizations has already played a crucial role in revealing the hidden connection between Tr$(\Phi^3)$ theory, pions, and gluons, and beyond \cite{zeros,NLSM,Gluons}. 

The origin of conventional factorizations near poles is a completely obvious and very general fact following from Lagrangians and Feynman diagrams. But what makes the new ``split'' factorization similarly obvious? We have provided such an understanding in this paper. The splits originate immediately and naturally from the ``surfaceology'' and the curve integral formalism. We simply consider the most natural ways of joining two surfaces ${\cal S}_{1,2}$ to produce the larger surface ${\cal S} = {\cal S}_1 \otimes {\cal S}_2$ without producing any new internal edges, together with the realization of $u$ variables for subsurfaces. Of course the conventional picture of factorization on poles is {\it also} obvious from surfaceology -- as beautifully captured in the ``binary'' nature of the $u$ variables -- and can also be phrased in the ``joining surfaces'' language. After all another way of taking two surfaces and joining them to produce a bigger surface is to simply take two external lines and glue them into an internal propagator. By itself, this can't give us factorization, since we've added an internal edge. So we have to take a residue in the new variable to split the surfaces apart again; that residue is forced by looking at the pole associated with the new internal propagator created by the gluing, and this is the ordinary picture of factorization. It is amusing that both the non-trivial new split factorizations and the old factorizations on poles are treated on an equal footing and trivialized in the surface picture, arising from closely related but different ways of joining surfaces to produce bigger surfaces. 

There are many obvious avenues for further investigation, and we end by listing some of the more interesting and immediate ones. 

To begin with, split kinematics have nothing per se to do with (multi) soft limits -- we simply have various loci in kinematic space where integrands and loop-integrated amplitudes naturally factorize. But as we have indicated the splits do naturally connect to multi-soft limits, and give both the leading soft behavior (which at any rate is more directly associated with ordinary factorization), as well as first subleading behavior, together with an infinite number of sum rules controlling the sub$^k$ soft limits, reflecting the fact that splits are really making a statement about the full amplitude away from the soft-limit but on a lower-codimension locus in kinematic space. It would be interesting to systematically explore these general subleading statements for pions, as well as more systematically explore all aspects of soft limits for gluons after taking scaffolding residues from splits of the shifted Tr$(\Phi^3)$ amplitudes. 

We have also seen that the splits are most naturally statements at the level of the surface loop integrands -- while we can specialize them to ``physical kinematics'' with e.g. $X_{i,j} = X_{j,i}$ at one-loop, they are clearly most naturally formulated at the level of the general surface integrand. It has already been experimentally observed at tree-level, that just the zeroes associated with factorizations are amazingly restrictive: starting very general ansatz for the amplitude, inputting only the correct units as well as mild assumptions about Regge behavior, then even imposing the ``skinny rectangle'' zeros of \cite{zeros} suffices to completely fix the amplitude. It is remarkable that none of the features of amplitudes conventionally thought to be most crucial to their characterization -- the pattern of allowed poles -- are needed; the zeros are essentially enough to fully determine the amplitude. Since we now know that the split factorization extends to the full integrand, it would be fascinating to explore whether loop integrands are similarly strongly constrained, and possibly uniquely fixed by patterns of splits/zeros. 

In particular, for the case of the non-linear sigma model, it is known that, at tree level, imposing the Adler zero together with the correct pole structure completely fixes the amplitude \cite{Uniqueness2,UniquenessNLSM}. At loop-level standard physical integrands do not make the Adler zero manifest directly at the level of the integrand, while the surface integrand does. So one could ask whether this latter object, which has both the Adler zero and correct cuts, is uniquely determined at loop-level. Of course we can further impose the split properties, which would be even more constraining. A similar question can be asked for the gluon, for which at tree-level, a similar statement has been made \cite{Uniqueness2,UniquenessNLSM}, now by imposing gauge-invariance. While for standard physical integrands, gauge invariance is lost, the surface integrand is manifestly gauge-invariant, so one might hope a similar statement exists at loop-level.  

Finally, in this paper, we largely concentrated on a tiny subset of all possible split patterns of the ``tree $\otimes$ loop'' type. This was motivated by the fact that for the join of general surfaces, if the kinematics for the two subsurfaces ${\cal S}_{1,2}$ transform nicely under their respective mapping class groups MCG$_{1,2}$, then the split kinematics transforms only under MCG$_1 \times $ MCG$_2$ which is in general much smaller of the MCG for the big surface. While these give us non-trivial identities for the stringy curve-integrals, they don't appear to have immediate physical relevance. It would nonetheless be interesting to see if variations of these ideas can give rise to joined surfaces with kinematics more naturally acted on by the full MCG. We saw a possible hint for this in our discussion of one-loop splits, with surfaces overlapping on two triangles, where the statement involved not a single integrand factorizing on splits, but rather the factorization for an interesting sum of integrands with different shifts for spiraling loop variables. Given the ease with which the simplest ``tree $\otimes$ loop'' splits expose and trivialize a large number of physically important facts from zeros to multi-soft limits, we expect a rich set of further surprises from the study of the most general way in which smaller surfaces can be joined into larger ones.

\acknowledgments

It is our pleasure to thank Jeffrey Backus, Freddy Cachazo, Sérgio Carrôlo, Nick Early, Hadleigh Frost, Song He,  Hugh Thomas, and Jaroslav Trnka for inspiring discussions and comments on the draft. The work of N.A.H. is supported by 
the DOE (Grant No. DE-SC0009988), by the Simons Collaboration on Celestial Holography, and further support was made possible by the Carl B. Feinberg cross-disciplinary program in innovation at the IAS. The work of C.F. is supported by FCT/Portugal
(Grant No. 2023.01221.BD).

\appendix
\section{Subsurface $u$ variables and generalized $u$ equations} 
\label{sec:UAppendix}

Borrowing from \cite{curvy}, we have already discussed how to see the $u_x$ variables for curves $x$ in a subsurface $s$ as a product over the $u_X$ of the parent surface ${\cal S}$ as 
\begin{equation}
u_x = \prod_X u_X^{\#[x \subset X]}
\label{eq:extendu}
\end{equation}
where $\#[x \subset X]$ is the number of different ways in which $x$ can occur as a subcurve of $X$. Here we spell out the concrete meaning of the exponent $\#[x \subset X]$ completely explicitly, and also illustrate with very simple examples the fact we referred to in the introduction, that we can think of $u$ equations (and also ``generalized'' u equations) reflections of this basic fact. 

To begin with, recall that we can specify ${\cal S}$ by giving a representative fatgraph, and the subsurface $s$
by some subgraph of the fatgraph. We can build $u$ variables for any surface, expressed as ratio of polynomials in the $y$ variables of the fat graph edges. The expression \ref{eq:extendu} tells us that a certain product over $u_X$'s, which in general depend on {\it all} the $y$'s of the surface, actually give the $u_x$ variables of the subsurface expressed only in terms of the $y$'s of the associated subgraph. 

Intuitively the exponent  $\#[x \subset X]$ is the simply the number of times $x$ occurs when we restrict the curve $X$ to $s$, as we saw in our one-loop examples in the text. This can be described completely precisely with no need for drawing pictures, in terms of the ``navigation on the fat-graph'' words that fundamentally define curves on surfaces \cite{curveint}.  The curve $X$ is associated with a word $W_X$, recording the roads the curve encounters the left/right turns it takes at vertices as it makes its way through the fatgraph. The curve $x$ also has a word $w_x$. Then $\#[x \subset X]$ simply counts the number of distinct ways in which the word $w_X$ is contained in $W_X$. 

\begin{figure}[t]
    \centering
    \includegraphics[width=\textwidth]{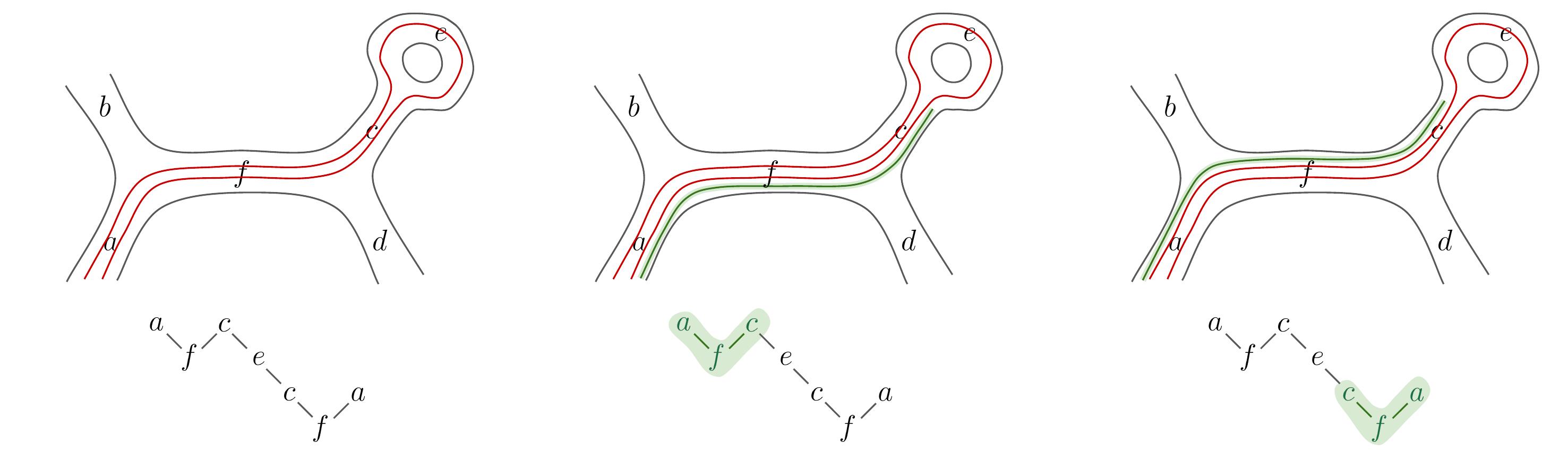}
    \caption{The green curve $x$ on the four point subsurface occurs in two different ways in the red curve $X$ on the full surface; seen both geometrically, as well as in the way the word for $x$ occurs in two ways in the word for $X$. Thus $\#[x \subset X] = 2$.}
    \label{fig:loopext}
\end{figure}
We illustrate with a simple one-loop example in figure \ref{fig:loopext}, where we consider a 4-pt tree subgraph of a 3-pt one-loop problem, and consider the curve $(ac)$ in the 4 pt problem that travels from the boundary $a$ to $c$. This curve is contained twice in the curve $X_{1,1}$ of the big one-loop surface, so $\#[(ac) \subset X_{1,1}] = 2$. This can be seen both pictorially in the two different ways it occurs inside the drawing of $X_{1,1}$, as well as more formally by the number of instances of of the word for $x$ contained in that for $X_{1,1}$. Note that the order in which the words are read forwards or backwards is immaterial, in our example the two occurrences of the words for $(ac)$ occur in opposite order in $X_{1,1}$. 

Note also that in this example, the curve $(bd)$ doesn't have any extensions other than the same curve $(bd)$ on the big surface. So the $u$ equations for the 4 pt problem, $u_{(bd)} + u_{(ac)} = 1$, simply turn into the $u$ equation for $u_{(bd)}$ in the full surface, with the exponents occurring in the expression for $u_{(ac)}$ correctly giving the intersection number with $(bd)$, that is,  ${\rm int}(X,(bd)) = \#[(ac) \subset X$. We can see this with our example of the $X_{1,1}$ curve, where the curve $(bd)$ clearly intersects $X_{1,1}$ twice on the surface. 

\begin{figure}[t]
    \centering
    \includegraphics[width=\textwidth]{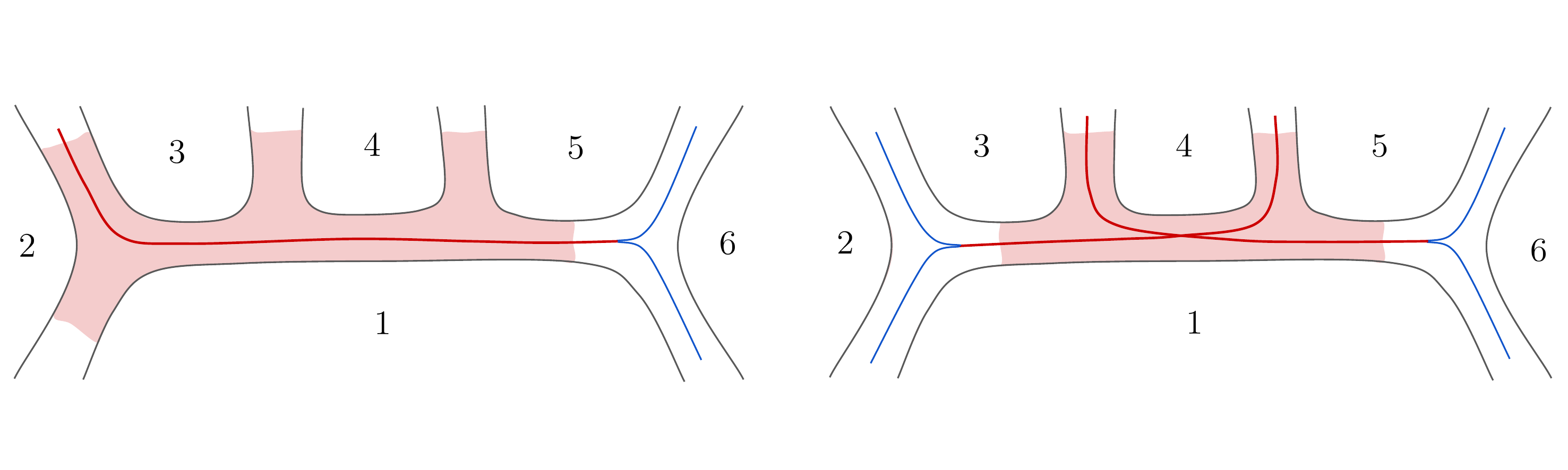}
    \caption{Examples of 5 point and 4 points subsurfaces of a 6 point tree fatgraph, together with representative extensions of curves from the subsurfaces into the full surface.}
    \label{fig:6ptext}
\end{figure}

We end with a simple example for 6pt trees, using the  ``ray triangulation'' or ``half-ladder'' fatgraph of figure \ref{fig:6ptext}. The 9 $u$ variables in this example are 
\begin{equation}
\begin{aligned}
&  u_{1,3}= \frac{y_{1,3}}{1 + y_{1,3}}, \\
&u_{1,4} = \frac{y_{1,4} (1 + y_{1,3})}{1 + y_{1,4}(1 + y_{1,3})},\\
&u_{1,5}= \frac{y_{1,5} (1 + y_{1,4}(1 + y_{1,3}))}{1 + y_{1,5}(1 + y_{1,4}(1 + y_{1,3}))},\\
&u_{2,4}= \frac{1 + y_{1,4}(1 + y_{1,3})}{(1 + y_{1,3})(1 + y_{1,4})}, \\
&u_{2,5} = \frac{(1 + y_{1,4})(1 + y_{1,5}(1 + y_{1,4}(1 + y_{1,3})))}{(1 + y_{1,4}(1 + y_{1,3})(1 + y_{1,5}(1 + y_{1,4})},
\end{aligned}\quad \quad 
\begin{aligned}
&  u_{2,6} = \frac{1 + y_{1,5}(1 + y_{1,4})}{1 + y_{1,5}(1 + y_{1,4}(1 + y_{1,3}))},\\
&u_{3,5} = \frac{1 + y_{1,5}(1 + y_{1,4})}{(1 + y_{1,4})(1 + y_{1,5})}, \\
&u_{3,6} = \frac{1+ y_{1,5}}{1 + y_{1,5}(1 + y_{1,4})},\\
&u_{4,6} = \frac{1}{1 + y_{1,5}}.
\end{aligned}
\end{equation}

The reader can readily verify that these solve the $u$ equations. 
Let's choose the 5-point subsurface in the second panel of figure \ref{fig:6ptext}. We'll denote the five $u$ variables for the subsurface as $\tilde{u}$'s, and we have $\tilde{u}_{1,3}=u_{1,3},\tilde{u}_{1,4}=u_{1,4},\tilde{u}_{2,4}=u_{2,4},\tilde{u}_{2,5} = u_{2,5} u_{2,6}, \tilde{u}_{3,5}=u_{3,5} u_{3,6}$, where for $\tilde{u}_{3,5},\tilde{u}_{3,6}$ we need the extension of the curves into the bigger surface. Then we find that 
\begin{equation}
\begin{aligned}
 \tilde{u}_{1,3} =  &\frac{y_{1,3}}{1 + y_{1,3}}, \quad  \tilde{u}_{1,4} = \frac{y_{1,4} (1 + y_{1,3})}{1 + y_{1,4}(1 + y_{1,3})}, \quad \tilde{u}_{2,4} = \frac{1 + y_{1,4}(1 + y_{1,3})}{(1 + y_{1,3})(1 + y_{1,4})}, \\
& \quad \quad \quad \tilde{u}_{2,5} = \frac{1+y_{1,4}}{1 + y_{1,4}(1 + y_{1,3})} , \quad \tilde{u}_{3,5} = \frac{1}{1 + y_{1,4}} .
\end{aligned}
\end{equation}
These are precisely the $u$ variables associated with the 5 point subgraph, and in particular, they only depend on the $y$ variables for the edges contained inside the 5pt subgraph; the dependence on $y_{1,5}$ present in $u_{2,5}, u_{2,6}$ disappears in the product $\tilde{u}_{2,5} = u_{2,5} u_{2,6}$, similarly for $\tilde{u}_{3,5} = u_{3,5} u_{3,6}$. This is of course fundamental to the  origin of the split factorization we are describing in this note. 

We can also use this example to illustrate ``generalized'' $u$ equations. We now choose a 4 pt subsurface bounded by $(1,3,4,5)$, with the property that 
the both it's ``diagonal'' curves $\tilde{u}_{1,4},\tilde{u}_{3,5}$ have non-trivial extensions into the full surface as shown in the third panel of figure \ref{fig:6ptext}. The $u$ variables for the 4-pt subgraph are the product of the two extensions for each of them, $\tilde{u}_{1,4} = u_{1,4} u_{2,4}$ and $\tilde{u}_{3,5} = u_{3,5} u_{3,6}$, so the $u$ equation for the this 4 pt subsurface tells us that 
\begin{equation}
\tilde{u}_{1,4} + \tilde{u}_{3,5} = 1 \to u_{1,4} u_{2,4} + u_{3,5} u_{3,6} = 1
\end{equation}
which is a ``generalized'' $u$ equation, where the two non-trivial monomials in $u$ sum to one. Note again that $\tilde{u}_{1,4} = u_{1,4} u_{2,4} = \frac{y_{1,4}}{1 + y_{1,4}}, \tilde{u}_{3,5} = u_{3,5} u_{3,6} = \frac{1}{1 + y_{1,4}}$ are exactly the $u$ variables for the 4 point problem depending only on the variable $y_{1,4}$ of our 4 point subsurface. 

\section{Case 3. one-loop split, general split kinematics}
\label{sec:TreeTreeSplit}

In this appendix we build explicitly the split kinematics, $X_{\mathcal{S}}\to X_{\mathcal{S}_1 \otimes \mathcal{S}_2}$, for which the n-point one-loop surface integrand splits into the product of two tree-level amplitudes. This type of split is achieved for the case in which the two subsurfaces, $\mathcal{S}_{1,2}$, are tree-level surfaces and overlap in two different triangles. Recalling the notation introduced in \ref{sec:TreeTreeSplit1loop} let's define our subsurfaces $\mathcal{S}_1$ and $\mathcal{S}_2$ to be:
\begin{equation*}
    \mathcal{S}_1 = (i, j, p,k,m,m+1\cdots, i-1), \quad \mathcal{S}_2 = (i, i+1, \cdots, j, j+1, \cdots, k, k+1, \cdots, m, p),
\end{equation*}
which overlap on triangles $\tau_1 =(p,i,j)$ and $\tau_2 = (p,k,m)$, where $p$ is labeling the puncture. It is natural to define the following four different subsets of indices:
\begin{equation}
\begin{aligned}
    &A=(i, i+1, \cdots, j-1), \\
    &B=(j,j+1, \cdots, k-1), 
\end{aligned} \quad \quad 
\begin{aligned}
    &C=(k, k+1, \cdots, m-1), \\
    &D=(m,m+1, \cdots, i-1). 
\end{aligned}
\end{equation}

Let's use $x_{i,j}/y_{i,j}$ to label the kinematics of $\mathcal{S}_1/\mathcal{S}_2$. Following the usual strategy we start by identify the curves on $\mathcal{S}$ that the curves on the smaller surfaces extend to, $i.e. $ the curves in the big surface that are decomposed into a single component in the split problem. In trying to do this for curves such as $x_{j,d},x_{j,k}$, we immediately run into the fact that we need the curves that spiral around the puncture clockwise (and not just those spiraling counter-clockwise). As explained previously, this is a consequence of the fact that in the tree-level problems the integration domain in $y$-space ranges from $0$ to $+\infty$, for $all$ $y$s. As explained in \cite{curveint}, to define the loop-integrand we consider all curves with a single type of spiraling and integrate over the region where the $u$-variables associated with these curves converge: for the case of counter-clockwise spiraling, this happens for $0<\prod_{i}y_{i,p}<1$, while the clockwise spiraling ones converge in the rest of the space, $i.e.$ $1<\prod_{i}y_{i,p}<\infty$. Therefore if we want the final object to reproduce the full integration domain of tree-surfaces, giving us the correct tree-level amplitudes, we need to sum over two loop integrands, one in which we include the counter-clockwise spiraling, integrated over the appropriate region, and one including the clockwise spirals, so that the sum is indeed covering the full $y$-space.

Bearing this in mind we obtain:
\begin{equation}
    \begin{aligned}
        &x_{k,d} = X_{m-1,d}, \\
        &x_{p,d} = X_{d,p}^{+}, \\
        &x_{d,i} = X_{d,i}, \\
        &x_{j,d} = X_{d,p}^{-},   
    \end{aligned} \quad \quad 
    \begin{aligned}
        &x_{i,p} = X_{i,p}^{+}, \\
        &x_{i,k} = X_{m-1,i}, \\
        &x_{j,k} = X_{m-1,p}^{-},\\  
    \end{aligned} 
\end{equation}
and as for the curves decomposing into single curves of $\Stwo$ we get:
\begin{equation}
    \begin{aligned}
        &y_{a,p} = X_{i-1,a}, \\
        &y_{a,c} = X_{a,c}, \\
        &y_{a,m} = X_{a,p}^{-}, \\
    \end{aligned} \quad \quad 
    \begin{aligned}
        &y_{a,b} = X_{a,b}, \\
        &y_{b,p} = X_{b,p}^{+}, \\
        &y_{b,c} = X_{b,c}, \\
    \end{aligned} 
     \quad \quad 
    \begin{aligned}
        &y_{b,m} = X_{b,p}^{-}, \\
        &y_{c,p} = X_{c,p}^{+}, \\
        &y_{c,m} = X_{c,m}, \\
    \end{aligned} 
\end{equation}

Just as in the previous cases, we now use these definitions to derive the split kinematic locus:
\begin{equation}
    \begingroup
    \addtolength{\jot}{0.8em}
    \begin{aligned}
        &X_{a,d} = x_{d,p} + y_{a,m}\\
        & X_{d,a} = x_{d,i} + y_{a,p} \\
        & X_{a,b} = y_{a,m} + y_{p,b} \\
        & X_{b,a} = x_{i,p}+y_{a,p} + y_{b,m} \\
        & X_{a,c} = y_{a,m} + y_{p,c} \\
        & X_{c,a} = x_{i,k}+y_{a,p} + y_{c,m} \\
        & X_{a,p}^{+} = x_{i,p}+ y_{a,p} \\
        & X_{a_1,a_2} = x_{i,p}+ y_{a_2,p} + y_{a_1,m} \\
        & X_{a_2,a_1} = x_{i,p}+ y_{a_1,p} + y_{a_2,m} \\
        & X_{b,c} = y_{b,m} + y_{c,p} \\
        & X_{c,b} = x_{k,j}+ y_{c,m} + y_{b,p} \\
        & X_{b,d} = y_{b,m} + y_{d,p} \\
        & X_{d,b} = x_{d,j} + y_{b,p}\\ 
         & X_{b_1,b_2} = y_{b_1,m} + y_{b_2,p} \\
        & X_{b_2,b_1} = y_{b_2,m} + y_{b_1,p} 
    \end{aligned} 
    \endgroup
    \begingroup
    \addtolength{\jot}{0.7em}
    \begin{aligned}
        & \quad \quad \to \quad X_{a,d} - X_{d,p}^{+} - X_{a,p}^{-}= 0\\
        & \quad \quad \to \quad X_{d,a} - X_{d,i} - X_{i-1,a}=0 \\
        & \quad \quad \to \quad X_{a,b} - X_{a,p}^{-} -X_{b,p}^{+}=0 \\
        & \quad \quad \to \quad X_{b,a} - X_{i,p}^{+}-X_{i-1,a} - X_{b,p}^{-}=0 \\
        & \quad \quad \to \quad X_{a,c} - X_{a,p}^{-} + X_{c,p}^{+} \\
        & \quad \quad \to \quad X_{c,a} - X_{m-1,i}-X_{c,m} - X_{i-1,a}=0 \\
        & \quad \quad \to \quad X_{a,p}^{+} - X_{i,p}^{+} -  X_{i-1,a}=0 \\
        & \quad \quad \to \quad X_{a_1,a_2} - X_{i,p}^{+}- X_{i-1,a_2} -X_{a_1,p}^{-} =0 \\
        & \quad \quad \to \quad X_{a_2,a_1} - X_{i,p}^{+}- X_{i-1,a_1} -X_{a_2,p}^{-} =0 \\
        & \quad \quad \to \quad X_{b,c} - X_{b,p}^{-} + X_{c,p}^{+}=0  \\
        & \quad \quad \to \quad X_{c,b} - X_{c,m} - X_{b,p}^{+}  -X_{m-1,p}^{+}=0 \\
        & \quad \quad \to \quad X_{b,d} - X_{b,p}^{-}  -X_{d,p}^{+}=0  \\
        & \quad \quad \to \quad X_{d,b} -X_{b,p}^{+}  -X_{d,p}^{-}=0 \\
         & \quad \quad \to \quad X_{b_1,b_2} -X_{b_1,p}^{-} - X_{b_2,p}^{+}=0\\
        & \quad \quad \to \quad X_{b_2,b_1} -X_{b_2,p}^{-} - X_{b_1,p}^{+}=0 
    \end{aligned}
    \endgroup
\end{equation}

\begin{equation}
    \begingroup
    \addtolength{\jot}{0.8em}
    \begin{aligned}
        & X_{c,d} = y_{c,m} + x_{k,d} \\
        & X_{d,c} = x_{j,d}+ y_{c,p} \\
        & X_{c,p}^{-} = x_{k,j}+ y_{c,m} \\
        & X_{c_1,c_2} = y_{c_1,m}+y_{c_2,p} + x_{j,k} \\
        & X_{c_2,c_1} = y_{c_2,m}+y_{c_1,p} + x_{j,k} \\
        & X_{d_1,d_2} =  x_{d_1,j} + x_{d_2,p} \\
        & X_{d_2,d_1} = x_{d_2,j} + x_{d_1,p} \\  
    \end{aligned} 
    \endgroup
    \begingroup
    \addtolength{\jot}{0.7em}
    \begin{aligned}
        & \quad \quad \to \quad X_{c,d} -X_{c,m} -X_{m-1,d} =0 \\
        & \quad \quad \to \quad X_{d,c} -X_{c,p}^{+} -X_{d,p}^{-}=0 \\
        & \quad \quad \to \quad X_{c,p}^{-} -X_{c,m}- X_{m-1,p}^{-} =0 \\
        & \quad \quad \to \quad X_{c_1,c_2} - X_{c_1,m} - X_{c_2,p}^{+}- X_{m-1,p}^{-}=0 \\
        & \quad \quad \to \quad X_{c_2,c_1} - X_{c_2,m} - X_{c_1,p}^{+}- X_{m-1,p}^{-}=0 \\
        & \quad \quad \to \quad X_{d_1,d_2}- X_{d_1,p}^{-} - X_{d_2,p}^{+} =0 \\
        & \quad \quad \to \quad X_{d_2,d_1} - X_{d_2,p}^{-} - X_{d_1,p}^{+} =0 \\  
    \end{aligned}
    \endgroup
\end{equation}

Finally, the mapping for the self-intersecting curves is not simply obtained by, $X^{(q)} \to x+y^{(q)}$, because neither subsurface as self-intesecting curves. Still for most the curves the mapping of $X^{(q)}$ precisely agrees with the mapping of $X$ presented above. The only exception are now for the self-intersecting versions of curves $X_{d,a}$,  $X_{c,a}$ and $X_{c,d}$, for which we get:
\begin{equation}
\begin{aligned}
&X_{d,a}^{(q)} \to y_{p,a} + x_{d,j}+x_{i,p},  \\
& X_{c,a}^{(q)} \to x_{i,p} + x_{j,k} + y_{c,m} + y_{a,p}, \\
& X_{c,d}^{(q)} \to  y_{c,m} + x_{k,j} +x_{p,d},
\end{aligned}
\end{equation}
with $q\neq0$. 

So in this case we have the following split pattern:
\begin{equation}
\left[\mathcal{I}_{\mathcal{S}}(\{X_{i,j},X_{i,p}^+\} \to X_{\mathcal{S}_1\otimes\mathcal{S}_2}) +  \mathcal{I}_{\mathcal{S}}(\{X_{i,j},X_{i,p}^-\} \to X_{\mathcal{S}_1\otimes\mathcal{S}_2})\right] \to \mathcal{A}_{\mathcal{S}_1}(y_{i,j}) \times \mathcal{A}_{\mathcal{S}_2}(x_{i,j}) ,
\end{equation}
where in the first term we have the curves spiraling anti-clockwise $(+)$ while in the second one those spiraling clockwise $(-)$.

\section{Non-planar splits for the infinite integrand -- the annulus}
\label{sec:annulusSplit}
\begin{figure}[t]
    \centering
    \includegraphics[width=\textwidth]{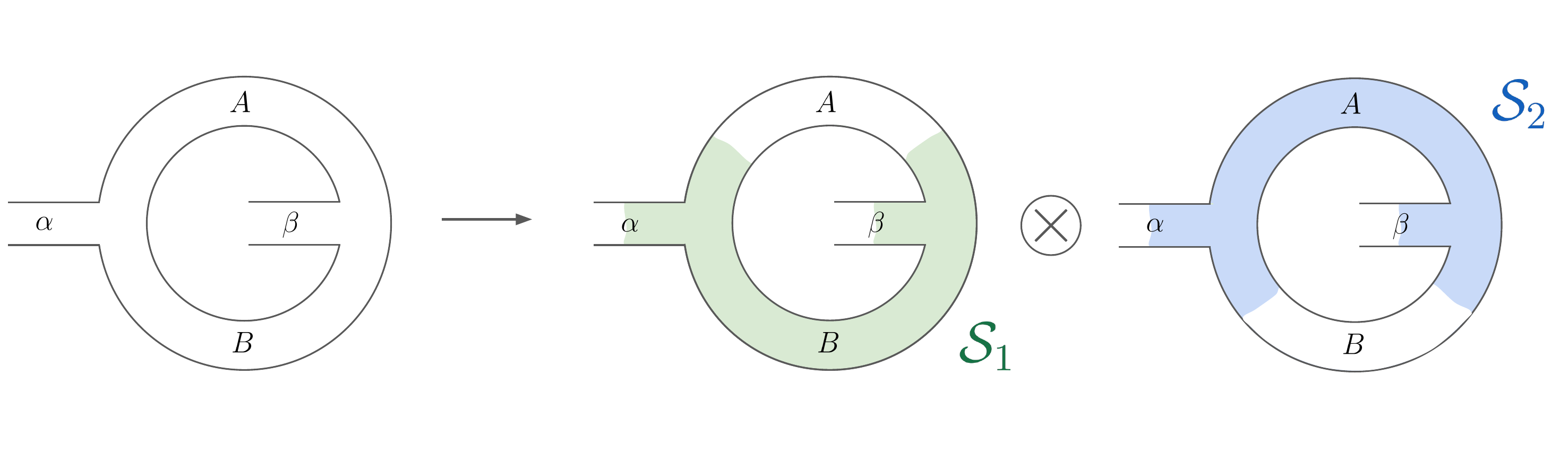}
    \caption{A one-loop annulus fatgraph represented as the join of two four-point trees.}
    \label{fig:annulus}
\end{figure}

Here we give an example of splits for a non-planar integrand, focusing on the very simplest case of the annulus with a single marke points on each boundary. The representative fat graph is shown in figure \ref{fig:annulus}, with external edges $\alpha, \beta$ and internal edges $A,B$.  

Consider the obvious choices for ${\cal S}_1, {\cal S}_2$ each four-point tree-surfaces, with internal roads $A,B$ respectively. We will also be interested in looking at the field theory limit so we will only consider the non-self-intersecting curves on ${\cal S}$ that start on $\alpha$ and end on $\beta$.  These are labeled by $X_w$, where $w$ is an integer that specifies the number of times the curve winds in going from one boundary to the other. We define the kinematics for the curves of the four-point problem to be $(x, \tilde x)$ and $(y, \tilde y)$ respectively, as shown in figure \ref{fig:annulus}. Then we can easily work out the split kinematics; this is a case where the $\#[x \subset X]$ factors are interesting and grow with $w$. We find that 
\begin{equation}
\begin{aligned}
& X_0 \to x,\\  
&X_{-1} \to \tilde{x},
\end{aligned}\quad \quad 
\begin{aligned}
&X_{1} \to \tilde{y}, \\
&X_{-2} \to y, 
\end{aligned}
\quad \quad 
\begin{aligned}
&X_{+j} \to \tilde{y} + j(y + \tilde y),\\
&X_{-(j+1)} \to y + j (\tilde{y} + y).
\end{aligned}
\end{equation}

Note that trivially there is no action on the mapping class group on the two four-point subsurfaces. The split kinematics we have defined is therefore {\it not} invariant under the MCG of the annulus, which would identify all the $X_w$. Of course, this is as must be, because with this kinematics the full integral is guaranteed to converge and indeed factorize into the product of the two four-point factors, without any need for modding out by the MCG. The ``infinite integrand'' for the annulus is trivial, since we have only one type of diagram/triangulation, and we have 
\begin{equation}
{\cal A}= \sum_{w=-\infty}^{+\infty} \frac{1}{X_w X_{w+1}}.
\end{equation}

We see again that if the kinematics is invariant under the MCG we get an infinite factor--coming from counting triangulations differing by winding an infinite number of times--which usually necessitates modding out by the MCG. But instead for our kinematics, the sum will be perfectly convergent. Indeed on the split kinematics, we have 
\begin{equation}
\begin{aligned}
&\frac{1}{X_0 X_1} + \frac{1}{X_1 X_2} + \cdots = \frac{1}{x \tilde{y}} + \sum_{j=0}^\infty \frac{1}{(\tilde{y} + j (y + \tilde{y}))(\tilde{y} + (j+1)(y + \tilde{y})}, \\ 
&\frac{1}{X_{-1} X_0} + \frac{1}{X_{-2} X_{-1}} + \cdots = \frac{1}{x \tilde{x}} + \frac{1}{y \tilde{x}} + \sum_{j=0}^\infty \frac{1}{(y + j(\tilde{y} + y)) (y + (j+1) (\tilde{y} + y)}.
\end{aligned}
\end{equation}

Now the infinite sum can trivially be performed telescopically 
\begin{equation}
\begin{aligned}
\sum_{j=0}^\infty \frac{1}{(\tilde{y} + j (y + \tilde{y}))(\tilde{y} + (j+1)(y + \tilde{y})} &= \frac{1}{y + \tilde y} \sum_j \left(\frac{1}{(\tilde{y} + j (y + \tilde{y}))} - \frac{1}{(\tilde{y} + (j+1) (y + \tilde{y}))} \right) \\
&= \frac{1}{y + \tilde y} \frac{1}{\tilde{y}},
\end{aligned}
\end{equation}
and similarly for the second sum. Thus we find 
\begin{equation}
\begin{aligned}
 {\cal A} &= \frac{1}{x \tilde x} + \frac{1}{(y + \tilde y)} \frac{1}{\tilde{y}} + \frac{1}{x \tilde{x}} + \frac{1}{y \tilde{x}} + \frac{1}{y + \tilde{y}} \frac{1}{y} \\ 
& = \frac{1}{x \tilde{y}} + \frac{1}{x \tilde{x}} + \frac{1}{y \tilde{x}} + \frac{1}{y \tilde y} = \left(\frac{1}{x} + \frac{1}{\tilde{x}}\right) \times \left( \frac{1}{y} + \frac{1}{\tilde{y}} \right),
\end{aligned}
\end{equation}
which factorizes just as it should.

\bibliographystyle{JHEP}\bibliography{Refs}

\end{document}